\documentclass[mnsc]{Support/informs3mod}

\usepackage{blindtext}
\usepackage{multicol}
\OneAndAHalfSpacedXI
\usepackage{natbib}
\usepackage{color}
\usepackage{amsmath}
\usepackage{amssymb}
\usepackage[utf8]{inputenc}
\usepackage{amsfonts}
\usepackage[normalem]{ulem}
 \bibpunct[, ]{(}{)}{,}{a}{}{,}%
 \def\bibfont{\small}%
\TheoremsNumberedThrough
\ECRepeatTheorems
\usepackage{natbib}

\EquationsNumberedThrough 

\def \qedsymbol{\hbox{ }\hfill$\Box$}

\usepackage{comment}
\excludecomment{ignore}

\usepackage{xcolor}
\usepackage{soul}

\usepackage{hyperref}
\newcommand{\tps}[1]{\texorpdfstring{#1}.}

\begin{document}

\TITLE{Price Experimentation and Interference}

\ABSTRACT{In this paper, we examine the biases that arise when firms run A/B tests on continuous parameters to estimate global treatment effects on performance metrics of interest; we particularly focus on {\em price experiments} to measure the price impact on quantity demanded, and on profit. In canonical A/B experimental estimators, biases emerge due to interference between market participants.  We employ structural modeling and differential calculus to derive intuitive characterizations of these biases.  We then specialize our general model to the standard revenue-management pricing problem. This setting highlights a fundamental risk innate to A/B pricing experiments: that the canonical estimator for the expected change in profits, counterintuitively, can have the {\em wrong sign} in expectation. In other words, following the guidance of canonical estimators may lead firms to move prices (or fees) in the wrong direction, inadvertently decreasing profits. We introduce a novel debiasing technique for these canonical experiments, requiring only that firms equally split units between treatment and control. We apply these results to a two-sided market model, and demonstrate how the ``change of sign" regime depends on market factors such as the supply/demand imbalance, and the price markup. We conclude by calibrating our revenue-management pricing model to published empirical estimates from Airbnb marketplaces, demonstrating that estimators with the wrong sign are not a knife-edge issue, and that they may be prevalent enough to be of concern to practitioners.\footnote{We would like to thank Wassim Dhaouadi for his valuable input during the beginning stages of the project, and Andrey Fradkin for his significant assistance on our Airbnb empirical case study. We are also grateful to Eric Aldrich, Alex Deng, Thu Le, Ruben Lobel, Ali Rauh, Dan Schmierer, and seminar participants at Boston University, University of Chile, the MIT Conference on Digital Experimentation, the INFORMS RMP Conference, the INFORMS Annual Meeting, the INFORMS MSOM Conference, and Airbnb for helpful feedback.  Orrie Page was supported by a National Science Foundation Graduate Research Fellowship.}}

\ARTICLEAUTHORS{

\AUTHOR{Ramesh Johari}
\AFF{Management Science and Engineering, 
Stanford University,
\EMAIL{rjohari@stanford.edu}}

\AUTHOR{Orrie B. Page}
\AFF{Management Science and Engineering, 
Stanford University,
\EMAIL{orrie@stanford.edu}}

\AUTHOR{Gabriel Y. Weintraub}
\AFF{Graduate School of Business, 
Stanford University,
\EMAIL{gweintra@stanford.edu}} 

}

\maketitle

\section{Introduction}

In this work, we analyze the biases that arise when a firm or platform conducts an A/B test by varying a continuous parameter, in order to determine the global treatment effect associated with a performance metric of interest. These biases appear when using canonical experimental designs and estimators, and are the result of interference between market participants.

More concretely, consider the example of a typical pricing experiment, commonly used by online platforms, where a fraction of the products are assigned a treatment price that is higher than the baseline (control) price.   The {\em global treatment effect} of this higher price compares two counterfactual universes: one where all products have the high price (global treatment), and one where all products have the low price (global control).  Observe that in the experiment, some consumers who would have purchased at the high price in global treatment instead purchase control units at the low price. When using canonical estimators (such as the difference-in-means estimator), this substitution within the experiment biases the estimation of the global treatment effect. This phenomenon is an {\em interference} effect
capturing the ``cannibalization" of treatment purchases by their control counterparts; see Related Work (Section \ref{sec:related}) for background.

Our paper considers the following question: Given the standard experimental designs (e.g., listing-side randomization, LR, or customer-side randomization, CR) and estimators (i.e., difference-in-means) that are employed by platforms, what are the consequences of interference bias in pricing experiments?  There are several key features of our approach. First, drawing inspiration from recent work in the operations literature, we employ structural modeling to gain valuable insights into biases. Furthermore, our analysis focuses on differential changes in continuous variables such as prices, enabling us to leverage calculus to derive straightforward and intuitive expressions. (In practical settings, this corresponds to  ``small'' changes around the control price.) Finally, unlike many preceding studies on online platform experimentation, we focus on {\em profits} as an outcome metric, in addition to {\em bookings} or transaction volume.

We first introduce in Section \ref{sec: general} a succinct yet general model of platform experimentation in which a firm wants to understand the impact of modifications to a continuous variable, on a metric of interest. The model is general, in the sense that it can represent various market arrangements and experimental designs. By leveraging differential calculus, we provide several structural results regarding the performance of canonical estimators and designs. Notably, we show that the bias corresponds to the cross-effects of the control and treatment group metrics, with respect to marginal changes in the other group's variable. These cross-effects are precisely an indicator of interference, providing an intuitive structural characterization of bias.  

Second, in Sections \ref{sec: Revenue Management Setting} and \ref{sec: bias_analysis}, we specialize our results to a canonical revenue management pricing problem. 
By using our bias characterizations, we show that the bias can be severe enough that experimenters can obtain the {\em wrong sign} when the metric of interest is profit (or revenue). In particular, we show that when treatment involves raising prices (resp., lowering prices), the global treatment effect may be positive (resp., negative), but due to interference, the estimator can be negative (resp., positive) in expectation. Interestingly, when treatment involves raising prices (resp., lowering prices), and the global treatment effect is negative (resp., positive), the experimental estimate always obtains the correct sign. We express the conditions under which an estimate with the wrong sign can be obtained with straightforward expressions that involve price elasticities.\footnote{The expressions are reminiscent of elasticity conditions in \cite{blake2014marketplace}, though their analysis differs from ours: they focus on estimating the relative bias in terms of supply and demand elasticities, and carry out a local change with respect to a multiplicative factor by which treatment demand changes relative to control.  By contrast, our estimand is local with respect to the magnitude of the price change, and of course, a significant emphasis of our study is the possibility of mismatched signs between the true $GTE$ and the estimated $GTE$, which they do not study.}  In the appendix, we also show that these results generalize to a setting where a platform with heterogeneous goods for sale sets a fixed {\em fee} as a percentage of overall transaction volume, as is common in online marketplaces.  Our results have important implications for price optimization based on experiments: following the suggestion of a canonical experiment may lead one to move in the wrong direction and {\em decrease} profits relative to the status-quo!

Taking advantage of our structural characterization of bias, in Section \ref{sec: debiasing}, we construct a novel debiasing method for canonical A/B experiments.  In particular, we show that if the experimenter equally allocates units between treatment and control, it is possible to fully debias the canonical estimator by taking advantage of symmetry in the structure of bias.  Notably, this allows firms to run the most simple version of an A/B pricing experiment, while still emerging with an unbiased estimator (under a set of assumptions that me make clear below). This is a contrast to other methods for reducing bias in A/B experiments such as clustered or switchback experiments, which require more complicated experimental designs and estimators \citep{holtz2025reducing,chamandy16,bojinov2022design, glynn2020adaptive,hu2022switchback}. 

To obtain additional insight, in Section \ref{sec: logit}, we apply the results of the revenue management setting to the two-sided market model of \cite{johari2022experimental}. We show how the biases of LR and CR experiments depend on model parameters, with a particular focus on regimes in which a ``change of sign" may occur. For example, for an LR experiment, a platform is more likely to obtain the incorrect sign in an experiment if the demand is low relative to supply.  We also investigate numerically how bias depends on both market balance and the baseline control price, and show  how canonical A/B methods for price experimentation may result in significant profit underperformance.

In Section \ref{sec: empirics}, we conclude the paper by calibrating our revenue-management pricing model to empirical estimates from \citet{Fradkin2022} for Airbnb markets in major U.S. cities. This exercise is useful for offering insights into the plausibility of estimators with the wrong sign in real-world markets. In our investigation, we find that such estimators are quite plausible in the Airbnb context. Notably, all markets in the \citet{Fradkin2022} study demonstrated a susceptibility to estimators with the wrong sign in LR experimentation, which would be the typical approach for testing pricing changes in practice. These results suggest that changes of sign are prevalent enough to be consequential and of concern for practitioners. We conclude in Section \ref{sec: conclusion}.


\section{Related Work}
\label{sec:related}

The phenomenon we study is an {\em interference} effect capturing the ``cannibalization" of treatment purchases by their control counterparts. More generally, this interference effect is a violation of the Stable Unit Treatment Value Assumption (SUTVA) \citep{ImbensRubin15}. Interference among users has been widely studied in social networks  \citep[see, e.g.]{Manski13,Athey18, Basse19, Saveski17} and more recently on two-sided platforms \citep{blake2014marketplace, fradkin2019simulation, liu2020trustworthy}, showing its importance in practice. Platforms have developed alternative designs and/or estimators to alleviate these biases such as cluster randomization \citep{Ugander13,chamandy16,holtz2025reducing,candogan2024correlated}, switchback testing \citep{chamandy16,bojinov2022design, glynn2020adaptive,hu2022switchback}, two-sided randomization and bipartite experiments \citep{bajari2023multiple,johari2022experimental,pouget2019variance}, shadow-price-based methods \citep{lobel2023}, Q-learning \citep{farias2022markovian,farias2023correcting}, rollout experiments \cite{han2023detecting}, and message passing \citep{shirani2023causal}.

A number of other papers prior to ours have considered price experiments specifically, including \cite{wager2021experimenting, munro2021treatment, li2023experimenting, aparicio2023algorithmic, roemheld2023interference, deng2023price, simchi2023pricing, DelarePricing2025}.  Some are less related to our work; for example, \cite{simchi2023pricing} focuses on designing price experiments to mitigate the tail risk of revenue loss from experimentation, or \cite{DelarePricing2025}, which proposes a method for reducing interference bias in matching markets. More related to our paper, both \cite{wager2021experimenting} and \cite{li2023experimenting} consider estimation of treatment effects from marginal price changes; in \cite{wager2021experimenting}, this is studied via a mean-field limit approach that admits a debiasing technique to optimize supply-side payments via gradient descent in a centralized marketplace; while in \cite{li2023experimenting}, the emphasis is on the estimation of such effects via model-free and model-based estimators in switchback experiments with stochastic congestion (i.e., due to queueing effects).  \cite{roemheld2023interference} uses an empirical study to suggest that false positives may be inflated due to interference in pricing experiments. 

More similar to our findings, \cite{deng2023price} also notes the possibility of the sign of the treatment effect estimator being incorrect; in their setting this is due to an effect they call {\em lead-day bias} in markets with expiring goods.  In an auction design setting, \cite{Chawla2016} points out that A/B testing different auction formats can also lead to the wrong sign.  On its part, \cite{LiWangWang2025} study A/B experiments comparing recommendation algorithms and shows how symbiosis bias—arising from shared training data between treatment and control—can result in flipping the sign of the estimated global treatment effect.  Finally, more broadly, we note that \cite{carneiro2010evaluating} refers to the kinds of treatment effects we consider in this paper---those where a continuous variable (such as price) is incrementally varied---as {\em marginal policy changes}, and develops a framework to study them.  

The closest of these works to ours is \cite{munro2021treatment}, which studies general estimands in an economic price equilibrium setting, and decomposes the treatment effect into direct and indirect effects; the former is obtained via ``canonical'' estimators, while the latter corresponds to our bias, and the authors show how it can be corrected using local price experiments.  In that work, the direct and indirect effects are obtained as marginal changes with respect to the treatment assignment probability (inspired by work in \cite{hu2022average}), rather than the price as in our present paper.  Further, our emphasis in applying our results is on structural insight regarding when changes of sign may
be obtained for objectives such as profits or revenue, while their paper is focused on design and estimation for general estimands in market equilibrium. (An earlier version of \cite{munro2021treatment} considers treatment changes in a continuous variable, with an emphasis on relating the resulting differential treatment effect to direct and indirect effects.)


\section{General Model Formulation} \label{sec: general}

This section presents our general model, which we subsequently adapt to specific settings.

\subsection{Market Model and Estimand}

We consider a market in which a firm or platform sets a single variable of interest, $x\in\mathbb{R}_+$, evaluated by a performance metric $T(x)\in\mathbb{R}, \forall  x\in\mathbb{R}_+$.  The performance metric can describe bookings, profits, or revenues of the platform. We make the following assumption that is kept throughout the paper. 
\begin{assumption} \label{as:metric}
The function $T(x)$ is differentiable for all $x\in\mathbb{R}_+$.
\end{assumption}

The firm is interested in optimizing the performance metric by adjusting the decision variable $x$. In particular, the firm is interested in the change in the target metric, when the decision variable takes on a new ``treatment'' value, denoted by $x_1$,
moving away from the status-quo ``control'' value, denoted by $x_0$. This difference, $T(x_1)-T(x_0)$, corresponds to the \emph{global treatment effect ($GTE$)}. 
We define a local version of the $GTE$ as the derivative of the target metric, representing the case in which the treatment value of the decision variable is ``close'' to the control value:
\begin{align} \label{eq: GTE}
GTE(x_0) &:= T'(x_0)  \ .
\end{align}
$GTE(x_0)$ is our estimand.

\subsection{Experiment: Design, Model, and Estimator}

Our work focuses on settings where the function $T$ may be unknown to the firm or platform. Therefore, an experiment is conducted to determine the direction in which the decision variable should be adjusted in order to improve the performance metric (a form of policy improvement). We assume that the platform conducts an experiment with a single treatment condition where a fraction $q$ of the units are randomized to treatment and a fraction $1-q$ to control.  We carry out our analysis in a formal ``mean field" or continuum regime, where a continuum mass of units are allocated to each group; this allows us to focus on the magnitude and sign of bias (see also \cite{johari2022experimental, wager2021experimenting} for related works using a similar approach).  
 
 It is worth noting that in the settings we are considering, randomization can occur on either side of the market, i.e., customers or products (or listings). In a customer-side randomization experiment (CR), a treatment (resp., control) customer is exposed to the entire market under the treatment (resp., control) condition. On the other hand, in a listing-side randomization experiment (LR), a customer is exposed to a mix of treatment and control products (or listings). As before, we denote by $x_0$ the baseline (control) value of the decision variable and by $x_1$ the treatment value. 
 
We denote the in-experiment performance metrics for control and treatment goods by the functions $T_0(x_0,x_1,q)$ and $T_1(x_0,x_1,q)$ respectively. The function $T_{0}(x_0,x_1,q)$  (resp., $T_1(x_0,x_1,q)$) corresponds to the performance metric in the control (resp., treatment) group  when the control and treatment values are given by $x_0$ and $x_1$, respectively, and a fraction $q$ of units are randomized into treatment. The nature of these functions will become more clear in the context of the  applications described  below.  Notably, this notation abstracts away from the specific experimental design (whether CR or LR).\footnote{At this level of abstraction, even other experimental designs like switchbacks could be analyzed.}
We  make the following assumption that is kept throughout the paper:
\begin{assumption} \label{as:diff_metric}
The functions $T_0(x_0,x_1,q)$, and $T_1(x_0,x_1,q)$ are differentiable for all $x_0,x_1\in \mathbb{R}_+$ and $0<q<1$. 
\end{assumption}

The canonical experimental designs at each side of the market have an associated natural ``canonical" local estimator. To define this estimator, consider a sequence of experiments for which the treatment value of the decision variable approaches the control value. Then, we compute the relative difference between the performance metrics appropriately normalized by the treatment and control fractions, respectively, which corresponds to the Horvitz-Thompson estimator, or equivalently in our formal mean-field setting, the difference-in-means estimator. More formally, we consider a sequence of experiments indexed by $n\in \mathbb{Z} _+$, in which the treatment value approaches the control value: $\lim_{n\to\infty }x_1^n=x_0$. We define a local version of the canonical estimator as:
\begin{align}
\widehat{GTE}(x_0,q) &:= \lim_{n\to\infty}  \frac{\frac{T_1(x_0,x_1^n,q)}{q}-\frac{T_0(x_0,x_1^n,q)}{(1-q)}}{x_1^n-x_0}  \ . \label{eq: EstGTEL}
\end{align}

We make the following assumption that is kept throughout the paper:

\begin{assumption} 
\label{as:regular_treat} 
 $T_0(x,x, q)/(1-q)=T_1(x,x, q)/q=T(x)$, for all $x \in \mathbb{R}_+$ and $0<q <1$.
\end{assumption}

The assumption states that if the values of control and treatment decision variables are the same, then both the scaled control and treatment performance metrics correspond to the market-wide metric. The assumption will play a key role in our analysis. To further understand it, consider $T_1$ with an extended domain $0<q\leq 1$ (the argument for $T_0$ is symmetric). Now, suppose $T_1$ is linear in its third argument over the relevant domain, that is, for any $0<q\leq 1$ and $0<\alpha< 1$, $T_1(x,x,\alpha q)=\alpha T_1(x,x,q)$. If we take $q=1$, then $T_1(x,x,\alpha)/\alpha= T_1(x,x,1)=T(x)$, where the second equation follows directly by definition. Hence, the assumption is implied by constant (i.e., linear) returns of the performance metric in the fraction randomized. Note that in practice, experimental estimates typically implicitly rely on the assumption of constant returns in order to extrapolate estimates for control and treatment groups to the entire market. For the marketplace models and metrics we study later, this assumption will be satisfied.

Now, we obtain:
\begin{align}
\widehat{GTE} (x_0,q)
&= \lim_{n\to\infty}  \frac{T_1(x_0,x_1^n,q)/q-T_1(x_0,x_0,q)/q+T_1(x_0,x_0,q)/q-T_0(x_0,x_1^n,q)/(1-q)}{x_1^n-x_0}  \notag \\
&= \lim_{n\to\infty}  \frac{T_1(x_0,x_1^n,q)/q-T_1(x_0,x_0,q)/q+T_0(x_0,x_0,q)/(1-q)-T_0(x_0,x_1^n,q)/(1-q)}{x_1^n-x_0}  \notag \\
&=T_{1y}(x_0,x_0,q)/q - T_{0y}(x_0,x_0,q)/(1-q) \ .\label{eq:expr_est}
\end{align} 

The first equation is direct from the definition of $\widehat{GTE} (x_0,q)$, the second follows by Assumption \ref{as:regular_treat}, and the third by Assumption \ref{as:diff_metric}.
The subscript $x$ (resp., $y$) denotes the partial derivative with respect to the first (resp., second) argument of the function. 

Finally, we define the estimation bias of this local estimator as:\footnote{Note that in statistics, bias is often defined as the estimator minus the estimand.  However, we make the preceding choice because (as we will see) the resulting sign of $Bias$ is typically nonnegative, making it easier to obtain practical interpretations of our results.}

\begin{align}
Bias(x_0,q)&:= GTE(x_0)-\widehat{GTE}(x_0,q) \label{eq:bias1} \\
&= T'(x_0) - \left(T_{1y}(x_0,x_0,q)/q - T_{0y}(x_0,x_0,q)/(1-q)\right) \ .\label{eq:bias2} \ 
\end{align}

Studying bias in this local regime proves fruitful, as it leads to the following simple characterization.
\begin{proposition} \label{pr:bias}
The bias is given by the following expression: $$ Bias(x_0,q)=T_{0y}(x_0,x_0,q)/(1-q)+T_{1x}(x_0,x_0,q)/q \ .$$ 
\end{proposition}
\proof{Proof.}
By Assumption \ref{as:regular_treat}, we have that:
$$ T(x_0)=T_1(x_0,x_0,q)/q \ .$$
By Assumption \ref{as:diff_metric} and application of the chain rule we get:
$$ T'(x_0)=T_{1x}(x_0,x_0,q)/q+T_{1y}(x_0,x_0,q)/q \ .$$
 Replacing in \eqref{eq:bias2}, we obtain the desired result:
 $$ Bias(x_0,q)=T_{0y}(x_0,x_0,q)/(1-q)+T_{1x}(x_0,x_0,q)/q \ .$$ 
\qedsymbol
 \endproof
The result provides an intuitive characterization of bias. Specifically, $T_{0y}(x_0,x_0,q)$ and $T_{1x}(x_0,x_0,q)$ correspond to the relative changes in performance for the control and treatment groups, respectively, when the decision variable of the {\em other} group changes marginally. The cross-partial derivatives represent the effect on the performance metric across experimental groups, making them an intuitive indicator of interference. As expected, as this interference effect increases, bias correspondingly increases.

\section{Revenue Management Setting}

\label{sec: Revenue Management Setting}

In this section we examine a traditional revenue management setting with monopoly pricing. A firm or platform offers products that, at least based on observable characteristics, are deemed homogeneous and are thus priced uniformly. The firm faces an (unknown to them) demand function given by $D(p)$ as a function of the price $p$. The firm has a constant marginal production cost, $c\geq 0$. We make the following assumption that is kept throughout the paper and will imply Assumption \ref{as:metric} for the performance metrics we study.

\begin{assumption} \label{as: demand}
The function $D(p)$ is differentiable and $D'(p)<0$, for all $p\in\mathbb{R}_+$.
\end{assumption}
At this level of abstraction, the demand function can describe a scenario with either an infinite supply or a limited supply, where a rationing mechanism determines the consumption. In the limit of a large market, even with supply constraints, we expect the demand function to be differentiable. 

To provide a concrete example of a limiting regime with infinite supply, the demand function can be understood as the fraction among a large population of customers buying the product at a given price. That is, suppose $x_i(p) \in \{0,1\}, i = 1, 2, \ldots$ are i.i.d.~Bernoulli random variables with mean $0 \leq D(p) \leq 1$; we interpret $x_i(p)$ as the purchasing decision of customer $i$ at price $p$. Then trivially by the law of large numbers $D(p)=\lim_{n\to\infty}\frac{1}{n}\sum_{i=1}^n x_i(p)$.

\subsection{Estimands}

We will consider two performance metrics: (1) demand itself, which serves as a measure of growth, and (2) the firm's profits (or revenue, in case costs are zero). 
The firm's decision variable is price, and it is seeking to gain an understanding of how changes in price impact demand (i.e., price elasticity of demand), and subsequently affect profit. To achieve this goal, the firm is interested in estimating the difference in demand or profits generated when products are priced at a new ``treatment'' price relative to the demand or profits generated when products are priced at the status-quo ``control'' price. These differences correspond to the global treatment effects of interest that we derive below. To do so, we first define the ``global'' profit function as:
$$ \pi(p) := (p-c)  D(p)  \ .$$
Following equation \eqref{as:metric}, we present our estimands.  These are the local versions of the global treatment effects associated to demand and profits: 
\begin{align} 
GTE_{D} (p) &:= D'(p) \label{eq: GTE_D} \ . \\
GTE_{\pi} (p) &:= \pi'(p)  \notag\\
&= D(p)+(p-c) D'(p)  \ .  \label{eq: GTE_Pi} 
\end{align}
Note that $GTE_{\pi}(p)$ is the marginal profit of the firm at price $p$.

\subsection{Experiment: Design, Model, and Estimator}

Given that the demand function is unknown, we assume the firm uses price experimentation to estimate the $GTE$s of interest. As mentioned above, canonical A/B-tests can randomize on either side of the market, and our model does not preclude either possibility. That said, we do note that when conducting price experiments in practice, companies prefer to refrain from randomizing prices on the customer side to adhere to the basic principle that each customer should be offered the same price for the same product \citep{Phil_book}.

We assume a single treatment condition in which a fraction $q$ of the subjects are randomized to treatment and a fraction $1-q$ to control. We denote by $p_0$ the baseline (control) price and by $p_1$ the treatment price. We assume that the experimental demands of control and treated subjects are given by the functions $Q_0(p_0,p_1,q)$ and $Q_1(p_0,p_1,q)$, respectively. We further define the functions:
 \[ D_0(p_0,p_1,q)=\frac{Q_0(p_0,p_1,q)}{1-q}; \quad D_1(p_0,p_1,q)=\frac{Q_1(p_0,p_1,q)}{q}, \]
 which represent the demand for control and treated units scaled by the control and treatment fractions, respectively. Henceforth, we refer to $D_0$ and $D_1$ as the control and treatment demand functions, respectively. Again, at this level of abstraction these demand functions could, for example, incorporate supply constraints and rationing mechanisms.
 
Recall that in a CR experiment treatment and control customers will observe all products priced at the treatment and control prices respectively, with the proportion of customers assigned to the treatment group being $q$. In contrast, in a LR experiment, a customer will face an array of products, a fraction $q$ of which are priced at $p_1$ and the rest at $p_0$. 
 
We  make the following assumption that implies Assumption \ref{as:diff_metric} and
 that is kept throughout the paper. 
 \begin{assumption} \label{as: diff_D}
The functions $D_0(p_0,p_1,q)$ and $D_1(p_0,p_1,q)$ are continuously differentiable 
in their first two arguments $(p_0,p_1)$, for all $p_0,p_1 \in \mathbb{R}_+$ and all 
$q \in (0,1)$. In addition, prices satisfy $p_0,p_1 \geq c$, where $c\geq0$, and we assume 
$0<q<1$.

\end{assumption} 
Note that the assumption precludes the case of perfect substitution  between control and treatment products. Under perfect substitution, a negligible undercut of the treatment relative to the control price would result in a discrete increase of the treatment demand; as a result, the demand function would not be continuous. Even though, as initially stated, the firm is selling a single type of product, we can rationalize the smoothness assumption in at least one of two ways, commonly assumed in pricing settings: (1) products can be identical in terms of observable characteristics but there may be latent horizontal differentiation; (2) there could be search frictions that do not allow customers to always buy the lowest-priced product.

We introduce the following additional assumption on the experimental demand functions which implies Assumption \ref{as:regular_treat}.
\begin{assumption} 
\label{as: control_treat_demand} 
The following statements all hold:
\begin{enumerate}
\item $D_0(p_0,p_1, q)$ is non-decreasing in $p_1$, for all $p_0 \in \mathbb{R}_+$ and  $0<q <1$.

\item $D_1(p_0,p_1, q)$ is non-decreasing in $p_0$, for all $p_1 \in \mathbb{R}_+$ and  $0<q <1$.

\item $D_0(p,p, q)=D_1(p,p, q)=D(p)$, for all $p \in \mathbb{R}_+$ and $0<q <1$.

\end{enumerate}
\end{assumption}

 The first two statements in Assumption \ref{as: control_treat_demand} are natural as they imply that control and treatment products are (imperfect) substitutes. 
The third assumption states that if the control and treatment prices are the same, then both the scaled control and treatment demand functions are equivalent to the market demand function; this assumption is implied by constant returns to scale. This assumption is satisfied by demand systems for which the control and treatment demand quantities are proportional to their shares in the marketplace. Indeed, this assumption will be satisfied for the model we later study in Section \ref{sec: logit}, as well for a number of other common demand systems.

We finish this subsection by defining the experimental estimators. To do so, we define the experimental control and treatment profit functions, respectively:
\begin{align}
\pi_0(p_0,p_1,q) &:= (p_0-c) D_0(p_0,p_1,q) \notag \ , \\
\pi_1(p_0,p_1,q) &:= (p_1-c) D_1(p_0,p_1,q) \notag \ .
\end{align}
Following Definition \eqref{eq: EstGTEL} we define canonical estimators for demand and profit associated with the canonical experimental designs:

\begin{align}
\widehat{GTE}_D (p_0,q) &:= \lim_{n\to\infty}  \frac{D_1(p_0,p_1^n,q)-D_0(p_0,p_1^n,q)}{p_1^n-p_0}  \ ,\notag \\
\widehat{GTE}_\pi (p_0,q) &:= \lim_{n\to\infty}  \frac{\pi_1(p_0,p_1^n,q)-\pi_0(p_0,p_1^n,q)}{p_1^n-p_0}  \ . \notag 
\end{align}
We note that analogs of these canonical estimators are frequently used in practice in estimating global treatment effects from experimental data; in particular, in our continuum model, these estimators naturally correspond to difference-in-means or Horvitz-Thompson type estimators, scaled by the price difference to estimate the target estimand (the derivative of demand or profit at the control price).

Now, by equation \eqref{eq:expr_est}, we obtain:
\begin{align}
\widehat{GTE}_D(p_0,q) &= D_{1y}(p_0,p_0,q) - D_{0y}(p_0,p_0,q)  \ ,\label{eq:expr_est_D} \\
\widehat{GTE}_\pi(p_0,q) &= \pi_{1y}(p_0,p_0,q) - \pi_{0y}(p_0,p_0,q)  \notag \\
&= D(p_0) + (p_0-c) \left(D_{1y}(p_0,p_0,q) -  D_{0y}(p_0,p_0,q)\right) \ ,  \label{eq:expr_est_Pi}
\end{align} 
where we used Assumption \ref{as: diff_D} and the last equation follows by Assumption \ref{as: control_treat_demand}.3.

\section{Biases: Analysis and Change of Sign \label{sec: bias_analysis}}

In this section we analyze the biases associated with the canonical estimators introduced above. Following Definition \eqref{eq:bias1}, we define the estimation biases for demand and profit as:
\begin{align}
Bias_D (p_0,q) &:= GTE_D(p_0)-\widehat{GTE}_D(p_0,q) \ ,\notag \\
Bias_\pi (p_0,q) &:= GTE_{\pi}(p_0)-\widehat{GTE}_\pi(p_0,q) \ .\notag
\end{align}

First, note that by Proposition \ref{pr:bias} we obtain the following bias characterizations:
\begin{align}
Bias_D(p_0,q)&=D_{0y}(p_0,p_0,q)+D_{1x}(p_0,p_0,q) \ ,\label{eq:bias_D}\\
Bias_\pi(p_0,q)&=\pi_{0y}(p_0,p_0,q)+\pi_{1x}(p_0,p_0,q) \notag \\
&=  (p_0-c) \left(D_{0y}(p_0,p_0,q) +D_{1x}(p_0,p_0,q) \right) \label{eq: bias_pi_alt} \\
&=  (p_0-c) Bias_D(p_0,q) \ .\label{eq:bias_pi}\
\end{align}
As discussed in Proposition \ref{pr:bias}, bias increases as interference effects between treatment and control groups increase. More specifically, the bias is influenced by the degree to which the control (or treatment) group cannibalizes demand from its counterpart in response to a marginal price change in the latter group.

Now, note that under Assumptions \ref{as: diff_D} and \ref{as: control_treat_demand}, both $Bias_D(p_0,q)$ and $Bias_\pi(p_0,q)$ are non-negative, meaning the canonical estimators always (weakly) underestimate the $GTE$:

\begin{proposition}
\label{pr: underestimate}
We have that $\widehat{GTE}_D(p_0,q)\leq GTE_D(p_0)$ and $\widehat{GTE}_\pi (p_0,q)\leq GTE_\pi(p_0)$, for all $p_0\in\mathbb{R}_+$ and $0<q<1$. 
\end{proposition} 

This result follows similar cannibalization arguments developed in previous work \citep{johari2022experimental}. To provide intuition, consider an LR experiment in which the treatment price is higher than control the price. In this case, experimental treatment demand underestimates global treatment demand because of substitution in the experiment from treatment units to control units. Similarly, experimental control demand overestimates global control demand.  Instead, let's consider a CR experiment where there are supply constraints, so customers compete for limited products. In such cases, the control customers may ``cannibalize'' supply from the treatment group, because of their lower control price, causing an underestimation of the global treatment effect again. Note that in a CR experiment without supply constraints, we anticipate no bias because there is no competition between the treatment and control groups.

It is important to observe that because $D'(p)<0$ by Assumption \ref{as: demand}, $GTE_D(p_0)< 0$, for all $p$. Moreover, since $\widehat{GTE}_D(p_0,q)$ is an underestimation, both the estimand and the canonical estimator will have the same (nonpositive) sign. 

However, there is a more intriguing result for profits. Suppose that the treatment price $p_1$ is greater than the control price $p_0$. In this scenario, even if the demand $D(p_1)$ at $p_1$ is lower than the demand $D(p_0)$ at $p_0$, the increase in price may offset the decrease in demand, leading to a positive true global treatment effect in profits. But, because the canonical estimator underestimates the treatment effect, the profit GTE estimator could be negative. Next, we will provide conditions under which the canonical estimator not only underestimates the global treatment effect, but actually obtains the wrong sign.

By definition, $GTE_\pi(p_0)\geq 0$ if and only if $\pi'(p_0)\geq 0$, which after some simple manipulation yields the following standard ``markup'' inequality: 
\begin{equation}
\label{eq:condA}
 \mbox{Condition (a): \ \ } A(p_0):=\frac{p_0-c}{p_0}\leq -\frac{1}{e_p(p_0)} \ ,
 \end{equation}
 where $e_p(p_0)=D'(p_0)p_0/D(p_0)$ is the price elasticity of demand. If $\pi(p)$ is strictly quasi-concave, the condition is equivalent to the price $p_0$ being below the optimal monopoly price. For the case  $c=0$, condition (a) naturally states that marginal revenue is increasing at $p_0$, or equivalently that the demand is relatively inelastic with respect to price. 
 
Now, consider the canonical estimator. Note that $\widehat{GTE}_\pi (p_0,q) \leq 0$ is equivalent to $Bias_\pi(p_0,q) \geq {GTE}_\pi (p_0)$. That is,  provided that $GTE_\pi(p_0)\geq 0$ holds, we have a change of sign when the bias is so severe that it becomes larger than the treatment effect. Furthermore, after some manipulation of equation \eqref{eq:expr_est_Pi}, we obtain that $\widehat{GTE}_\pi (p_0,q) \leq 0$ if and only if:
\begin{equation}
\label{eq:condB}
\mbox{Condition (b): \ \ } -\frac{1}{e_p(p_0)}\leq \frac{(p_0-c)}{p_0} \frac{\left(D_{1y}(p_0,p_0,q) -  D_{0y}(p_0,p_0,q)\right)}{D'(p_0)} :=B(p_0,q)\ .
\end{equation}
Note that condition (b) is a ``modified markup" inequality. We have the following result.
\begin{proposition} \label{pr: change of sign}
\begin{enumerate}
\item Condition (a) holds if and only if $GTE_\pi(p_0)\geq 0$.
\item Condition (b) holds if and only if $\widehat{GTE}_\pi(p_0,q) \leq 0$.
\item $A(p_0)\leq B(p_0,q)$, for all $p_0\in\mathbb{R}_+$ and $0<q<1$.
\end{enumerate}
\end{proposition}
\proof{Proof.}
Points (1) and (2) in the proposition are direct from the arguments above. To prove point (3), note that $D_{1y}(p_0,p_0,q)\leq  D'(p_0)$ (by applying Assumptions \ref{as: control_treat_demand}.2 and \ref{as: control_treat_demand}.3 together with the chain rule) and $D_{0y}(p_0,p_0,q)\geq 0$ (by Assumption \ref{as: control_treat_demand}.1). 
Hence, 
$$
     \frac{p_0-c}{p_0}\left(D_{1y}(p_0,p_0,q) -  D_{0y}(p_0,p_0,q)\right)\leq \frac{p_0-c}{p_0} D'(p_0)  \ ,
  $$
and thus the result follows because $D'(p_0)<0$ by Assumption \ref{as: demand}. \qedsymbol{} 
\endproof

The proposition states a sign change is obtained, $\widehat{GTE}_\pi(p_0,q) \leq 0 \leq GTE_\pi(p_0)$, if and only if conditions (a) and (b) hold. This  implies lower and upper bounds on the inverse of the price elasticity of demand. Note that we use weak inequalities in conditions (a) and (b) for a change of sign; strictly speaking of course, a change of sign can only occur if at least one of the two inequalities is strict.\footnote{If exactly one of these inequalities is strict (and both hold), the canonical estimator will still instruct firms to move in a non-optimal direction; either by telling them to lower prices when they were already at the optimal price, or by suggesting they keep their price constant when they were underpricing.}  We suppress this edge case in the sequel in favor of simplicity of presentation.\footnote{In our two-sided market model, introduced in section \ref{sec: logit}, the scenario $\widehat{GTE}_\pi(p_0,q) = GTE_\pi(p_0) = 0$ is impossible, meaning this edge case does not exist.}

To provide intuition, note that by using equation \eqref{eq:expr_est_D} we can define 
the {\em ``experimental'' price elasticity of demand} as: \begin{equation}
\label{eq: exp_elast}
    \widehat{e_p}(p_0,q)=\frac{p_0( D_{1y}(p_0,p_0,q) -  D_{0y}(p_0,p_0,q) )}{D(p_0)}.
\end{equation}
Hence, we can re-write condition (b) as:
$$ -\frac{1}{\widehat{e_p}(p_0,q)}\leq \frac{(p_0-c)}{p_0}  \ ,$$ 
which can be understood as the experimental analogue of a markup inequality. It states that the experimental elasticity suggests the current price is above the optimal monopoly price (under an appropriate concavity assumption). Hence, as expected, a change of sign occurs when the canonical estimator indicates that the platform is overpricing relative to the optimal monopoly price (and therefore prefers to decrease prices), when in reality it is underpricing (and therefore should raise prices). This result has important consequences for experiment-driven price optimization: the prescription implied an the experiment (reduce price) may be the opposite of the one implied by the $GTE$ (increase price). Therefore, if a firm moves in the direction suggested by the experiment, profits have the potential to {\em decrease} relative to the status-quo.

To provide further intuition, let us assume that condition (a) holds and that profits are quasiconcave, so that the firm is underpricing relative to the optimal monopoly price. Hence, we  have a change of sign if and only if condition (b) holds. If 
 $|D_{1y}(p_0,p_0,q)| +  D_{0y}(p_0,p_0,q)$ is large compared to $|D'(p_0)|$, condition (b) will hold.  Note that $|D_{1y}(p_0,p_0,q)|$ represents the magnitude of the decrease in treatment demand given a small increase in treatment price, which captures two effects: (i) demand lost to the outside option; and (ii) cannibalization to the control group. Note that (i) is the global demand effect we aim to estimate. On the other hand, the term $D_{0y}(p_0,p_0,q)$ represents the increase in control demand given a small increase in treatment price, which captures cannibalization from the treatment to the control group (equivalent to effect (ii)). A change of sign happens as the two cannibalization effects just mentioned become larger.  

We conclude with two remarks regarding our estimators, our findings on changes of sign, and guidance for practitioners. First, note that the estimators $\widehat{GTE}_D$ and $\widehat{GTE}_\pi$ that we defined are estimates of {\em derivatives} of the demand and profit functions, respectively, as is natural given our estimands.  In practice, a platform may use a difference in estimated treatment and control demand, or estimated treatment and control profit, {\em without} normalizing by the difference in prices.  This can have some bearing in interpreting our results, depending on the sign of the difference between the treatment and control price.

Concretely, suppose that treatment involves raising the price relative to control.  Then our results show that a change of sign can only occur if the unnormalized global treatment effect on profits is positive, but the treatment effect is estimated to be negative in the experiment.  On the other hand, suppose the treatment involves lowering the price relative to control.  Then our results show that a change of sign can only occur if the unnormalized global treatment effect on profits is negative, but the treatment effect is estimated to be positive in the experiment. In either case, the platform risks either failing to move prices in a profitable direction, or worse, moving them in an unprofitable direction.

Second, we note that in our case, we have studied a model where the platform changes the {\em price} directly.  In practice, another relevant platform parameter that affects profits can be a platform-wide {\em fee}.  Specifically, in Appendix \ref{sec: fee_model}, we present a model where a platform offers a multitude of heterogeneous goods (potentially at different prices), and earns profit by charging a percentage fee on each transaction. This is a common monetization model for online marketplaces. Viewing the fee as the variable of interest for the platform's profit optimization, we obtain analogous results to those in this section, including the bias characterization, the possibility of a change of sign, and interpretation of the conditions for a change of sign in terms of appropriate derivatives.  This extension suggests that our insights in this paper also have implications for practitioners in platform settings with many heterogeneously differentiated goods, where fees are the key variable for monetization.


\section{Debiasing Canonical Estimators}
\label{sec: debiasing}

We take advantage of our structural characterization of bias to introduce a novel approach to debiasing $\widehat{GTE}_D$ and $\widehat{GTE}_\pi$. The key to this approach, and a necessary step for it to be applied, is for the experimenter to equally allocate units between treatment and control, i.e., setting $q = 1/2$.

Recall that equations \eqref{eq:bias_D} and  \eqref{eq: bias_pi_alt} quantify bias in terms of the derivatives of demand.  We will now show how we can use observables to estimate this bias.  First, we show that $D_{0y}(p_0,p_0,q)$ can be computed (approximately) using observed data. To see this, note that $D_{0y}(p_0,p_0,q)$ is defined by:
\[D_{0y}(p_0,p_0,q) = \lim_{p_1\to p_0^+}\frac{D_0(p_0,p_1,q) - D_0(p_0,p_0,q)}{p_1-p_0}.\]
By Assumption 6 we have that $D_0(p_0,p_0,q) = D(p_0)$. To make progress, we make the assumption that since the firm was pricing at $p_0$ before the experiment began, the pre-experiment observation of demand is a reasonable proxy for the control demand $D(p_0)$.  The quantity $D_0(p_0,p_1,q)$ is observed as well: it represents the control demand during the experiment. This means that in the limit as $p_1$ approaches $p_0$, observables can be used to approximately compute the derivative $D_{0y}(p_0,p_0,q)$.

We now note that by symmetry, we have $D_0(p_0,p_1,1/2) = D_1(p_1,p_0,1/2)$.  This holds because both terms represent demand for the half of the goods that are priced at $p_0$, when the other half of goods are priced at $p_1$. Taking the derivative of both sides of this equality with respect to $p_1$, we obtain $D_{0y}(p_0,p_1,1/2) =D_{1x}(p_1,p_0,1/2)$, so that when $p_1=p_0$, there holds $D_{0y}(p_0,p_0,1/2) =D_{1x}(p_0,p_0,1/2)$.  We have thus shown the following result.

\begin{proposition}
Suppose that $q = 1/2$.  Then there holds:
\begin{align}
Bias_D(p_0, 1/2) &= 2 D_{0y}(p_0,p_0,1/2); \\
Bias_\pi(p_0, 1/2) &= 2 (p_0 - c) D_{0y}(p_0, p_0, 1/2).
\end{align}
In particular, there holds:
\begin{align}
GTE_D &= \widehat{GTE}_D(p_0, 1/2) + 2 D_{0y}(p_0,p_0,1/2); \label{eq:debiased_D}\\
GTE_\pi &= \widehat{GTE}_\pi(p_0, 1/2) + 2 (p_0 - c) D_{0y}(p_0, p_0, 1/2). \label{eq:debiased_pi}
\end{align}
\end{proposition}

Thus if the firm is able to estimate $D_{0y}(p_0,p_0,q)$ using observables (as described above), then the right hand sides of \eqref{eq:debiased_D} and \eqref{eq:debiased_pi} are debiased estimators of $GTE_D$ and $GTE_\pi$ respectively.

We note three limitations of this method.  First, of course, the symmetry property being exploited only applies when the market is equally allocated to treatment and control.  Second, the method is not sample efficient: only half the experimental samples, those from the control group, are used. Finally, our view that the platform can accurately estimate $D(p_0)$ depends on the market remaining reasonably stationary during the experiment; if the market is nonstationary, this adjusted estimator may not capture the true treatment effect during the experiment. 


\section{A Two-Sided Market Model} \label{sec: logit}

In this section, we illustrate the application of our results from the  general revenue management setting (introduced in Section \ref{sec: Revenue Management Setting}) to the two-sided market model  developed in {\cite{johari2022experimental}}. That paper develops a stochastic dynamic market model with supply constraints in which customers book listings for a finite period of time. The paper also introduces a mean-field limit that greatly simplifies the analysis. We focus on the mean-field model here and describe its main elements.

\subsection{Preliminaries} 

We begin by recapping the mean-field model formulation in \cite{johari2022experimental}, adapted to our setting with price experiments. There is a two-sided platform evolving in continuous time with {\em listings} on the supply side and {\em customers} on the demand side. 
There is a continuum of listings of mass $\rho$ that are homogenous with respect to their observable characteristics. Having a limited supply will play a key role in the analysis. We assume all listings have the same price $p$. At each time $t$, listings can be available or booked. The state of the system at time $t$ is given by $s_t$, which denotes the mass of available listings available at time $t$. 

On the demand side, there is a continuum of customers arriving to the system at rate $\lambda$. When a customer arrives, they form a consideration set consisting of all available listings at that point in time.\footnote{It is straightforward to add a process in which customers sample listings to consider with the same probability and independently. We omit this probability to simplify notation.}
Given the consideration set, customers decide whether or not to book a listing according to a multinomial logit choice model with an outside option of value $\varepsilon > 0$.\footnote{Note that by assumption we will constrain the value of the outside option $\varepsilon$ to be strictly positive.} The value of booking a listing is given by the function $v(p)$. Under this specification, the probability that a customer who arrives in state $s$ books a listing is equal to:
\begin{equation}
\frac{s v(p)}{\varepsilon+sv(p)}  \notag .
\end{equation}
Finally, a listing remains occupied for an expected time $\frac{1}{\tau}$ when it gets booked; in our analysis we normalize $\tau$ to be 1 without loss of generality. Note that in our model, the parameter $\lambda$ mediates market balance: when $\lambda\rightarrow 0$ the market becomes {\em demand constrained}, when $\lambda\rightarrow \infty$ the market becomes {\em supply constrained}.

The general model in \cite{johari2022experimental} allows for heterogeneity in both listings and consumers.  To simplify the analysis and exposition, in the model we consider in this section, we make both consumers and listings homogeneous.

As in \cite{johari2022experimental} we focus on the steady-state of the system. Let $s^*(\lambda,p)$ be the steady-state mass of available listings for a given market balance parameter  $\lambda$ and price $p$. It is shown in \cite{johari2022experimental} that $s^*(\lambda,p)$ is the unique solution to the following conservation of flow equation:
\begin{equation}
    \label{eq: market_balance}
    \rho-s^*(\lambda,p) =\lambda \frac{s^*(\lambda,p)v(p)}{\varepsilon+s^*(\lambda,p)v(p)}:=D(\lambda,p) \ .
\end{equation}
The left-hand side of the equation corresponds to the rate at which listings become available. The right-hand side corresponds to the booking (or demand) rate, defined as $D(\lambda,p)$. In the steady-state both rates are equal. 

\begin{assumption}
\label{as: valuation_function}
    We assume the following for the utility function $v(p)$:
\begin{enumerate}
    \item $v(p)$ is strictly positive for $p \in [c,+\infty);$
    
    \item $v(p)$ is strictly decreasing for $p\in [c,+\infty);$

    \item $v(p)$ is differentiable for $p \in [c,+\infty)$, and $v'(p)$ is continuous;
    
    \item  $ -pv'(p)/v(p)$ is strictly increasing in $p$ for $p \in [c,+\infty).$ 
\end{enumerate}
\end{assumption}

Assumptions \ref{as: valuation_function}.1-\ref{as: valuation_function}.3 are natural for a utility function. We assumed strict positivity to avoid degeneracy. Assumption  \ref{as: valuation_function}.4 is reminiscent  to the standard increasing generalized failure rate assumption for distributions \citep{lariviere2006note} and is satisfied for standard specifications such as linear and exponential valuation functions, among others. 
Indeed,  we consider the standard specification of $v(p)=e^{V-p}$ in our numerics. In Appendix \ref{sec: two_side_demand_assumption_4} we show that  Assumption $\ref{as: valuation_function}$ implies Assumption \ref{as: demand}.\footnote{In addition, when $v(p)$ satisfies Assumption $\ref{as: valuation_function}.1$  the market balance equation has a closed form solution given by:
\begin{equation} 
s^*(\lambda,p)=\frac{(\rho-\lambda) v(p)-\varepsilon+\sqrt{4 \rho \varepsilon  v(p)+((\rho -\lambda)  v(p) -\varepsilon )^2}   }{2  v(p)}
\notag
\end{equation}
}

\subsection{Steady-State and Estimand}

To gain some initial intuition, we start this section by providing some analysis on the steady-state and the associated demand. Then, we define our estimands.

First, we note that listing availability $s^*(\lambda,p)$ is monotonically increasing with price $p$ and decreasing with the market balance parameter $\lambda$, which is natural. These monotonicity results can be derived by differentiating $s^*(\lambda,p)$ in equation \eqref{eq: market_balance}.\footnote{Indeed, we obtain $s^*_p(\lambda,p)=-\frac{ s^*(\lambda,p) v'(p) \varepsilon}{\frac{1}{\lambda}(\varepsilon+s^*(\lambda,p)v(p))^2+ v(p) \varepsilon}>0$ and $s^*_\lambda(\lambda,p)=-\frac{v(p) s(\lambda,p) (v(p) s(\lambda,p)+\varepsilon )}{\varepsilon  v(p) (2 s(\lambda,p)+\lambda)+v(p)^2
   s(\lambda,p)^2+\varepsilon ^2}<0$.}  Figure \ref{fig1} in the Appendix illustrates the steady-state listing availability as a function of price ($p$) and market balance ($\lambda$).

Naturally, $D(\lambda,p)$ is monotonically decreasing in price ($p$) and increasing in market balance ($\lambda$). Again, we can prove these monotonicity results by differentiating.\footnote{This can be seen by rewriting the demand as the rate at which the listings become available. Using the market balance equation, we have $D(\lambda,p)=\lambda \frac{s^*(\lambda,p)v(p)}{\varepsilon+s^*(\lambda,p)v(p)}=\rho-s^*(\lambda,p)$. Hence, $D_{p}(\lambda,p)=-s^*_p(\lambda,p)$ and $D_{\lambda}(\lambda,p)=-s^*_\lambda(\lambda,p)$. Since $s^*_p(\lambda,p)>0$ and $s^*_\lambda(\lambda,p)<0$ we obtain that  $D_{p}(\lambda,p)<0$ and $D_{\lambda}(\lambda,p)>0$.}  Figure \ref{fig2} in the Appendix shows the dependence of the demand function on the market balance parameter and price. 

Now, we define our estimands. Following the definitions in equations \eqref{eq: GTE_D} and \eqref{eq: GTE_Pi}:
\begin{align}
GTE_{D} (\lambda,p) &= D_p(\lambda,p) \notag \\
GTE_{\pi}(\lambda,p)&=D(\lambda,p)+(p-c)D_p(\lambda,p)\notag\\
&=\rho-s^*(\lambda,p)- (p-c) s^*_p(\lambda,p) \label{eq: GTE_MarketBalance},
\end{align}
where in the last equation we used equation \eqref{eq: market_balance}.

\subsection{Experimental Designs}

In this section we analyze two canonical experimental designs in platforms: randomization across listings (LR experimentation) and randomization across customers (CR experimentation). 

\subsubsection{LR Experimentation}

We assume that randomization occurs on the listing side, where a fraction $q$ of the listings are assigned to treatment.
Hence, a customer is exposed to a mix of treatment and control listings. In this experimental design, the balance equations are given by: 
\begin{equation}
\begin{aligned}
\label{eq: exp_avail_list}
(1-q)\rho-s_0^*(\lambda,p_0,p_1,q)&=\lambda \frac{s_0^*(\lambda,p_0,p_1,q)v(p_0)}{\varepsilon+s_0^*(\lambda,p_0,p_1,q)v(p_0)+s_1^*(\lambda,p_0,p_1,q)v(p_1)}\\
q\rho-s_1^*(\lambda,p_0,p_1,q) &=\lambda \frac{s_1^*(\lambda,p_0,p_1,q)v(p_1)}{\varepsilon+s_0^*(\lambda,p_0,p_1,q)v(p_0)+s_1^*(\lambda,p_0,p_1,q)v(p_1)} \ ,
\end{aligned}
\end{equation}
where $s^*_0(\lambda,p_0,p_1,q)$ and $s^*_1(\lambda,p_0,p_1,q)$ refer to the steady-state available control and treated listings, respectively. 

The control and treatment demand functions $D^{LR}_0(\lambda,p_0,p_1,q)$ and $D_1^{LR}(\lambda,p_0,p_1,q)$ are given by:
\begin{equation}
\begin{aligned}
\label{eq: LR_Demand_Functions}
D_0^{LR}(\lambda,p_0,p_1,q)&= \rho - s_0^*(\lambda, p_0, p_1, q)/(1-q); \\
D^{LR}_1(\lambda,p_0,p_1,q)&=\rho - s_1^*(\lambda, p_0, p_1, q)/q.
\end{aligned}
\end{equation}
We can show that $D^{LR}_0(\lambda,p_0,p_1,q)$ and $D^{LR}_1(\lambda,p_0,p_1,q)$ satisfy Assumption \ref{as: control_treat_demand} (see Appendix \ref{sec: LR_assmption_6}). 

Following the characterization of bias from equations \eqref{eq:bias_D} and \eqref{eq:bias_pi}, we obtain that the bias in this setting, $Bias_\pi^{LR}(\lambda,p_0,q)$ , is given by:
\begin{align}
Bias^{LR}_{\pi}(\lambda,p_0,q)&=(p_0-c)(D^{LR}_{1x}(\lambda,p_0,p_0,q)+D^{LR}_{0y}(\lambda,p_0,p_0,q)) \notag \\
&=(p_0-c)\frac{-\lambda  s^*(\lambda,p_0)^2 v(p_0) v'(p_0) (\varepsilon+s^*(\lambda,p_0) v(p_0) )}{[\varepsilon+s^*(\lambda,p_0) v(p_0) +\lambda v(p_0)]
   [(s^*(\lambda,p_0)v(p_0) +\varepsilon)^2 +\lambda v(p_0) \varepsilon ]} \label{eq: LR_Bias_formula}
\end{align}
Note that the bias is independent of the treatment fraction $q$.

\subsubsection{CR Experimentation}
In this setting, we randomize over customers, with a fraction $q$ of customers being placed into treatment, and the remainder in control. 
In this case, a treatment (resp., control) customer is exposed to the entire market under the treatment (resp., control) condition. The 
steady-state is given by:
\begin{equation}
\label{eq: CR_Market_Bal}
    \rho-s^*(\lambda,p_0,p_1,q)= \left[ q \frac{s^*(\lambda,p_0,p_1,q) v(p_1)}{\varepsilon+s^*(\lambda,p_0,p_1,q)v(p_1)} +(1-q) \frac{s^*(\lambda,p_0,p_1,q) v(p_0)}{\varepsilon+s^*(\lambda,p_0,p_1,q)v(p_0)} \right].
\end{equation}

Observe that the two terms on the right hand side above are rescaled versions of the control and treatment demand functions:
\begin{equation}
\begin{aligned}
\label{eq: CR_Demand_Functions}
D^{CR}_0(\lambda,p_0,p_1,q)&=\lambda \frac{s^*(\lambda,p_0,p_1,q)v(p_0)}{\varepsilon+s^*(\lambda,p_0,p_1,q)v(p_0)},\\
D^{CR}_1(\lambda,p_0,p_1,q)&=\lambda \frac{s^*(\lambda,p_0,p_1,q)v(p_1)}{\varepsilon+s^*(\lambda,p_0,p_1,q)v(p_1)}.
\end{aligned}
\end{equation}

One can show that the demand functions $D^{CR}_0(\lambda,p_0,p_1,q)$ and $D^{CR}_1(\lambda,p_0,p_1,q)$ satisfy Assumption \ref{as: control_treat_demand} (see Appendix \ref{sec: CR_assmption_6}).
   
Using the characterization in equation (\ref{eq:bias_pi}) we obtain the following expression for $Bias_\pi^{CR}$:

\begin{align}
Bias^{CR}_{\pi}(\lambda,p_0,q)&=(p_0-c)(D^{CR}_{1x}(\lambda,p_0,p_0,q)+D^{CR}_{0y}(\lambda,p_0,p_0,q))\notag\\
&= -(p_0-c)\frac{\lambda^2 \varepsilon ^2 v(p_0) v'(p_0) s^*(\lambda,p_0)}{(v(p_0) s^*(\lambda,p_0)+\varepsilon )^2 \left[  (s^*(\lambda,p_0)v(p_0)+\varepsilon)^2+\lambda v(p_0) \varepsilon \right]}\label{eq: CR_Bias_formula}
\end{align}
Note that the bias in the CR-setting is independent of the treatment fraction $q$.

\subsection{Properties of Biases with Market Balance \label{sec: bias_formulas}}

\cite{johari2022experimental} studies the bias of CR and LR experiments as the market becomes heavily demand constrained or supply constrained, respectively. Following the analysis of that paper, we obtain analogous insights, with proofs similar to Johari et al. (2022) but also leveraging our characterization of biases through cross-partial derivatives. In particular, as the market becomes more demand constrained ($\lambda \rightarrow 0$), competition between customers declines, resulting in less interference in a CR experiment; in the limit, the bias in a CR experiment approaches zero.  On the other hand, in the same regime the competition between listings is high, resulting in more interference for an LR experiment, so in the limit the bias of an LR experiment remains positive.  By contrast, as the market becomes more supply constrained ($\lambda \to \infty$), the opposite results hold: in this regime, competition between customers is stronger and competition between listings is weaker, so in the limit the bias of a CR experiment remains positive while the bias of a LR experiment approaches zero.  We derive these results in Propositions \ref{pr: Bias_beta_zero} and \ref{pr: Bias_beta_infty} in Appendix \ref{sec: logit_model_limit_bias}.

\begin{figure}[!ht]
    \centering
    \includegraphics[width=0.9\textwidth]{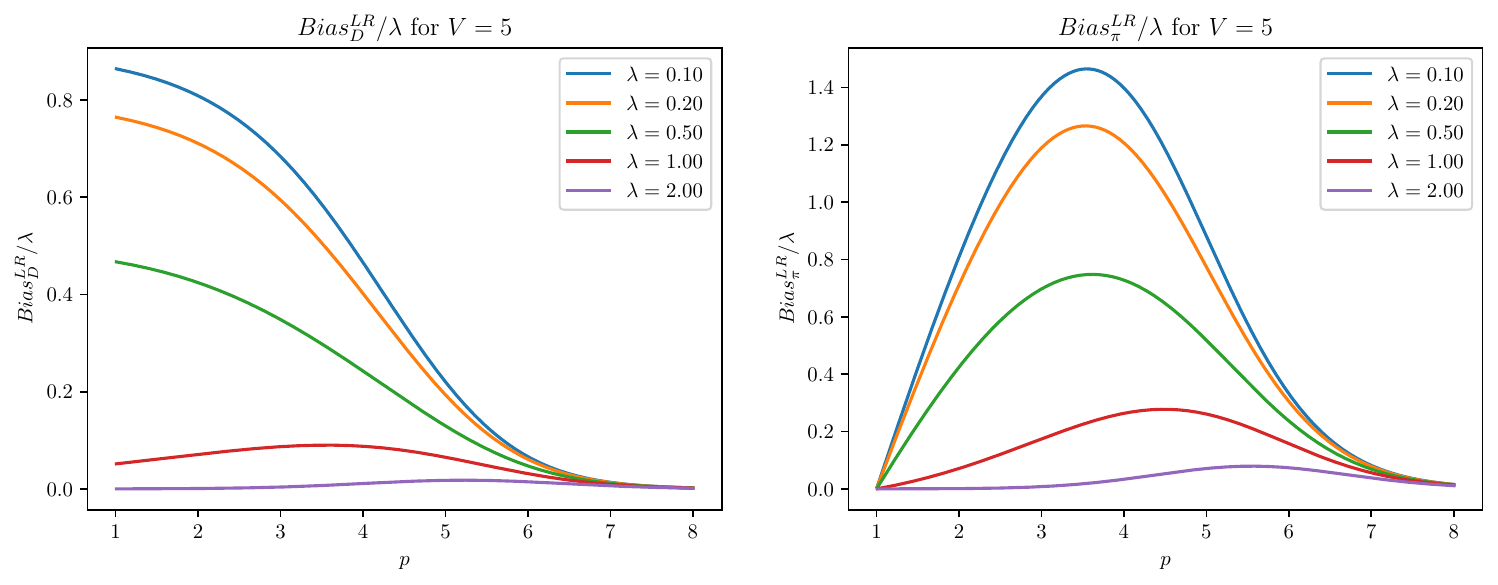} 
    \caption{Illustration of  $Bias^{LR}_D/\lambda$ and $Bias^{LR}_\pi/\lambda$ as a function of price ($p$) and market balance ($\lambda$). Instance is given by $v(p)=e^{V-p}$, $\epsilon = 1$, $c = 1$, $V=5$, and $\rho = 1$.}
    \label{Fig_Bias_LR}
\end{figure}

\begin{figure}[!ht]
    \centering
    \includegraphics[width=0.9\textwidth]{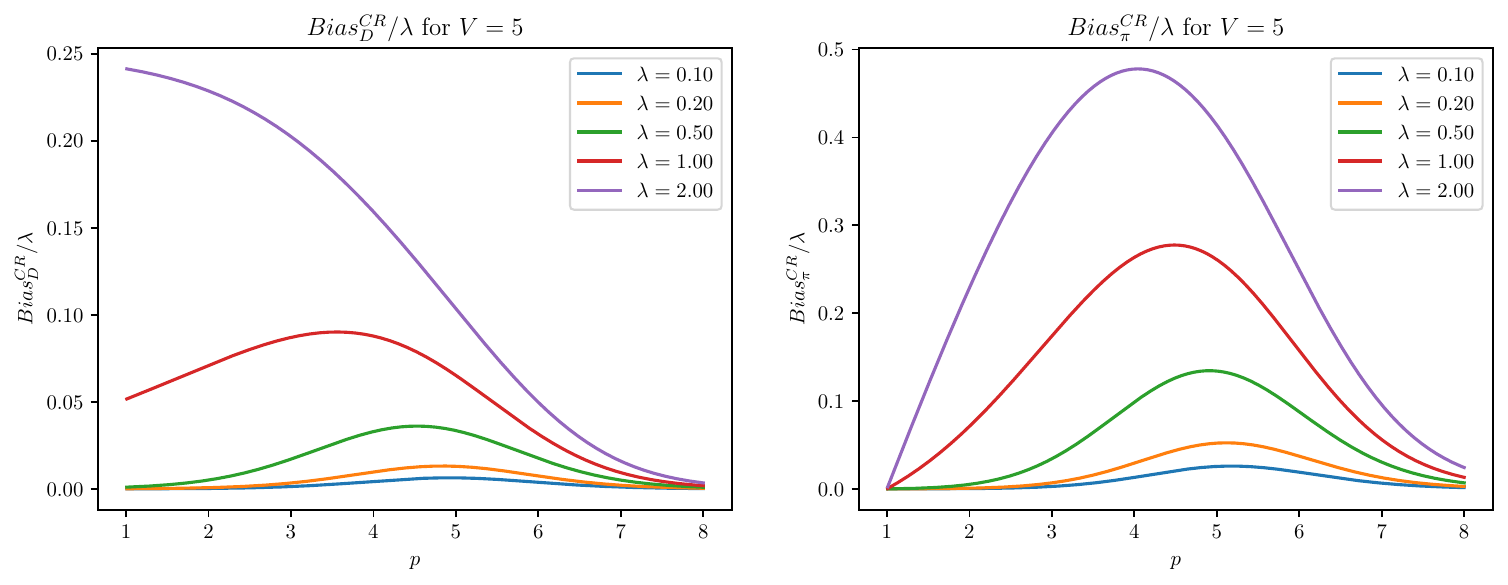} 
    \caption{Illustration of  $Bias^{CR}_D/\lambda$ and $Bias^{LR}_\pi/\lambda$ as a function of price ($p$) and market balance ($\lambda$). Instance is given by $v(p)=e^{V-p}$, $\epsilon = 1$, $c = 1$, $V=5$, and $\rho = 1$.}
    \label{Fig_Bias_CR}
\end{figure}

Below we provide two figures to complement our limit results. In Figures \ref{Fig_Bias_LR} and \ref{Fig_Bias_CR} we present bias normalized by the arrival rate $\lambda$. We normalize by $\lambda$ instead of the GTE because the GTE for profits converges to 0 around the optimal price.  As $\lambda$ increases, the market becomes more supply-constrained and bias for an LR experiment decreases as competition between listings dissipates. Figure \ref{Fig_Bias_CR} shows the normalized bias of demand and profits in a CR experiment and this dependence on $\lambda$ is naturally reversed. Both figures also show interesting behavior of biases with respect to the baseline price $p$, a novel aspect of our setting relative to \cite{johari2022experimental}.  As the price $p$ changes, the sensitivity of demand to price changes as well.  In particular, in both CR and LR experiments, both $D_{0y}$ and $D_{1x}$ tend to be smaller at larger prices (i.e., demand becomes more inelastic as the price becomes large); this behavior drives the shape of the bias curve with respect to price, and subsequently profit. Note in particular that LR experiments exhibit interesting behavior when prices are large: despite the fact that we nominally expect bias to increase when the market is demand constrained, and higher prices reduce demand in the market, nevertheless the bias {\em drops} when prices are high.  This is because of the fact that elasticities change with price, and illustrates the nuanced nature of interference in price experiments.

\subsection{Change of Sign}

We recall the two conditions for a change of sign to occur:
$$\text{Condition (a)} \quad GTE_{\pi}\geq 0; \qquad \text{Condition (b)} \quad Bias_{\pi}\geq GTE_{\pi}.  $$
Interestingly, observe that the bias terms derived in Section \ref{sec: bias_formulas} do not depend on the treatment fraction $q$ for either LR or CR experiments. In other words, the treatment fraction is not a lever that can be used to prevent---or even reduce---the possibility of a change of sign in the treatment effect estimator.

Using the structure of the model we can provide the following insightful characterizations for the change of sign regions.

\begin{theorem} 
\label{thm: LR_interval}
For all $\lambda\in(0,\infty)$, there exists a non-empty, closed and convex set (i.e., an interval) $P^{LR}_{\lambda}\subseteq [c,\infty)$ such that the sign of an LR experiment's canonical estimator for $GTE_\pi(\lambda,p)$ at price $p\in[c,\infty)$ will be wrong if and only if $p\in P^{LR}_{\lambda}$.  
\end{theorem}

\begin{theorem} 
\label{thm: CR_interval}
For all $\lambda\in(0,\infty)$, there exists a non-empty, closed and convex set (i.e., an interval) $P^{CR}_{\lambda}\subseteq [c,\infty)$ such that the sign of a CR experiment's canonical estimator for $GTE_\pi(\lambda,p)$ at price $p\in[c,\infty)$ will be wrong if and only if $p\in P^{CR}_{\lambda}$.
\end{theorem}

Theorems \ref{thm: LR_interval} and \ref{thm: CR_interval} state that regardless of the market balance parameter $\lambda$, there is always an interval of prices $P_\lambda$, such that the sign of the canonical estimator will be wrong, regardless of whether an LR or CR experiment is run.  The proofs are provided in Appendix \ref{sec: interval_proof} and repeatedly exploit the market balance equations for CR and LR respectively.

The right hand side of both the intervals $P^{LR}_\lambda$ and $P^{CR}_\lambda$ is the optimal monopoly price $p^*(\lambda)$, i.e., where $GTE_\pi(\lambda,p^*)=0$, for each $\lambda$.  In particular, this reveals that the sets $P^{LR}_\lambda$ and $P^{CR}_\lambda$ are not comprised of unrealistic prices: if a firm is currently pricing optimally or very slightly below optimally, the sign of their canonical estimator will be wrong, regardless of their experimental setup.

It is not true, however, that the two experimental setups are equally fallible in all settings. Note that Condition (b) in 
\ref{eq:condB} is different for LR and CR because the elasticities and demand sensitivities differ,  meaning that the left endpoints of $P^{LR}_{\lambda}$ and $P^{CR}_{\lambda}$ are not the same. As expected based on the results of the previous section, they behave quite differently as a function of market balance, as we illustrate through the following numerical study.

\subsubsection{Numerical Results}
\label{sec: numerics}

\begin{figure}[!ht]
    \centering
    \includegraphics[width=0.44\textwidth]{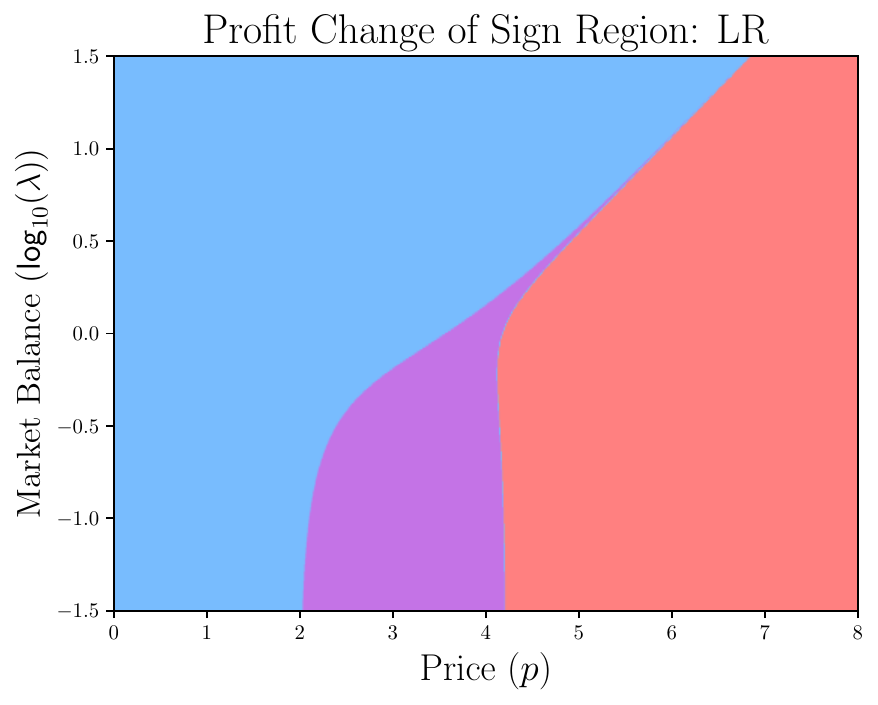} \includegraphics[width=0.44\textwidth]{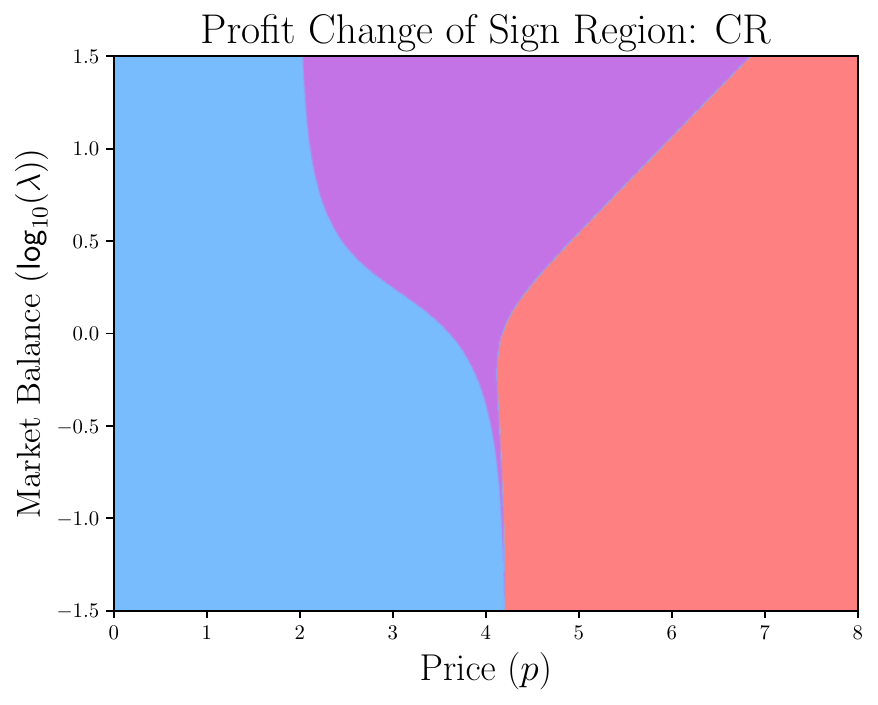}
     \caption{Illustration of the change of sign region (purple) in an LR-experiment (left) and CR-experiment (right), respectively, with $v(p)=e^{V-p}$, $V=5$, $c=0.5$, $\rho=1$, and $\epsilon=1$.  The vertical axis is the (log of) the arrival rate, or market balance, parameter $\lambda$; the horizontal axis is the (baseline) control price $p$, at which the experiment is conducted.  The region where condition (a) is not satisfied ($GTE_{\pi}< 0$), so there is no change of sign, is in red.  The region where condition (b) is not satisfied ($Bias_{\pi}^{LR} < GTE_\pi$ or $Bias_{\pi}^{CR} < GTE_\pi$, respectively), so there is no change of sign, is in blue. }
         \label{fig:change_of_sign}
\end{figure}

Figure (\ref{fig:change_of_sign}) shows numerical results on the set of prices where a change of sign occurs. 

 Consistent with Propositions \ref{pr: Bias_beta_zero} and \ref{pr: Bias_beta_infty}, in Figure \ref{fig:change_of_sign} we observe that the change of sign region becomes very small in the demand-constrained regime for a CR experiment and in the supply constrained regime for an LR experiment. At intermediate market balance regimes, we can more immediately observe that the change of sign region corresponds to a single continuous interval on prices, as Theorems \ref{thm: LR_interval} and \ref{thm: CR_interval} suggest. We observe that in the demand-constrained regime Condition (b) for a change of sign is more likely to be satisfied by the LR experiments, while in the supply-constrained regime Condition (b) is more likely to be satisfied by the CR experiments as expected. 
We also see that the region where a change of sign occurs is, overall, increasing with market balance for the CR experiments and decreasing for the LR experiment.

\subsubsection{Gap from Optimality}
One important implication of Figure \ref{fig:change_of_sign} is the impact that the change of sign region has on the net profits of firms who use canonical price experimentation methods. Consider the practical importance of the change of sign region's borders. The right border of the change of sign region (where purple meets red), represents the optimal price for that particular market balance parameter $\lambda$. This is the point where $GTE_{\pi} = 0$. The left border of the change of sign region (where blue meets purple) represents the final price a firm would reach, were it to repeatedly update it's price in response to canonical A/B experimentation. This is the point where $\widehat{GTE}_\pi = 0$, or equivalently, where $Bias_\pi = GTE_\pi$. Note that this price corresponds to the optimum suggested by repeated canonical A/B experiments. This is because, to the right of that price point, the experiment suggests decreasing the price, and to the left, increasing it.

\begin{figure}[!ht]
    \centering
    \includegraphics[width=0.44\textwidth]{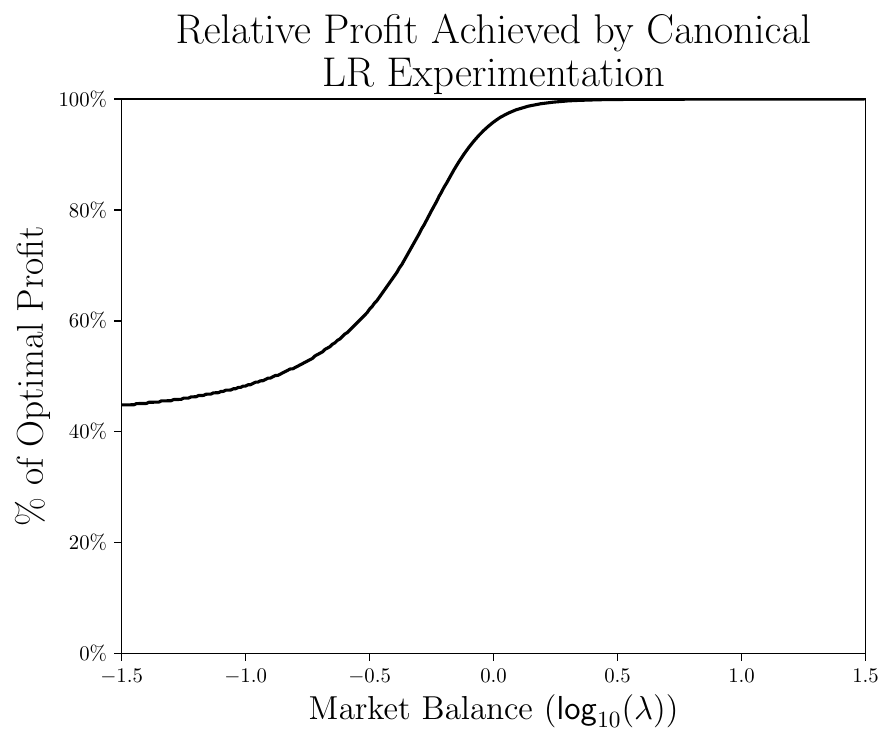} \includegraphics[width=0.44\textwidth]{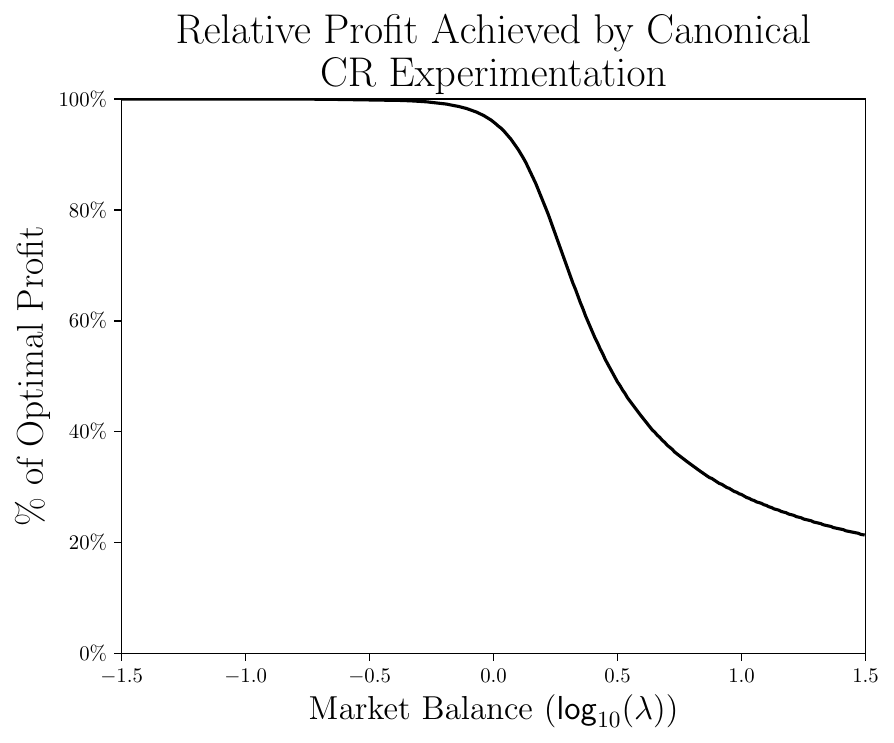}
      \caption{Illustration of the gap from optimal profits that canonical LR, and CR A/B experimentation yields, with $v(p) = e^{V-p}$, $V=5$, $c=1$, $\rho=1$, and $\epsilon=1$.}
         \label{fig:optimality_gap}
\end{figure}

The difference between these two prices (the true optimal price and the experimentally optimized price) can correspond to significant losses in overall profit. In Figure \ref{fig:optimality_gap}, we see the percent of true optimal profit that a firm will achieve if it uses canonical LR (left) or CR (right) experimentation, for a range of market balance scenarios. 

What we find is that the profit achieved from optimization via  experimentation can be far below the optimal profit, with the impact worsening in highly demand-constrained markets (for LR experiments) or demand-constrained markets (for CR experiments).  Even when markets are reasonably balanced (i.e., $\lambda \approx 1$), both LR and CR experimentation will lead to a profit that is more than 4\% lower than optimal. Overall, these findings suggest that firms who use canonical A/B methods for price experimentation might be operating well below their potentially optimal profits, even if they are in well balanced markets. Furthermore, these losses can increase significantly as markets become more imbalanced,  getting to $20-40\%$ of optimal.


\section{An Empirical Case Study}
\label{sec: empirics}

In this section, we investigate the feasibility of changes of sign in price experments in a calibrated, real-world setting. In particular, we consider the lodging marketplace Airbnb, using market conditions for a number of cities and market segments estimated by \citet{Fradkin2022}.  We leverage their estimates to calibrate our change of sign conditions \eqref{eq:condA} and \eqref{eq:condB}.  As we show, the estimates of \citet{Fradkin2022} allow us to obtain interpretable conditions under which a change of sign might occur in a price experiment in each city and market segment.  Notably, we demonstrate that for a reasonable range of market conditions, changes of sign would be expected in practice.

We begin by rewriting our change of sign conditions to be more suited to the \citet{Fradkin2022} data that estimates own price elasticities for (city, market) segments. Recall the following two necessary and sufficient conditions for a change of sign to occur for the canonical price experiment we study, cf. \eqref{eq:condA}-\eqref{eq:condB}:
\[\begin{aligned}
    \mbox{Condition (a): \ \ }& GTE_\pi(p)\geq 0;\\
    \mbox{Condition (b): \ \ }& Bias_\pi(p,q)\geq GTE(p).
\end{aligned}\]

Note that $e_p(p)$ is the price elasticity of demand and is defined as $e_p(p) := pD'(p) /D(p)$. The first of these conditions, Condition (a), ensures that the true global treatment effect of a sufficiently small price increase is positive. In other words, this condition is only satisfied when the firm is currently (locally) ``underpricing" relative to the optimal price . We can rewrite this condition using the monopoly pricing optimality condition as follows:
\begin{align} GTE_\pi(p)\geq 0
\iff & \frac{p}{p-c} +e_p(p) \geq 0. \label{eq: condA_simple}\end{align}

The second of these conditions, Condition (b), is satisfied when the interference bias is at least as large as the GTE itself. For the purposes of this exercise, we suppose that $q=1/2$, i.e., that half of all units are treated, and half are left in control.  As discussed in Section \ref{sec: debiasing}, under this equal allocation experimental design, there holds $D_{1x}(p,p,q)  = D_{0y}(p,p,q)$.

In order to rewrite Condition (b), we take advantage of \eqref{eq: bias_pi_alt}:
\[ Bias_\pi(p,q) = (p-c)\left(D_{0y}(p,p,q) + D_{1x}(p,p,q)\right). \]
Combining this with the assumption that $q = 1/2$, we can rewrite bias as:
\[ Bias_\pi(p,1/2) = 2(p-c)D_{0y}(p,p,1/2). \]
To proceed, define $e_{0y}(p,p,q)$ as the {\em cross-price elasticity} of control demand, i.e., the percentage change in control demand for a one percent increase in the treatment price. Recalling that $D_0(p,p,q)=Q_0(p,p,q)/(1-q)$, we define $e_{0y}(p,p,q)$ as follows:
\[ e_{0y}(p,p,q) := \frac{Q_{0y}(p,p,q)p}{Q_0(p,p,q)} = \frac{D_{0y}(p,p,q)p}{D_0(p,p,q)}. \]
We can then write Condition (b) as:
\begin{align}
    Bias_\pi(p,1/2)\geq GTE_\pi(p) \iff& 2(p-c) D_{0y}(p,p,1/2)  \geq D(p) + (p-c)D'(p)\notag\\
    \iff& \frac{2(p-c)}{p} \frac{pD_{0y}(p,p,1/2)}{D(p)}  \geq 1 + \frac{(p-c)}{p}\frac{pD'(p)}{D(p)}\notag\\
    \iff& \frac{2(p-c)}{p} e_{0y}(p,p,1/2)  \geq 1 + \frac{(p-c)}{p}e_p(p)\notag\\
    \iff&  2e_{0y}(p,p,1/2)  \geq \frac{p}{p-c} + e_p(p)\label{eq: condB_simple}
\end{align}
Note that $D_0(p,p,q) = D(p)$ by Assumption \ref{as: control_treat_demand}. Now, combining \eqref{eq: condA_simple}-\eqref{eq: condB_simple} allows us to write the change of sign conditions as the following chain of inequalities:
\begin{equation}
2e_{0y}(p,p,1/2)  \geq \frac{p}{p-c} + e_p(p)\geq 0 \label{eq: sign_flip_chain}
\end{equation}

Next, we investigate the feasibility of this chain of inequalities holding in real world markets. Observe that in \eqref{eq: sign_flip_chain} there are four unknown variables: the price $p$, the cost $c$, the own price elasticity of control demand $e_p(p)$, and the cross-price elasticity of demand $e_{0y}(p,p,1/2)$. \citet{Fradkin2022} reports estimates of the costs and own price elasticities of demand for Airbnb market segments (economy, midscale, upscale, luxury) in 10 cities around the United States. This leaves us with two missing variables: the price and the cross-price elasticity. 

It is important to note that the cross-price elasticity, $e_{0y}(p,p,1/2)$, has a different interpretations in a two-sided market like Airbnb when running LR and CR experiments. In an LR experiment, $e_{0y}(p,p,1/2)$ quantifies the change in demand for control listings when the price of treated listings is raised. The experiment effectively creates two categories of products with higher (treatment) and lower (control) prices. If treatment raises the price, then control demand rises because price sensitive customers move away from the higher priced treatment listings, towards the control listings.  The cross-price elasticity captures this effect, exactly as usual in demand modeling.

In a CR experiment, $e_{0y}(p,p,1/2)$ has a  different interpretation. Now, $e_{0y}(p,p,1/2)$ quantifies the degree to which demand from control customers changes when the price is raised for treatment customers. The mechanism by which this occurs depends on an {\em availability} effect: if the treatment raises the price, then as demand from treatment customers falls, product availability increases, which in turn positively impacts control demand. To demonstrate this more formally, with a slight abuse of notation, let $D(p,s)$ represent demand when price is $p$, and available inventory is $s$ (out of total inventory $\rho$). If we run a CR experiment on a fraction $q$ of customers, the following balance equation must hold:
\[(1-q) D(p_0, s) + q D(p_1,s) = \rho - s\]
This equation implicitly defines $s$ in terms of $p_0, p_1$, and $q$.  Indeed, observe that this a more general version of the CR steady-state condition given for the logit demand model in \eqref{eq: CR_Market_Bal}. If we now differentiate with respect to $p_1$, we obtain:
\[(1-q) D_s(p_0, s)\frac{\partial s}{\partial p_1} + q\left[D_p(p_1,s) + D_s(p_1,s)\frac{\partial s}{\partial p_1}\right]  = -\frac{\partial s}{\partial p_1}\]
Setting $p_1=p_0$ gives:
\begin{equation}
    D_s(p_0, s)\frac{\partial s}{\partial p_1}  = -\frac{\partial s}{\partial p_1} - q D_p(p_0,s)\label{eq: CR_cross_eq}
\end{equation}

In \eqref{eq: CR_cross_eq}, the left hand side represents the change in rescaled control demand as the treatment price is raised; this is $D_{0y}(p,p,q)$ for CR experiments. The right hand side shows that this quantity depends on two effects: a direct price effect on demand, $D_p(p_0,s)$; but also the impact of the price change on equilibrium availability, $\partial s/\partial p_1$. Measuring this quantity requires empirical calibration of the change in supply rationing when prices are changed for a subset of customers.  With published estimates such as those in \cite{Fradkin2022}, it is difficult to estimate any reasonable range for this availability effect.

We therefore focus on LR experimentation in our presentation, rather than CR experimentation.  Although $e_{0y}(p,p,1/2)$ is not observed, it is possible to consider a reasonable range for this cross elasticity, by comparing it to the underlying own price elasticity of demand in each market segment as reported in \citet{Fradkin2022}.
As we have discussed above, LR price experimentation is the industry standard.  Therefore this approach also captures the regime of most interest to practitioners.

Returning now to the chain of inequalities which produce a sign flip (\ref{eq: sign_flip_chain}), we can use the data reported in \citet{Fradkin2022} to plot ranges of prices and cross-price elasticities for which we would expect to see changes of sign in canonical LR price experiments. For price, we consider a range beginning at cost, and ending slightly above the optimal monopoly price implied by $\hat e_p$ and $\hat c$. For the cross-price elasticity we plot beginning at 0, and end at $2\times -\hat e_p$. We choose this range of cross-price elasticities because in general we expect the cross-price elasticity to be commensurate with the magnitude of the own price elasticity in LR experiments. Since the true cross-price elasticity is unknowable given our data, we select this range to be conservative. We should note that while it is possible to have a cross-price elasticity that exceeds $2\times -\hat e_p$, in these settings the interference bias will be so large that a sign flip is even more likely if the firm is underpricing.

\begin{figure}[ht]
\centering
\includegraphics[height=0.56\textwidth]{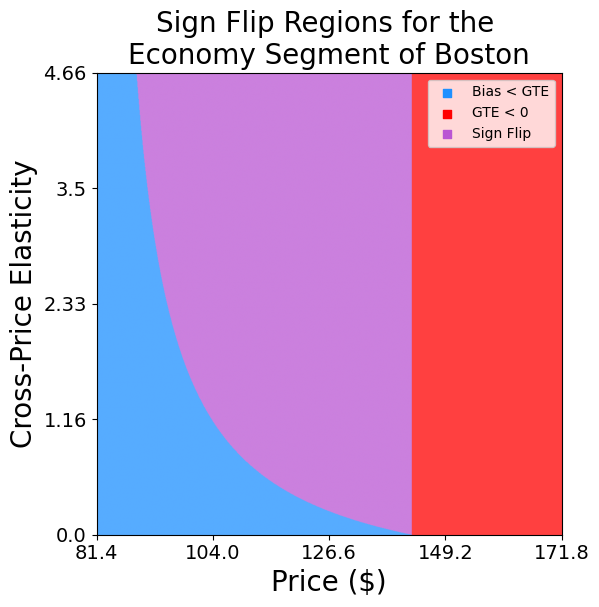}
\caption{Change of sign region for price and cross-price elasticity in the Airbnb Boston Economy Segment. \citet{Fradkin2022} estimates $\hat c = \$81.4$, and $\hat e_p(p): -2.33$.}
\label{fig: E_bound_region}
\end{figure}

Figure \ref{fig: E_bound_region} shows the change of sign region plot for the economy segment of Airbnb listings in Boston.  (See Appendix \ref{sec: add_figures_8} for similar plots for the other market segments, as well as the other cities we consider; the results are similar to Figure \ref{fig: E_bound_region}.) In the blue region the bias is too low to cause a change of sign (Condition (b) is violated), while in the red region the current price is already too high (Condition (a) is violated).  For parameters in the purple region, our analysis suggests that changes of sign will occur. It is important to note that since we do not observe the market price or the cross-price elasticity of demand, the estimates in \citet{Fradkin2022} are insufficient to identify the exact state of the world on this plot.

Nevertheless, one can reasonably conclude from Figure \ref{fig: E_bound_region} that changes of sign are plausible.  In particular, if listings are even mildly underpriced, and the cross-price elasticity of control demand is comparable to the magnitude of the own-price elasticity of demand (i.e., approximately $2$), then our results show that changes of sign would arise: the experiment will incorrectly suggest that the current price is too high, and the experimenter will be encouraged to move the price in the {\em wrong} direction.  We view these  results as compelling evidence that changes of sign are a phenomenon that can have significant consequence for practitioners.


\section{Conclusion and Future Work}
\label{sec: conclusion}

In this paper we studied the biases that arise due to interference in platform pricing experiments, with a particular emphasis on profit and revenue metrics. Notably, we find that in certain settings, these biases can be so severe that they introduce a ``change of sign'' into the canonical estimators for profit, suggesting to firms that they should move prices (or fees) in the wrong direction. We then derive a novel method for debiasing canonical A/B experiments, by splitting units equally between treatment and control. We conclude by giving empirical evidence for the plausibility of estimators with the wrong sign in practical settings, arguing that firms should be more cautious when using canonical A/B pricing experiments.

We propose multiple direction for future work. First, we believe that more analysis needs to be done on the debiasing of canonical estimators. Our debiasing method, while novel, relies on the treatment/control split being precisely 50/50. A debiasing method that functions in all treatment split settings would be both more flexible, and would allow us to mitigate the sampling inefficiency of our current debiasing method. Our current debiasing approach only uses half of the available data, namely only the control demand, which may be a source of increased variance in practice.

A second, and related direction for future work involves quantifying the variance of these estimators themselves. As we note in Section \ref{sec: logit}, an interval of prices always exists in which a sign flip will occur in expectation (for a logit demand function), but this interval is bordered by the optimal price. In close-to-optimal settings, it is possible for estimators to be assigned the wrong sign entirely by fault of natural variance in the experiment. Typically, practitioners avoid this by dismissing results that are not statistically significant, meaning many of these incorrectly signed estimators may be ignored in practice because they fail to reach some critical p-value. Identifying conditions under which we would expect to see a statistically significant estimator with the wrong sign would be helpful to fully quantify how these sign flips actually impact practitioners. That said, our view is that practitioners should be more concerned with the impact of bias as studied in our paper, since variance vanishes with large sample sizes while bias does not.

Finally, our paper motivates the need for further empirical study into this phenomenon.  Our empirical results suggest that estimators with the wrong sign may be more prevalent than previously expected. One approach might be for practitioners to employ a "meta-experiment" similar to \cite{holtz2025reducing}, with both a canonical LR experiment and a cluster-randomized experiment, as a way to compare the usual approach to a relatively less biased alternative.  More generally, our work suggests there is benefit in developing empirical tests to identify {\em when} debiasing is needed to avoid making wrong decisions.

\clearpage
\bibliography{Sections/bibliography.bib}

\begin{thebibliography}{}

\bibitem[Aparicio et~al., 2023]{aparicio2023algorithmic}
Aparicio, D., Eckles, D., and Kumar, M. (2023).
\newblock Algorithmic pricing and consumer sensitivity to price variability.
\newblock {\em Available at SSRN 4435831}.

\bibitem[Athey et~al., 2018]{Athey18}
Athey, S., Eckles, D., and Imbens, G.~W. (2018).
\newblock Exact p-values for network interference.
\newblock {\em Journal of the American Statistical Association}, 113(521):230--240.

\bibitem[Bajari et~al., 2023]{bajari2023multiple}
Bajari, P., Burdick, B., Imbens, G.~W., Masoero, L., McQueen, J., Richardson, T., and Rosen, I.~M. (2023).
\newblock Multiple randomization designs.
\newblock {\em Statistical Science}, 38(3):458–476.

\bibitem[Basse et~al., 2019]{Basse19}
Basse, G.~W., Feller, A., and Toulis, P. (2019).
\newblock {Randomization tests of causal effects under interference}.
\newblock {\em Biometrika}, 106(2):487--494.

\bibitem[Blake and Coey, 2014]{blake2014marketplace}
Blake, T. and Coey, D. (2014).
\newblock Why marketplace experimentation is harder than it seems: The role of test-control interference.
\newblock In {\em Proceedings of the fifteenth ACM Conference on Economics and Computation}, pages 567--582.

\bibitem[Bojinov et~al., 2022]{bojinov2022design}
Bojinov, I., Simchi-Levi, D., and Zhao, J. (2022).
\newblock Design and analysis of switchback experiments.
\newblock {\em Management Science}.

\bibitem[Bright et~al., 2023]{lobel2023}
Bright, I., Delarue, A., and Lobel, I. (2023).
\newblock Reducing marketplace interference bias via shadow prices.
\newblock {\em ACM Conference on Economics and Computation (EC), 2023}.

\bibitem[Candogan et~al., 2024]{candogan2024correlated}
Candogan, O., Chen, C., and Niazadeh, R. (2024).
\newblock Correlated cluster-based randomized experiments: Robust variance minimization.
\newblock {\em Management Science}, 70(6):4069--4086.

\bibitem[Carneiro et~al., 2010]{carneiro2010evaluating}
Carneiro, P., Heckman, J.~J., and Vytlacil, E. (2010).
\newblock Evaluating marginal policy changes and the average effect of treatment for individuals at the margin.
\newblock {\em Econometrica}, 78(1):377--394.

\bibitem[Chamandy, 2016]{chamandy16}
Chamandy, N. (2016).
\newblock Experimentation in a ridesharing marketplace.

\bibitem[Chawla et~al., 2016]{Chawla2016}
Chawla, S., Hartline, J.~D., and Denis, N. (2016).
\newblock A/b testing of auctions.
\newblock {\em ACM EC Conference, 2016}.

\bibitem[Delarue and Kleanthis, 2025]{DelarePricing2025}
Delarue, A. and Kleanthis, K. (2025).
\newblock Pricing experiments in matching marketplaces under interference: Designs and estimators.
\newblock {\em arXiv preprint https://arxiv.org/abs/2502.18839}.

\bibitem[Deng and Le, 2023]{deng2023price}
Deng, A. and Le, T. (2023).
\newblock The price is right: Removing a/b test bias in a marketplace of expirable goods.
\newblock Working paper.

\bibitem[Farias et~al., 2023]{farias2023correcting}
Farias, V., Li, H., Peng, T., Ren, X., Zhang, H., and Zheng, A. (2023).
\newblock Correcting for interference in experiments: A case study at douyin.
\newblock In {\em Proceedings of the 17th ACM Conference on Recommender Systems}, pages 455--466.

\bibitem[Farias et~al., 2022]{farias2022markovian}
Farias, V.~F., Li, A.~A., Peng, T., and Zheng, A. (2022).
\newblock Markovian interference in experiments.

\bibitem[Farronato and Fradkin, 2022]{Fradkin2022}
Farronato, C. and Fradkin, A. (2022).
\newblock The welfare effects of peer entry: The case of airbnb and the accommodation industry.
\newblock {\em American Economic Review}, 112(6):1782--1817.

\bibitem[Fradkin, 2019]{fradkin2019simulation}
Fradkin, A. (2019).
\newblock A simulation approach to designing digital matching platforms.
\newblock {\em Boston University Questrom School of Business Research Paper Forthcoming}.

\bibitem[Glynn et~al., 2020]{glynn2020adaptive}
Glynn, P.~W., Johari, R., and Rasouli, M. (2020).
\newblock Adaptive experimental design with temporal interference: A maximum likelihood approach.
\newblock {\em Advances in Neural Information Processing Systems}, 33:15054--15064.

\bibitem[Han et~al., 2023]{han2023detecting}
Han, K., Li, S., Mao, J., and Wu, H. (2023).
\newblock Detecting interference in online controlled experiments with increasing allocation.
\newblock In {\em Proceedings of the 29th ACM SIGKDD Conference on Knowledge Discovery and Data Mining}, pages 661--672.

\bibitem[Holtz et~al., 2025]{holtz2025reducing}
Holtz, D., Lobel, F., Lobel, R., Liskovich, I., and Aral, S. (2025).
\newblock Reducing interference bias in online marketplace experiments using cluster randomization: Evidence from a pricing meta-experiment on airbnb.
\newblock {\em Management Science}, 71(1):390--406.

\bibitem[Hu et~al., 2022]{hu2022average}
Hu, Y., Li, S., and Wager, S. (2022).
\newblock Average direct and indirect causal effects under interference.
\newblock {\em Biometrika}, 109(4):1165--1172.

\bibitem[Hu and Wager, 2022]{hu2022switchback}
Hu, Y. and Wager, S. (2022).
\newblock Switchback experiments under geometric mixing.
\newblock {\em arXiv preprint arXiv:2209.00197}.

\bibitem[Imbens and Rubin, 2015]{ImbensRubin15}
Imbens, G.~W. and Rubin, D.~B. (2015).
\newblock {\em Causal Inference for Statistics, Social, and Biomedical Sciences: An Introduction}.
\newblock Cambridge University Press, USA.

\bibitem[Johari et~al., 2022]{johari2022experimental}
Johari, R., Li, H., Liskovich, I., and Weintraub, G.~Y. (2022).
\newblock Experimental design in two-sided platforms: An analysis of bias.
\newblock {\em Management Science}, 68(10):7069--7089.

\bibitem[Lariviere, 2006]{lariviere2006note}
Lariviere, M.~A. (2006).
\newblock A note on probability distributions with increasing generalized failure rates.
\newblock {\em Operations Research}, 54(3):602--604.

\bibitem[Li et~al., 2023]{li2023experimenting}
Li, S., Johari, R., Kuang, X., and Wager, S. (2023).
\newblock Experimenting under stochastic congestion.
\newblock {\em arXiv preprint arXiv:2302.12093}.

\bibitem[Li et~al., 2025]{LiWangWang2025}
Li, S., Wang, C., and Wang, J. (2025).
\newblock Choosing the better bandit algorithm under data sharing: When do a/b experiments work?
\newblock {\em arXiv preprint arXiv:2507.11891}.

\bibitem[Liu et~al., 2020]{liu2020trustworthy}
Liu, M., Mao, J., and Kang, K. (2020).
\newblock Trustworthy online marketplace experimentation with budget-split design.
\newblock {\em arXiv preprint arXiv:2012.08724}.

\bibitem[Manski, 2013]{Manski13}
Manski, C.~F. (2013).
\newblock Identification of treatment response with social interactions.
\newblock {\em The Econometrics Journal}, 16(1):S1--S23.

\bibitem[Munro et~al., 2021]{munro2021treatment}
Munro, E., Wager, S., and Xu, K. (2021).
\newblock Treatment effects in market equilibrium.
\newblock {\em arXiv preprint arXiv:2109.11647}.

\bibitem[Phillips, 2021]{Phil_book}
Phillips, R. (2021).
\newblock {\em Pricing and Revenue Optimization}.
\newblock Stanford Business Books.

\bibitem[Pouget-Abadie et~al., 2019]{pouget2019variance}
Pouget-Abadie, J., Aydin, K., Schudy, W., Brodersen, K., and Mirrokni, V. (2019).
\newblock Variance reduction in bipartite experiments through correlation clustering.
\newblock In {\em Advances in Neural Information Processing Systems}, pages 13288--13298.

\bibitem[Roemheld and Rao, 2024]{roemheld2023interference}
Roemheld, L. and Rao, J. (2024).
\newblock {Interference Produces False-Positive Pricing Experiments}.
\newblock (2402.14538).

\bibitem[Saveski et~al., 2017]{Saveski17}
Saveski, M., Pouget-Abadie, J., Saint-Jacques, G., Duan, W., Ghosh, S., Xu, Y., and Airoldi, E.~M. (2017).
\newblock Detecting network effects: Randomizing over randomized experiments.
\newblock In {\em Proceedings of the 23rd ACM SIGKDD International Conference on Knowledge Discovery and Data Mining}, KDD ’17, page 1027–1035, New York, NY, USA. Association for Computing Machinery.

\bibitem[Shirani and Bayati, 2024]{shirani2023causal}
Shirani, S. and Bayati, M. (2024).
\newblock Causal message-passing for experiments with unknown and general network interference.
\newblock {\em Proceedings of the National Academy of Sciences}, 121(40):e2322232121.

\bibitem[Simchi-Levi and Wang, 2023]{simchi2023pricing}
Simchi-Levi, D. and Wang, C. (2023).
\newblock Pricing experimental design: causal effect, expected revenue and tail risk.
\newblock In {\em International Conference on Machine Learning}, pages 31788--31799. PMLR.

\bibitem[Ugander et~al., 2013]{Ugander13}
Ugander, J., Karrer, B., Backstrom, L., and Kleinberg, J. (2013).
\newblock Graph cluster randomization: Network exposure to multiple universes.
\newblock In {\em Proceedings of the 19th ACM SIGKDD International Conference on Knowledge Discovery and Data Mining}, KDD ’13, page 329–337, New York, NY, USA. Association for Computing Machinery.

\bibitem[Wager and Xu, 2021]{wager2021experimenting}
Wager, S. and Xu, K. (2021).
\newblock Experimenting in equilibrium.
\newblock {\em Management Science}, 67(11):6694--6715.

\end{thebibliography}
\clearpage
\begin{APPENDICES}
\section{Additional Figures for Section \ref{sec: logit}}
\label{sec: add_figures_7 }

\begin{figure}[!ht]     
    \centering
    \includegraphics[width=0.5\textwidth]{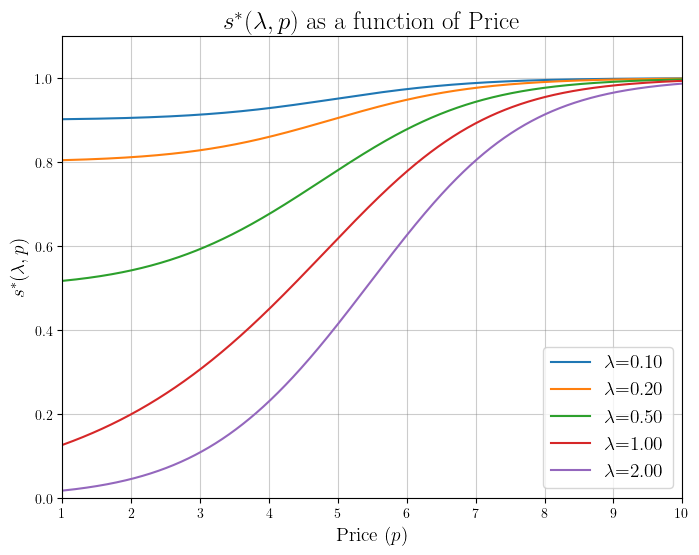}
    \caption[]{Illustration of the listing availability in the mean field limit as a function of price ($p$) and market balance ($\lambda$). Model instance is given by $v(p)=e^{V-p}$, $\varepsilon = 1$, $c=1$, $V = 5$, and $\rho = 1$.
    }
     \label{fig1}
\end{figure}

\begin{figure}[!ht] 
    \centering
    \includegraphics[width=0.5\textwidth]{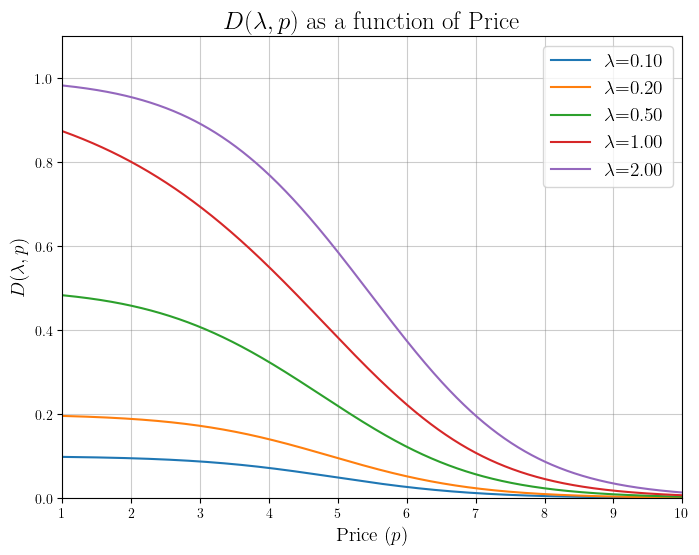} 
    \caption[]{Illustration of demand function in the mean field limit as a function of price ($p$) and market balance ($\lambda$). Same instance as in Figure \ref{fig1}.}
     \label{fig2}
\end{figure}

\vspace{5mm}

\section{Additional Figures for Section \ref{sec: empirics}}
\label{sec: add_figures_8}

In this Section we present figures with results that are analogous to Figure \ref{fig: E_bound_region} in Section \ref{sec: empirics}, for all four segments in each of ten cities studied in \cite{Fradkin2022}.

\newcommand{\plotset}[1]{
\includegraphics[width=0.23\textwidth]{Figures/Elast_bound_graphs/#1_Airbnb_Economy.pdf} &
\includegraphics[width=0.23\textwidth]{Figures/Elast_bound_graphs/#1_Airbnb_Midscale.pdf} &
\includegraphics[width=0.23\textwidth]{Figures/Elast_bound_graphs/#1_Airbnb_Upscale.pdf} &
\includegraphics[width=0.23\textwidth]{Figures/Elast_bound_graphs/#1_Airbnb_Luxury.pdf} \\
\hspace{0pt} &
\hspace{0pt} \\
}

\begin{figure}[p]
\centering
\setlength{\tabcolsep}{2pt}
\renewcommand{\arraystretch}{0}
\begin{tabular}{cccc}
\plotset{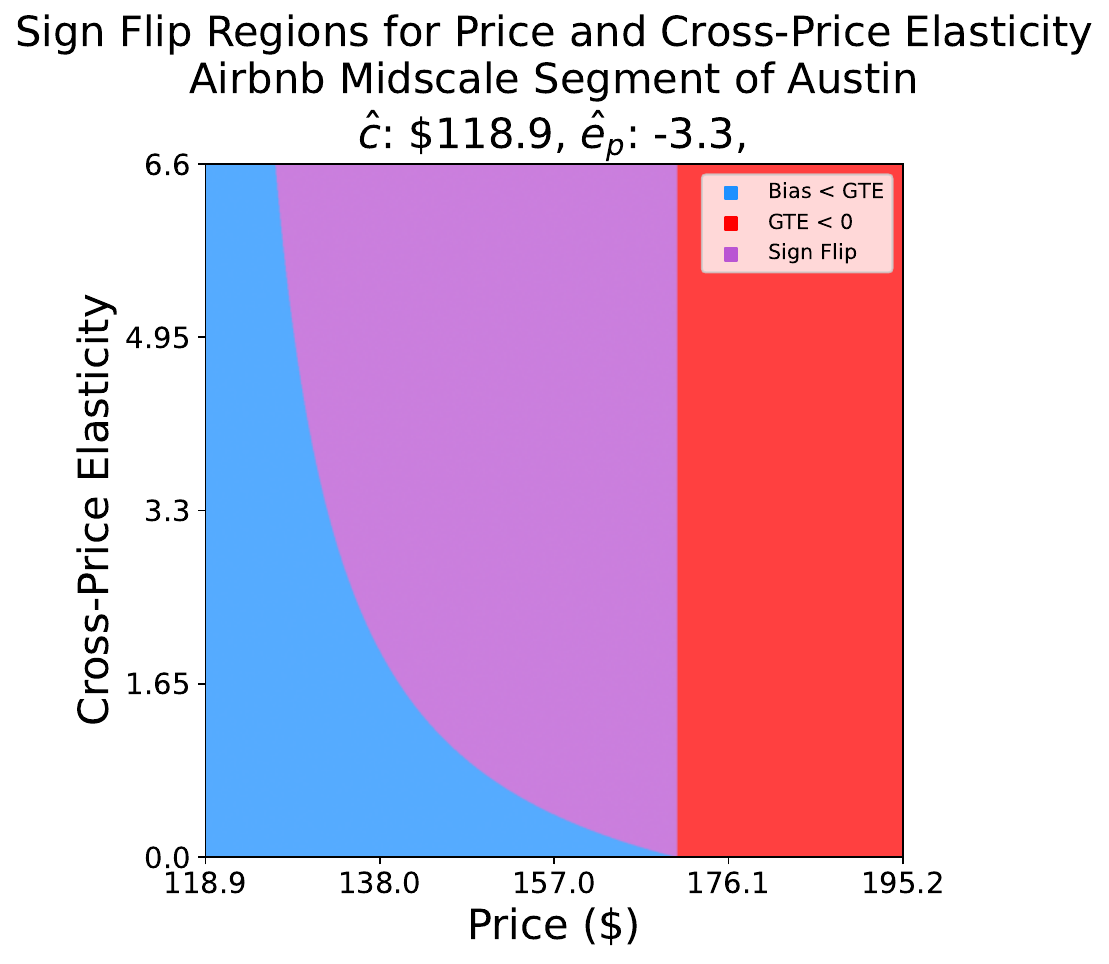}
\plotset{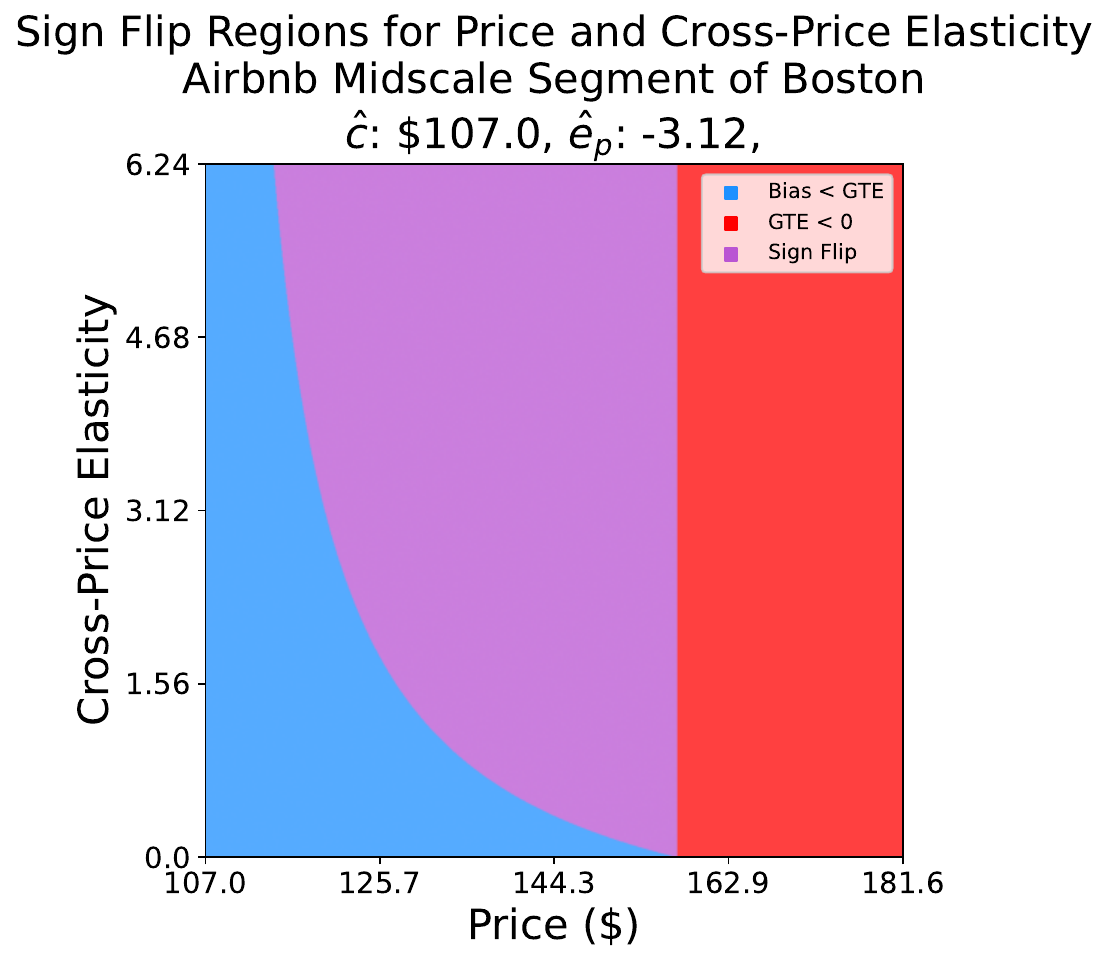}
\plotset{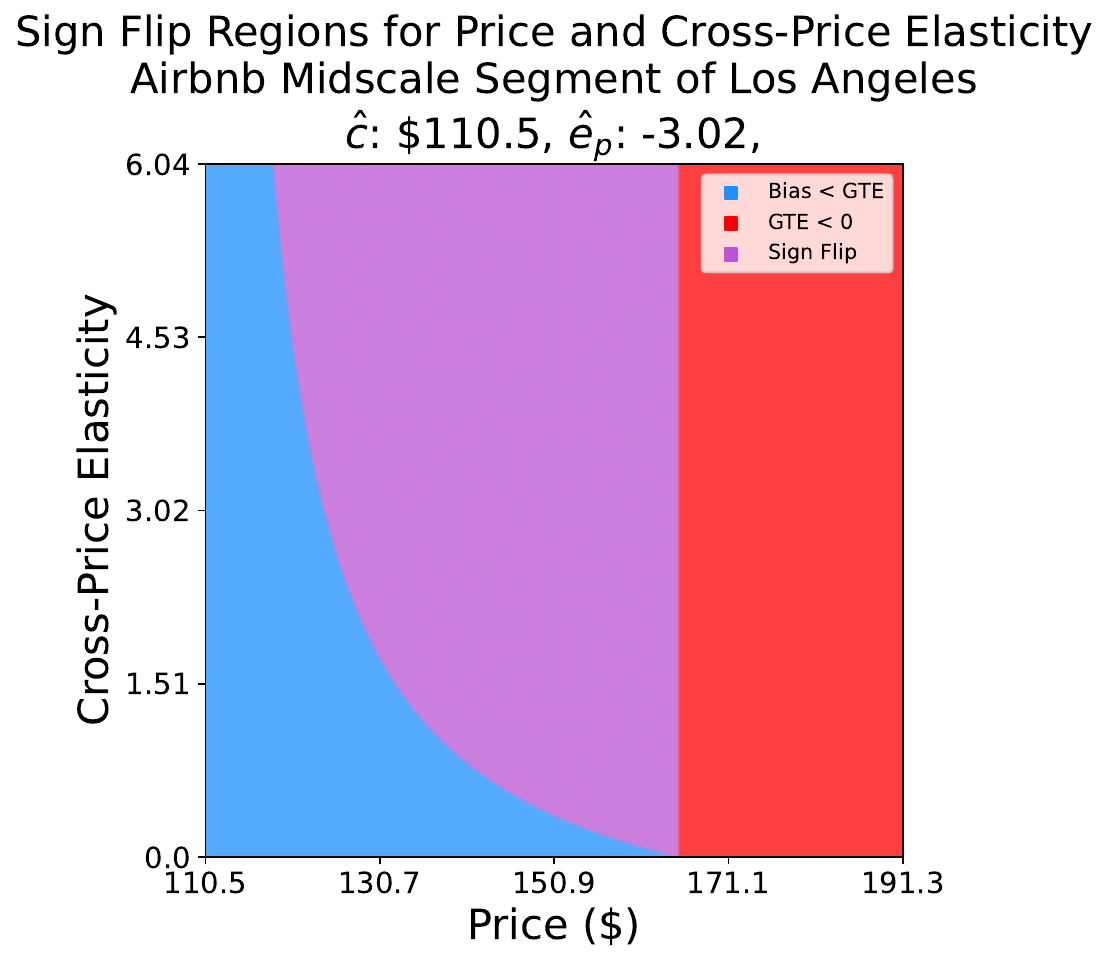}
\plotset{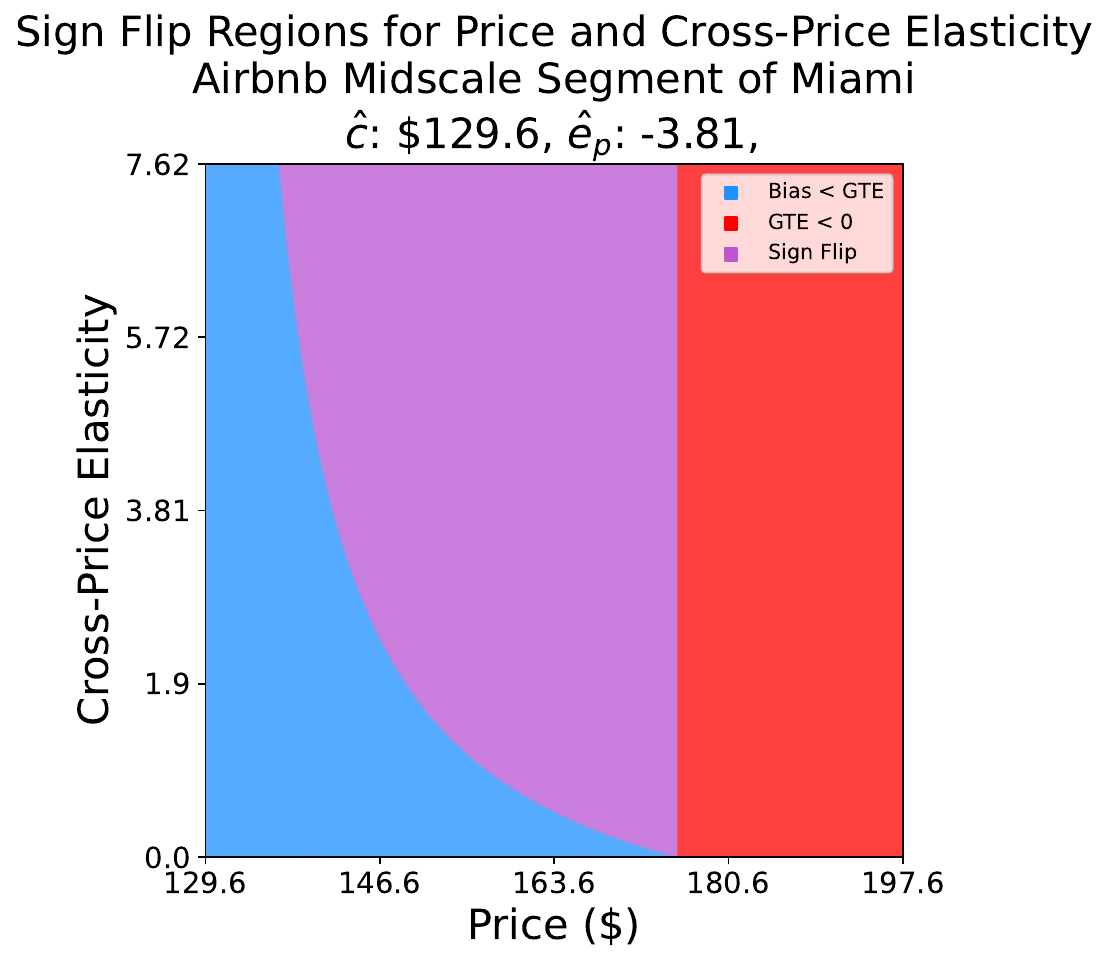}
\plotset{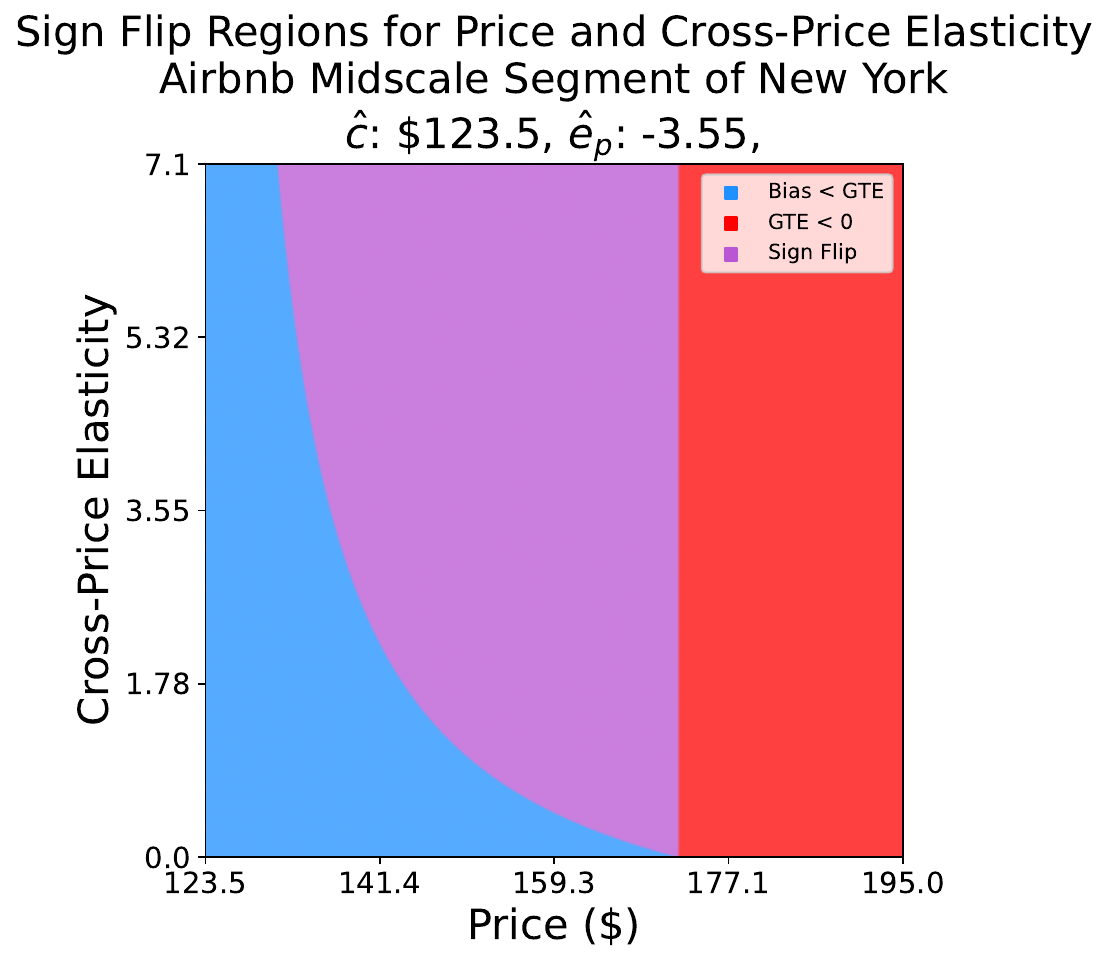}
\end{tabular}
\caption{Set 1 of 2: Change of sign region (purple) for each Airbnb segment, for Austin, Boston, Los Angeles, Miami, and New York}
\end{figure}

\clearpage

\begin{figure}[p]
\centering
\setlength{\tabcolsep}{2pt}
\renewcommand{\arraystretch}{0}
\begin{tabular}{cccc}
\plotset{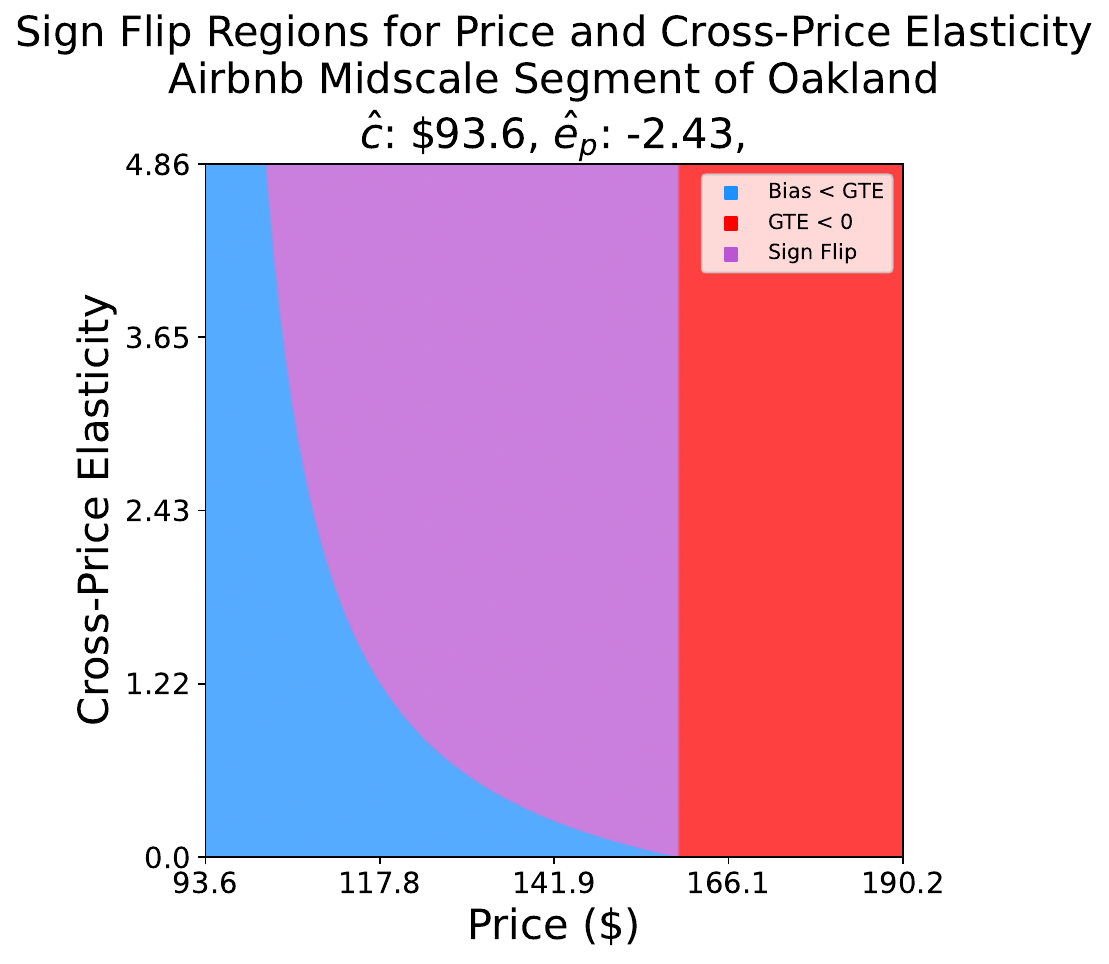}
\plotset{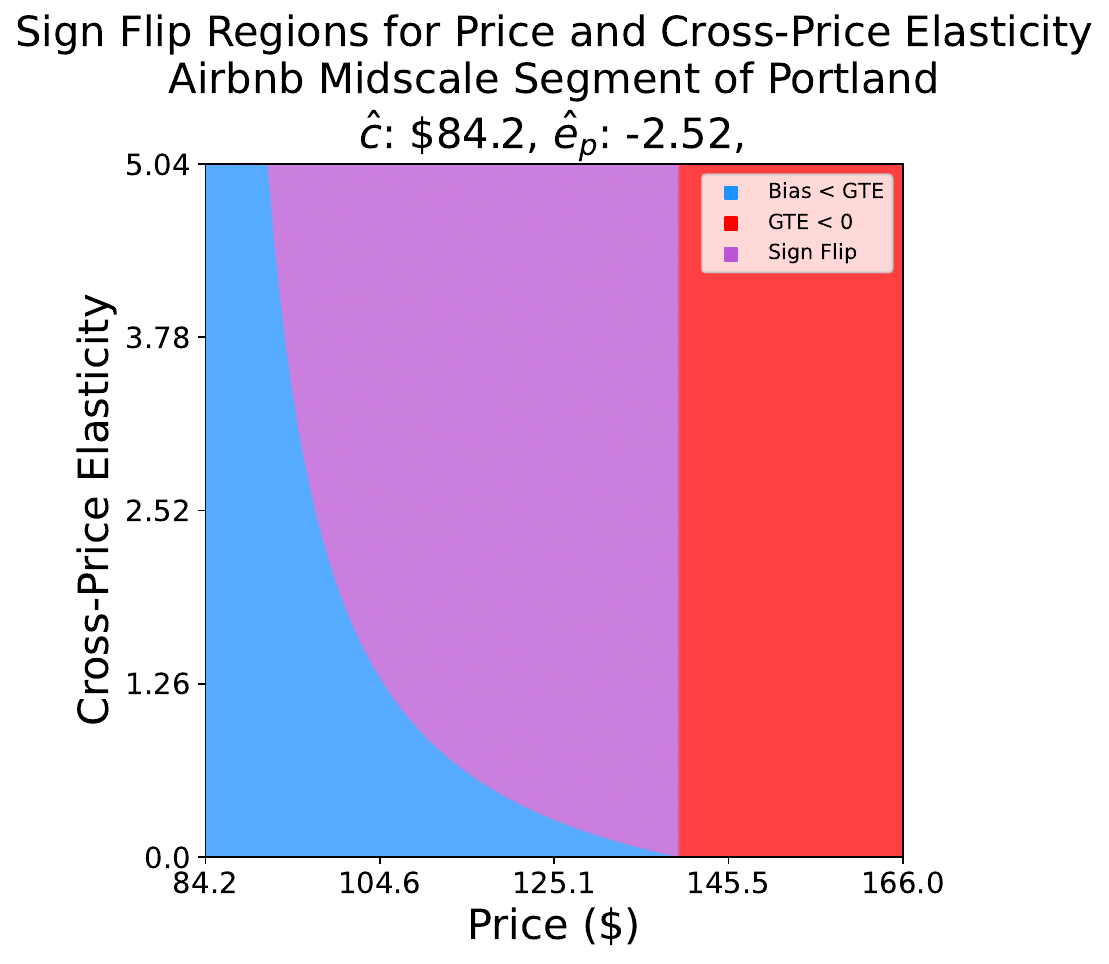}
\plotset{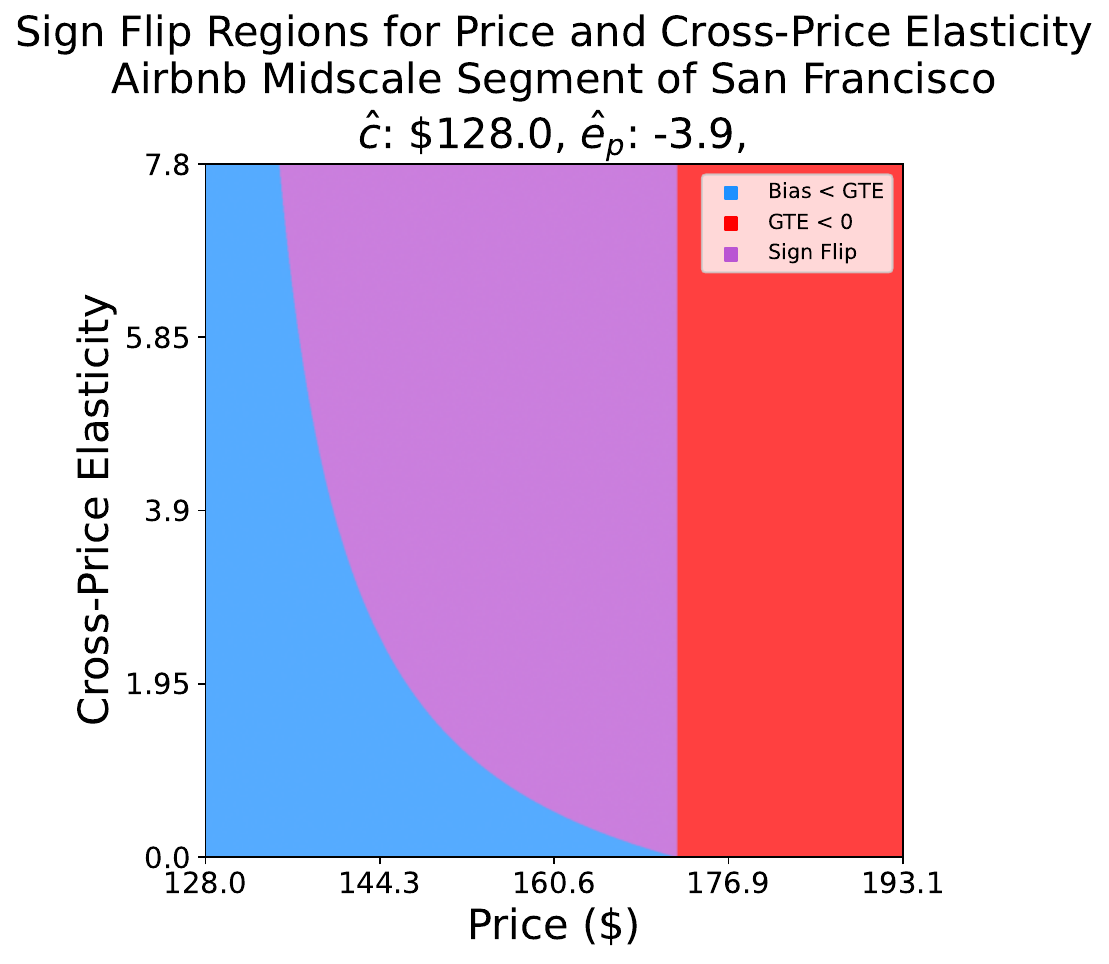}
\plotset{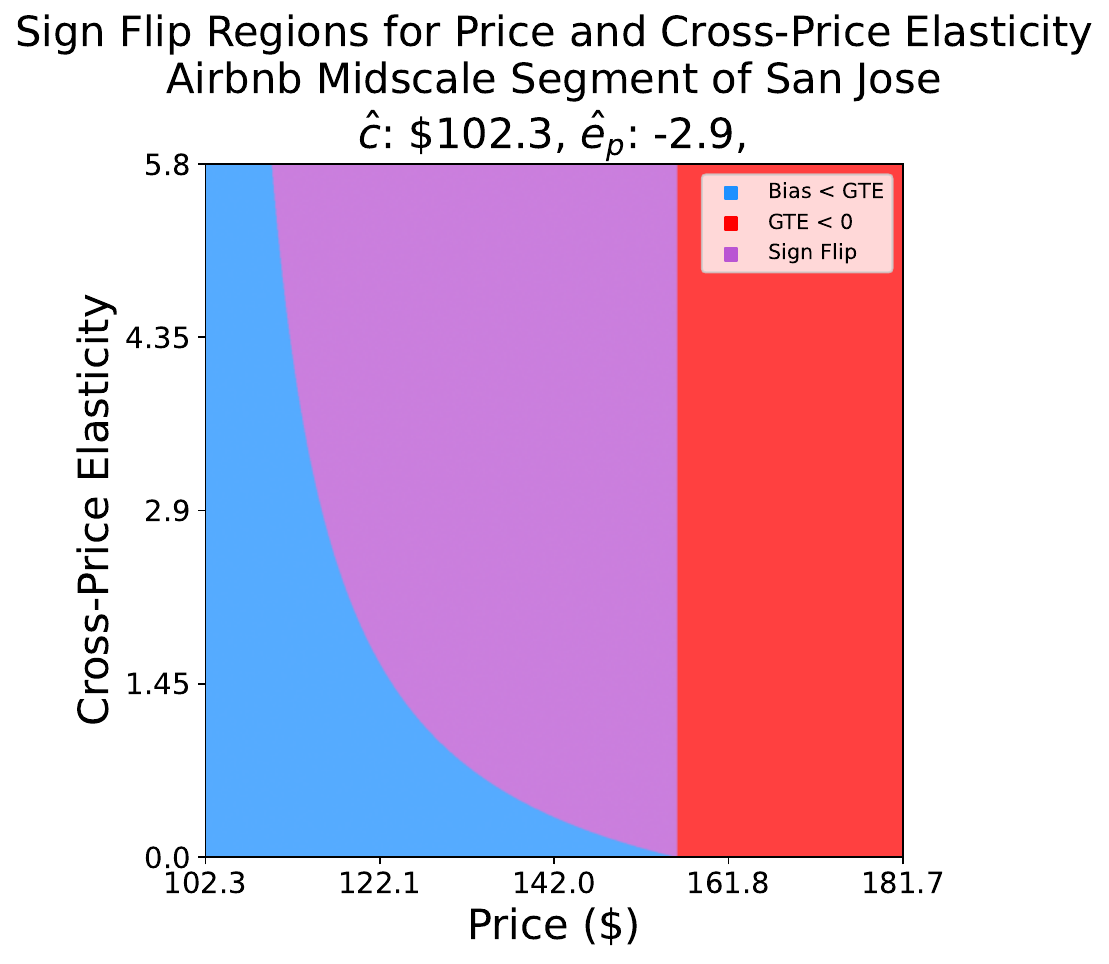}
\plotset{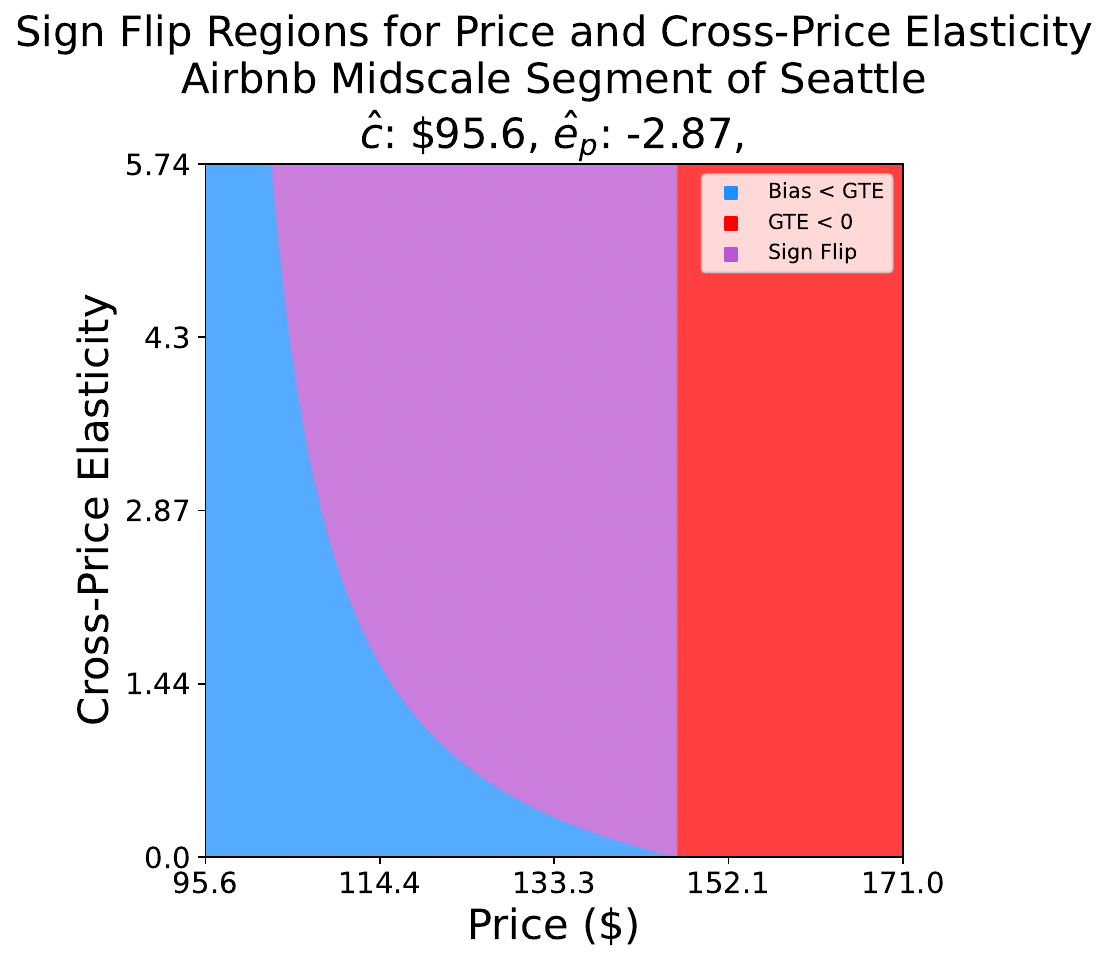}
\end{tabular}
\caption{Set 2 of 2: Change of sign region (purple) for each Airbnb segment, for Oakland, Portland, San Francisco, San Jose, and Seattle}
\end{figure}

\clearpage

\clearpage

\section{Fee Model }
\label{sec: fee_model}

In this section, we consider an environment in which a firm charges a fixed sales fee, as opposed to setting a price and incurring a marginal cost for each purchase. This represents the common situation--particularly in two-sided online marketplaces--where the platform does not directly control the prices of the goods on their platforms, but instead collect a fixed fee for the sales of heterogeneous goods. We include this section to make clear that our mathematical results still apply in the practically relevant case where the continuous variable being set is a fee.

Consider $N$ products, with a fixed vector of prices $\vec{p}\in\mathbb{R}^N$. The demand function $\vec{D}:\mathbb{R}^N\to \mathbb{R}^N$, takes in a vector of prices and outputs a vector in $\mathbb{R}^N$ that represents the quantity demanded for each good. We assume that the platform selects a single fee for all products $f\in\mathbb{R}^+$; this fee represents a percentage of all money transacted that the platform will collect. We assume that this fee is passed on entirely to the customer and that the platform's marginal cost is zero. Therefore,
we can write the platform's profit $\Pi(f,\vec{p})$ as follows:\footnote{For example, our specification corresponds to a setting in which the fee is charged directly to buyers by the platform and where we ignore that prices may be a function of the fee, since firms may pass-through fee changes. However, since experiments tend to be short-lived, we ignore these longer-term re-equilibration effects.}

\begin{equation}
    \Pi(f,\vec{p}) = f\cdot\vec{p}^\top \vec{D}((1+f)\vec{p}) \ .
    \label{eq: plat_profit}
\end{equation}

Let $D'(\cdot)$ represent the matrix of demand derivatives (the Jacobian) in price. We can then write the platform's profit derivative with respect to the fee as:
\begin{equation}
    \Pi_f(f,\vec{p}) = \vec{p}^\top \vec{D}((1+f)\vec{p}) + 
    f\cdot \vec{p}^\top D'((1+f)\vec{p}) \vec{p} \ . \label{eq: profit_f}
\end{equation}

Let $R(f,\vec{p})$ represent the total revenue associated to all goods sold on the platform when the fee is $f$ and the prices are $\vec{p}$. We can write $R(f,\vec{p})$ as:
\[R(f,\vec{p}) = \vec{p}^\top \vec{D}((1+f)\vec{p}) \ .\]

Let $R_f(f,\vec{p})$ represent the derivative of this quantity with respect to the fee. This allows us to rewrite (\ref{eq: profit_f}) as:
\begin{equation}
    \label{eq: fee_profit_rev}
    \Pi_f(f,\vec{p}) = R(f,\vec{p}) + fR_f(f,\vec{p}).
\end{equation}

\begin{assumption}
     We assume that $\Pi_f$ exists and is well defined for all $f\in\mathbb{R}^+$ and all $\vec{p}\in\mathbb{R}^N$, or in other words, that $\Pi$ is differentiable over its domain. (Equivalently, $\vec{D}$, and thus $R$, is differentiable over its domain.) 
\end{assumption}

When running an A/B experiment, the platform will compare a control fee $f_0$ to a treatment fee $f_1$. To keep the presentation succinct, in this section we focus only on the global treatment effect of profits, $\Pi(f_1,\vec{p})-\Pi(f_0,\vec{p})$.  In particular, we define the local $GTE$ at the control fee $f_0$ as:
\begin{equation}
    \label{eq: fee_GTE}
    GTE_\Pi(f_0,\vec{p}) = \Pi_f(f_0,\vec{p}).
\end{equation}

We assume that a fraction $q$ of all subjects are put in treatment and $1-q$ into control. Note that as in the monopoly pricing example, this randomization can occur on either the customer or listing side. For simplicity, in an LR experiment, we assume that for each of the $n$ products, a fraction $q$ is in treatment. Let $R_0(f_0,f_1,\vec{p},q)$ (resp., $R_1(f_0,f_1,\vec{p},q)$) be the treatment fraction adjusted revenue for control (resp., treatment) subjects in the experiment. Formally: $R_0(f_0,f_1,\vec{p},q) =  \vec{p}^\top \vec{D_0}(f_0,f_1,\vec{p},q)/(1-q)$ and $R_1(f_0,f_1,\vec{p},q) =  \vec{p}^\top \vec{D_1}(f_0,f_1,\vec{p},q)/q$, where $\vec{D_0}$ and $\vec{D_1}$ represent demand for control and treatment units respectively during the experiment.

Let $\Pi_0(f_0,f_1,\vec{p},q)$ (resp., $\Pi_1(f_0,f_1,\vec{p},q)$) represent the platform's treatment fraction adjusted profit from control (resp., treatment) subjects during the experiment. That is, $\Pi_0(f_0,f_1,\vec{p},q) = f_0R_0(f_0,f_1,\vec{p},q)$ and $\Pi_1(f_0,f_1,\vec{p},q) = f_1R_1(f_0,f_1,\vec{p},q)$. We make the following assumptions for these functions, analogous to Assumptions \ref{as: diff_D} and \ref{as: control_treat_demand}.3.

\begin{assumption} 
\label{ass: fee_rev}
The following holds:
\begin{enumerate}
\item $R_0(f_0,f_1,\vec{p},q)$ and $R_1(f_0,f_1,\vec{p},q)$ are both differentiable with respect to $f_0, f_1$ over their domain. 
\item $R_1(f,f,\vec{p},q) = R_0(f,f,\vec{p},q) = R(f,\vec{p}).$
\end{enumerate}
\end{assumption}

Let $f_1^n$ be a sequence such that $\lim_{n\to\infty} f_1^n=f_0$.  We then write the canonical estimator $\widehat{GTE}_\Pi$ as follows:
\[\widehat{GTE}_\Pi(f_0,\vec{p},q) = \lim_{n\to\infty}\frac{\Pi_1(f_0,f_1^n,\vec{p},q)- \Pi_0(f_0,f_1^n,\vec{p},q)}{f_1^n-f_0}.\]

This is equivalent to:
\begin{equation}
\label{eq: fee_HatGTE}
    \widehat{GTE}_\Pi(f_0,\vec{p},q) = \Pi_{1y}(f_0,f_0,\vec{p},q)-\Pi_{0y}(f_0,f_0,\vec{p},q),
\end{equation}
where $\Pi_{0y}$ and $\Pi_{1y}$ represent the derivative of the control and treatment platform profit functions with respect to the second argument respectively.

We define the bias as $Bias_\Pi(f_0,\vec{p},q) = GTE_\Pi(f_0,\vec{p},q) - \widehat{GTE}_\Pi(f_0,\vec{p},q)$. Using Assumption \ref{ass: fee_rev} we get that $\Pi(f_0,\vec{p},q) = \Pi_1(f_0,f_0,\vec{p},q)$. We can then use an analogous argument to our derivation of \eqref{eq: bias_pi_alt} to write bias as:
\begin{equation}
\label{eq: fee_bias}
    Bias_\Pi(f_0,\vec{p},q) = \Pi_{0y}(f_0,f_0,\vec{p},q) +\Pi_{1x}(f_0,f_0,\vec{p},q).
\end{equation}

Using our revenue functions and (\ref{eq: fee_profit_rev}) we can rewrite the preceding expressions to be more useful. First we rewrite bias as:
\[Bias_\Pi(f_0,\vec{p},q) = f_0R_{0y}(f_0,f_0,\vec{p},q)+ f_0R_{1x}(f_0,f_0,\vec{p},q).\]

Then, we  rewrite $\widehat{GTE}_\Pi$ from equation (\ref{eq: fee_HatGTE}) as:
\[\widehat{GTE}_\Pi(f_0,\vec{p},q) = R_1(f_0,f_0,\vec{p},q) + f_0R_{1y}(f_0,f_0,\vec{p},q)-f_0R_{0y}(f_0,f_0,\vec{p},q).\]

As before, there are two necessary and sufficient conditions for the canonical estimator to have the wrong sign in a fee experiment. Note that as was true in the pricing problem, bias is always positive when firms test a small fee increase. This means that the conditions required for a sign flip to occur are $GTE_\Pi(f_0,\vec{p})\geq 0$, meaning the true global treatment effect of a small fee increase is non-negative, and $\widehat{GTE}_\Pi(f_0,\vec{p},q)\leq 0$, that is the estimate of GTE is negative. 

We can now conclude this section by stating precisely the necessary and sufficient conditions, required for a change of sign in a fee experiment. First:
\[\text{Condition (a):} \quad R(f,\vec{p}) + fR_f(f,\vec{p}) \geq 0. \]

And second:
\[\text{Condition (b):} \quad 2 R_{0y}(f_0,f_0,\vec{p},q) + R_{1x}(f_0,f_0,\vec{p},q) - R_{1y}(f_0,f_0,\vec{p},q) \geq \frac{ R(f_0,\vec{p})}{f_0}.\]

\clearpage

\section{Two-Sided Market Model: Analysis of Bias}
\label{sec: logit_model_limit_bias}

In this section we study structural properties of biases with a particular focus on extreme regimes of market balance. The analysis in this subsection closely follows \cite{johari2022experimental}. 

In an LR experiment,  cannibalization between control and treated listings creates bias. Upon arrival, customers consider both treatment and control listings while choosing which listing to book, creating competition between listings.  In contrast, in CR experiments there is competition between treatment and control customers for the same limited supply. A listing booked by, e.g., a control customer will not be available to a treatment customer.\footnote{$Bias^{CR}$ would be zero in a setting with unlimited supply.} $Bias^{LR}$ and $Bias^{CR}$ capture these respective competition effects and both of them are naturally characterized by the  cross derivatives between groups. 

As the market becomes more demand constrained ($\lambda \rightarrow 0$), competition between customers declines resulting in less interference in a CR experiment. In the limit, there is no competition between customers and we expect the bias in a CR experiment to be zero. On the other hand, when demand is constrained competition between listings is high, resulting in more interference for an LR experiment. Because of this, in the limit as $\lambda \rightarrow 0$, we expect the bias in an LR experiment strictly positive. 

Conversely as the market becomes supply constrained ($\lambda \rightarrow \infty$), competition between listings declines resulting in less interference in a LR experiment. In the limit, there is no competition between listings and so we expect the bias in an LR experiment to be zero. Meanwhile, competition between customers increases resulting in more interference in a CR experiment. Because of this, in the limit as $\lambda \rightarrow \infty$, we expect the bias in a CR experiment to be strictly positive.

The following propositions formalize this intuition about the behavior of biases in the extremes of market balance:

\begin{proposition} 
\label{pr: Bias_beta_zero}

For all $p\in [c,+\infty)$, we have that:
\[ \lim_{\lambda\to 0} GTE_{D}(\lambda,p)/\lambda =\frac{  \rho  v'(p) \varepsilon}{(\varepsilon+\rho v(p))^2}; \ \ \lim_{\lambda\to 0} GTE_{\pi}(\lambda,p)/\lambda =\frac{\rho v(p)}{\varepsilon+\rho v(p)}+  \frac{  (p-c)\rho  v'(p) \varepsilon}{(\varepsilon+\rho v(p))^2}. \]

We have that $\lim_{\lambda\to 0} Bias_{\pi}^{CR}(\lambda,p)/\lambda=(p-c)\cdot\lim_{\lambda\to 0} Bias_{D}^{CR}(\lambda,p)/\lambda = 0$, while 
\[ \lim_{\lambda\to 0} Bias_{\pi}^{LR} (\lambda,p)/\lambda = (p-c)\cdot \lim_{\lambda \to 0} Bias_{D}^{
LR}(\lambda,p)/\lambda = -\frac{(p-c)\rho^2 v(p) v'(p) }{(\varepsilon+\rho v(p))^2}>0. \]
\end{proposition}

\begin{proposition}
\label{pr: Bias_beta_infty}
For all $p\in [c,+\infty)$, we have that: $\lim_{\lambda \to \infty} GTE_D(\lambda, p) = 0$ and that
$\lim_{\lambda\to \infty} GTE_{\pi}(\lambda,p) = \rho$. We have $\lim_{\lambda\to \infty} Bias_{\pi}^{LR}(\lambda,q)= (p-c) \cdot\lim_{\lambda\to \infty} Bias_{D}^{LR}(\lambda,q) =0$, while
\[ \lim_{\lambda\to \infty} Bias_{\pi}^{CR}(\lambda,p)= (p-c)\cdot \lim_{\lambda\to \infty} Bias_{\pi}^{CR}(\lambda,p) = -\frac{\rho (p-c) v'(p)}{v(p)}>0. \]
\end{proposition}

In the limit where $\lambda \rightarrow 0$, Proposition  \ref{pr: Bias_beta_zero} suggests the canonical estimator in the CR experiment is unbiased while the estimator in the LR experiment is biased.  Similarly, in the limit where $\lambda \rightarrow \infty$, Proposition  \ref{pr: Bias_beta_infty} suggests that the canonical estimator in the CR experiment is biased while the estimator in the LR experiment is unbiased. 
Proofs of both propositions are presented in Appendix \ref{app:proofs}.  The proofs are similar to the analysis in \cite{johari2022experimental} but we also leverage our characterization of biases based on cross-partial derivatives to obtain our closed-form expressions.

\proof{Proof of Proposition \ref{pr: Bias_beta_zero}}
First, consider the market equilibrium equation (\ref{eq: market_balance}):
$$ \rho-s^*(\lambda,p)=\lambda \frac{s^*(\lambda,p)v(p)}{\varepsilon+s^*(\lambda,p)v(p)}.$$

By dividing both sides by $s^*(\lambda,p)$, the left hand side becomes $\rho/s^*(\lambda,p)-1$. The right-hand side remains bounded, and therefore, it converges to zero as $\lambda\to 0$. Therefore, $\lim \limits_{\lambda\rightarrow 0} \rho/s^*(\lambda,p)-1=0$, which implies that $\lim \limits_{\lambda\rightarrow 0} s^*(\lambda,p)=\rho$. Next, consider the limit of $Bias_{\pi}^{CR}/\lambda$ as $\lambda\to0$:
\[\lim \limits_{\lambda\rightarrow 0}Bias_{\pi}^{CR}(\lambda,p)/\lambda=\lim \limits_{\lambda\rightarrow 0} \lambda\frac{ \varepsilon ^2 (c-p) v(p) v'(p) s^*(\lambda,p)}{(v(p) s^*(\lambda,p)+\varepsilon )^2 \left[  (s^*(\lambda,p)v(p)+\varepsilon)^2+\lambda v(p) \varepsilon \right]}.\]

Since the quotient in this expression is bounded, this will converge to 0 as $\lambda\to0$, giving us our first desired claim that $\lim_{\lambda \to 0} Bias_{\pi}^{CR}(\lambda,p)/\lambda=0$; since demand bias is proportional to profit bias, the result for $Bias_D^{CR}$ follows as well. The next step in our proof is to evaluate the limit of $Bias_{\pi}^{LR}/\lambda$ as $\lambda$ goes to 0: 
\[
\begin{aligned}
\lim \limits_{\lambda\rightarrow 0} Bias_{\pi}^{LR}(\beta,p)/\lambda&=
\lim \limits_{\lambda\rightarrow 0}
-(p-c)\frac{s^*(\lambda,p)^2 v(p) v'(p) (\varepsilon+s^*(\lambda,p) v(p) )}{[\varepsilon+s^*(\lambda,p) v(p) +\lambda v(p)]
   [(s^*(\lambda,p)v(p) +\varepsilon)^2 +\lambda v(p) \varepsilon ]}\\
&=
-(p-c)\frac{\rho^2 v(p) v'(p) (\varepsilon+\rho v(p) )}{[\varepsilon+\rho v(p) +0]
[(\rho v(p) +\varepsilon)^2 + 0 ]}\\
&=\frac{-  (p-c) \rho^2 v(p) v'(p) }{
     (\varepsilon+\rho v(p))^2} \ ,
\end{aligned}
\]
which is exactly as stated in Proposition \ref{pr: Bias_beta_zero}. Finally, we want to prove the stated result on $GTE_{\pi}$. From equation (\ref{eq: GTE_MarketBalance}), we have that $GTE_{\pi}=\rho-s^*(\lambda,p)- (p-c) s_p^*(\lambda,p)$. If we then substitute in our formula for $s_p^*(\lambda,p)$ we can solve as follows:
\[
\begin{aligned} 
\lim \limits_{\lambda\rightarrow 0}GTE_{\pi}/\lambda&=\lim \limits_{\lambda\rightarrow 0} \frac{\rho-s^*(\lambda,p)}{\lambda}+ \frac{ (p-c)  s^*(\lambda,p) v'(p) \varepsilon}{(\varepsilon+s^*(\lambda,p)v(p))^2+ \lambda v(p) \varepsilon}\\
&=\lim \limits_{\lambda\rightarrow 0} \frac{s^*(\lambda,p)v(p)}{\varepsilon+s^*(\lambda,p)v(p)}+ \frac{(p-c)   s^*(\lambda,p) v'(p) \varepsilon}{(\varepsilon+s^*(\lambda,p)v(p))^2+ \lambda v(p) \varepsilon}\\
&=\frac{\rho v(p)}{\varepsilon+\rho v(p)}+  \frac{  (p-c)\rho  v'(p) \varepsilon}{(\varepsilon+\rho v(p))^2}, \\ 
\end{aligned}
\]
where the second equation follows from the market balance equation (\ref{eq: market_balance}). This concludes our proof of Proposition  \ref{pr: Bias_beta_zero}.
 \qedsymbol{} 
\endproof

\proof{Proof of Proposition \ref{pr: Bias_beta_infty}}

First we prove that $\lim \limits_{\lambda\rightarrow \infty} s^*(\lambda,p)=0$. By dividing both sides of the market balance equation (\ref{eq: market_balance}) by $\lambda$, and then taking the limit as $\lambda$ approaches $\infty$, we arrive at:
\[\lim\limits_{\lambda\rightarrow \infty}\frac{s^*(\lambda,p)v(p)}{\varepsilon+s^*(\lambda,p)v(p)} = \lim\limits_{\lambda\rightarrow \infty}\frac{\rho  - s^*(\lambda,p)}{\lambda} \leq  \lim\limits_{\lambda\rightarrow \infty}\frac{\rho}{\lambda} =0.\]

Since $\varepsilon$ and $v(p)$ are both strictly positive by assumption, and $s(\lambda,p)\geq 0$, this implies $\lim \limits_{\lambda\rightarrow \infty} s^*(\lambda,p)=0$. Now, note the limit of the LR bias as $\lambda\to\infty$ is given by: 
\[
\begin{aligned}
\lim\limits_{\lambda\rightarrow \infty}Bias_{\pi}^{LR}(\lambda,p) &= \lim\limits_{\lambda\rightarrow \infty}(p-c)\frac{-\lambda  s^*(\lambda,p)^2 v(p) v'(p) (\varepsilon+s^*(\lambda,p) v(p) )}{[\varepsilon+s^*(\lambda,p) v(p) +\lambda v(p)][(s^*(\lambda,p)v(p) +\varepsilon)^2 +\lambda v(p) \varepsilon ]
}\\
&= \lim\limits_{\lambda\rightarrow \infty}(p-c)\frac{-s^*(\lambda,p)^2 v(p) v'(p) (\varepsilon+s^*(\lambda,p) v(p) )}{
[\varepsilon+s^*(\lambda,p) v(p) +\lambda v(p)][(s^*(\lambda,p)v(p) +\varepsilon)^2/\lambda +v(p) \varepsilon ]
   }\\
&=0 \ ,
\end{aligned}
\]
because the numerator converges to zero and the denominator diverges to infinity. Hence, we obtain the desired result. Now, recall that CR bias is given by:
\[
\begin{aligned}
\lim \limits_{\lambda\rightarrow \infty} Bias_{\pi}^{CR}(\lambda,p)
&=
\lim \limits_{\lambda\rightarrow 0} \frac{ \lambda^2\varepsilon ^2 (c-p) v(p) v'(p) s^*(\lambda,p)}{(v(p) s^*(\lambda,p)+\varepsilon )^2 \left[  (s^*(\lambda,p)v(p)+\varepsilon)^2+\lambda v(p) \varepsilon \right]}
\\
&= \lim \limits_{\lambda\rightarrow \infty} \frac{[\lambda s^*(\lambda,p)v(p)]  \varepsilon ^2 (c-p) v'(p) }{(v(p) s^*(\lambda,p)+\varepsilon )^2 \left[ \frac{1}{\lambda} (s^*(\lambda,p)v(p)+\varepsilon)^2+ v(p) \varepsilon \right]}\\
&= \lim \limits_{\lambda\rightarrow \infty} \frac{  [\rho-s^*(\lambda,p)]\varepsilon ^2 (c-p) v'(p)}{(v(p) s^*(\lambda,p)+\varepsilon ) \left[ \frac{1}{\lambda} (s^*(\lambda,p)v(p)+\varepsilon)^2+ v(p) \varepsilon \right]} \quad\text{using equation (\ref{eq: market_balance})}\\
&= \frac{ \rho(c-p) v'(p)}{ v(p) } \ ,
\end{aligned}
\]
because $\lim \limits_{\lambda\rightarrow \infty} s^*(\lambda,p)=0$.

From equation (\ref{eq: GTE_MarketBalance}), we have that $GTE_{\pi}=\rho-s^*(\lambda,p)- (p-c) s_p^*(\lambda,p)$. If we then substitute in our formula for $s_p^*(\lambda,p)$ we can solve as follows:
\begin{equation}
\lim\limits_{\lambda\rightarrow \infty}GTE_{\pi} = \lim\limits_{\lambda\rightarrow \infty} \rho-s^*(\lambda,p)+ (p-c) \frac{  s^*(\lambda,p) v'(p) \varepsilon}{\frac{1}{\lambda}(\varepsilon+s^*(\lambda,p)v(p))^2+ v(p) \varepsilon} =\rho  \notag ,
\end{equation}

again because $\lim \limits_{\lambda\rightarrow \infty} s^*(\lambda,p)=0$.  The results for demand bias and $GTE_D$ follow analogously.  This concludes our proof of Proposition (\ref{pr: Bias_beta_infty}). \qedsymbol{} 
\endproof

\clearpage

\section{Derivations}

\subsection{Derivation of $s^*_{0y}(\lambda,p,p,q)$ and $s^*_{1y}(\lambda,p,p,q)$ in LR settings \label{sec: s0y_s1y_deriv}}

We begin by starting with the listing-side market balance equations (\ref{eq: exp_avail_list}). In the interest of mathematical brevity we will define $\Psi$ as $\Psi = s_0^*(\lambda,p_0,p_1,q)v(p_0)+s_1^*(\lambda,p_0,p_1,q)v(p_1)$. This allows us to write equations (\ref{eq: exp_avail_list}) as:
\[(1-q)\rho-s_0^*(\lambda,p_0,p_1,q)=\lambda \frac{s_0^*(\lambda,p_0,p_1,q)v(p_0)}{\varepsilon+\Psi},\]
\[
q\rho-s_1^*(\lambda,p_0,p_1,q) =\lambda \frac{s_1^*(\lambda,p_0,p_1,q)v(p_1)}{\varepsilon+\Psi}. \]

Our next step is to be take the derivative of each of these equations with respect to $p_1$. First note that $\frac{\partial}{\partial y}\Psi = s_{0y}^*(\lambda,p_0,p_1,q)v(p_0)+s_{1y}^*(\lambda,p_0,p_1,q)v(p_1)+s_{1}^*(\lambda,p_0,p_1,q)v'(p_1)$. Proceeding now by differentiating we get the following:
\begin{equation}
\label{eq: ApB_s0y}
-s_{0y}^*(\lambda,p_0,p_1,q)=
\lambda \frac{(\varepsilon+\Psi) s_{0y}^*(\lambda,p_0,p_1,q)v(p_0)}
{(\varepsilon+\Psi)^2} - 
\lambda\frac{(\frac{\partial}{\partial y}\Psi) s_{0}^*(\lambda,p_0,p_1,q)v(p_0)}
{(\varepsilon+\Psi)^2},
\end{equation}
\begin{equation}
\label{eq: ApB_s1y}-s_{1y}^*(\lambda,p_0,p_1,q)=
\lambda \frac{(\varepsilon+\Psi) (s_{1y}^*(\lambda,p_0,p_1,q)v(p_1)+s_{1}^*(\lambda,p_0,p_1,q)v'(p_1))}
{(\varepsilon+\Psi)^2} - 
\lambda\frac{(\frac{\partial}{\partial y}\Psi) s_{1}^*(\lambda,p_0,p_1,q)v(p_1)}
{(\varepsilon+\Psi)^2}.
\end{equation}

Adding these two equations together we get that:
\[\begin{aligned}
s_{0y}^*(\lambda,p_0,p_1,q) + s_{1y}^*(\lambda,p_0,p_1,q) &= 
-\lambda\frac{(\varepsilon+\Psi)(\frac{\partial}{\partial y}\Psi) - (\frac{\partial}{\partial y}\Psi)(\Psi)}{(\varepsilon+\Psi)^2} = -\lambda\varepsilon\frac{\frac{\partial}{\partial y}\Psi}{(\varepsilon+\Psi)^2}\\
&= \lambda\varepsilon\frac{s_{0y}^*(\lambda,p_0,p_1,q)v(p_0)+s_{1y}^*(\lambda,p_0,p_1,q)v(p_1)+s_{1}^*(\lambda,p_0,p_1,q)v'(p_1)}{(\varepsilon+s_0^*(\lambda,p_0,p_1,q)v(p_0)+s_1^*(\lambda,p_0,p_1,q)v(p_1))^2} 
\end{aligned}\]

Our next step is to evaluate at $p_1=p_0$. Note that this step makes all arguments implied so we will drop them to simplify the formulas. We can now rewrite the above equation as follows and begin simplifying:
\begin{align}s_{0y}^* + s_{1y}^* &= 
-\lambda\varepsilon\frac{s_{0y}^*v+s_{1y}^*v+s_{1}^*v'}{(\varepsilon+s_0^*v+s_1^*v)^2},\notag\\
(s_{0y}^* + s_{1y}^*)(\varepsilon+s_0^*v+s_1^*v)^2 &=
-\lambda\varepsilon v(s_{0y}^*+s_{1y}^*)-\lambda\varepsilon s_{1}^*v'),\notag\\
s_{0y}^* + s_{1y}^* &= -\frac{\lambda\varepsilon s_{1}^*v'}{(\varepsilon+s_0^*v+s_1^*v)^2+\lambda\varepsilon v}. \label{eq: ApB_s0y_p_s1y}
\end{align}

We will now use (\ref{eq: ApB_s0y}) and  (\ref{eq: ApB_s0y_p_s1y}) to solve for $s_{0y}^*(\lambda,p_0,p_0,q)$. Again here we will be dropping arguments, this means we can rewrite (\ref{eq: ApB_s0y}) as:
\[\begin{aligned}
-s_{0y}^* &=
\lambda \frac{(\varepsilon+s_0^*v+s_1^*v) s_{0y}^*v}
{(\varepsilon+s_0^*v+s_1^*v)^2} - 
\lambda\frac{(s_{0y}^*v+s_{1y}^*v+s_{1}^*v') s_{0}^*v}
{(\varepsilon+s_0^*v+s_1^*v)^2},\\
  s_{0y}^* &=
- \frac{ \lambda v}
{(\varepsilon+s_0^*v+s_1^*v)}s_{0y}^* + 
\frac{\lambda s_{0}^*v}
{(\varepsilon+s_0^*v+s_1^*v)^2}[s_{1}^*v'+v(s_{0y}^*+s_{1y}^*)].\end{aligned}\]

Plugging in (\ref{eq: ApB_s0y_p_s1y}) on the far right of the above line we can simplify as follows:
\[\begin{aligned}
s_{0y}^* &=
- \frac{ \lambda v}
{(\varepsilon+s_0^*v+s_1^*v)}s_{0y}^* + 
\frac{\lambda s_{0}^*v}
{(\varepsilon+s_0^*v+s_1^*v)^2}
\left[s_{1}^*v'-\frac{\lambda\varepsilon vs_{1}^*v'}{(\varepsilon+s_0^*v+s_1^*v)^2+\lambda\varepsilon v}\right] ,
\\
s_{0y}^*\left[1+\frac{ \lambda v}
{(\varepsilon+s_0^*v+s_1^*v)}\right] &=
\frac{\lambda s_{0}^*v}
{(\varepsilon+s_0^*v+s_1^*v)^2}
\left[\frac{(\varepsilon+s_0^*v+s_1^*v)^2s_{1}^*v'+\lambda\varepsilon vs_{1}^*v' - \lambda\varepsilon vs_{1}^*v'}{(\varepsilon+s_0^*v+s_1^*v)^2+\lambda\varepsilon v}\right] ,
\\
s_{0y}^*\left[1+\frac{ \lambda v}
{(\varepsilon+s_0^*v+s_1^*v)}\right] &=
\frac{\lambda s_{0}^*s_{1}^*v'v}{(\varepsilon+s_0^*v+s_1^*v)^2+\lambda\varepsilon v} ,
\\
s_{0y}^*\left[(\varepsilon+s_0^*v+s_1^*v)+\lambda v
\right] &=
\frac{\lambda s_{0}^*s_{1}^*v'v(\varepsilon+s_0^*v+s_1^*v)}{(\varepsilon+s_0^*v+s_1^*v)^2+\lambda\varepsilon v} ,
\\
\end{aligned}\]

which by dividing across simplifies to:
\[
    s_{0y}^* =  \frac{\lambda s_{0}^*s_{1}^*v'v(\varepsilon+s_0^*v+s_1^*v)}{[(\varepsilon+s_0^*v+s_1^*v)^2+\lambda\varepsilon v][\varepsilon+s_0^*v+s_1^*v+\lambda v]}.
\]

Our final step is to note that when $p_1=p_0=p$ we have that $s_{0}^*(\lambda,p,p,q) = (1-q) s^*(\lambda,p)$ and $s_{1}^*(\lambda,p,p,q) = q s^*(\lambda,p)$. Plugging in these identities to the above line gives us the following final formula for $s_{0y}^*(\lambda,p,p,q)$:
\[\boxed{s_{0y}^*(\lambda,p,p,q) =  \frac{\lambda q(1-q){s^*(\lambda,p)}^2v'(p)v(p)(\varepsilon+s^*(\lambda,p)v(p))}{[(\varepsilon+s^*(\lambda,p_0)v(p))^2+\lambda\varepsilon v(p)][\varepsilon+s^*(\lambda,p_0)v(p)+\lambda v(p)]}
.}\]

\hfill \break

We now want to solve for $s_{1y}^*(\lambda,p,p,q)$. We will again use  (\ref{eq: ApB_s0y_p_s1y}), this time with (\ref{eq: ApB_s1y}) to solve for $s_{1y}^*(\lambda,p_0,p_0,q)$. Once again we will be dropping implied arguments, allowing us to rewrite (\ref{eq: ApB_s1y}) as:
\[\begin{aligned}
-s_{1y}^*&=
\lambda \frac{(\varepsilon+s_0^*v+s_1^*v) (s_{1y}^*v+s_{1}^*v')}
{(\varepsilon+s_0^*v+s_1^*v)^2} - 
\lambda\frac{(s_{0y}^*v+s_{1y}^*v+s_{1}^*v') s_{1}^*v}
{(\varepsilon+s_0^*v+s_1^*v)^2},\\
s_{1y}^*\left[1+\frac{\lambda v}
{(\varepsilon+s_0^*v+s_1^*v)}\right]&=
 -\frac{ \lambda s_{1}^*v'}
{(\varepsilon+s_0^*v+s_1^*v)} + 
\frac{ \lambda s_{1}^*v}
{(\varepsilon+s_0^*v+s_1^*v)^2}[s_{1}^*v'+v(s_{0y}^*+s_{1y}^*)].
  \end{aligned}\]

Plugging in (\ref{eq: ApB_s0y_p_s1y}) on the far right of the above line we can simplify as follows:
\[\begin{aligned}
s_{1y}^*\left[1+\frac{\lambda v}
{(\varepsilon+s_0^*v+s_1^*v)}\right]&=
-\frac{ \lambda s_{1}^*v'}
{(\varepsilon+s_0^*v+s_1^*v)} + 
\frac{ \lambda s_{1}^*v}
{(\varepsilon+s_0^*v+s_1^*v)^2}
\left[s_{1}^*v'-\frac{\lambda\varepsilon s_{1}^*v'v}{(\varepsilon+s_0^*v+s_1^*v)^2+\lambda\varepsilon v}\right],
\\
s_{1y}^*\left[1+\frac{\lambda v}
{(\varepsilon+s_0^*v+s_1^*v)}\right]&=
-\frac{ \lambda s_{1}^*v'}
{(\varepsilon+s_0^*v+s_1^*v)} + 
\frac{ \lambda s_{1}^*v}
{(\varepsilon+s_0^*v+s_1^*v)^2}
\left[\frac{s_{1}^*v'(\varepsilon+s_0^*v+s_1^*v)^2 +\lambda\varepsilon s_{1}^*v'v - \lambda\varepsilon s_{1}^*v'v}{(\varepsilon+s_0^*v+s_1^*v)^2+\lambda\varepsilon v}\right],
\\
s_{1y}^*\left[1+\frac{\lambda v}
{(\varepsilon+s_0^*v+s_1^*v)}\right]&=
-\frac{ \lambda s_{1}^*v'}
{(\varepsilon+s_0^*v+s_1^*v)} + 
\left[\frac{\lambda {s_{1}^*}^2v'v }{(\varepsilon+s_0^*v+s_1^*v)^2+\lambda\varepsilon v}\right],
\\
s_{1y}^*\left[(\varepsilon+s_0^*v+s_1^*v)+\lambda v\right]&=
-\lambda s_{1}^*v'
 + 
\left[\frac{\lambda {s_{1}^*}^2v'v(\varepsilon+s_0^*v+s_1^*v) }{(\varepsilon+s_0^*v+s_1^*v)^2+\lambda\varepsilon v}\right],\\
  \end{aligned}\]
which by dividing across simplifies to:
\[s_{1y}^*=
\frac{\lambda {s_{1}^*}^2v'v(\varepsilon+s_0^*v+s_1^*v) }{[(\varepsilon+s_0^*v+s_1^*v)^2+\lambda\varepsilon v][\varepsilon+s_0^*v+s_1^*v+\lambda v]} - 
\frac{\lambda s_{1}^*v'}{\varepsilon+s_0^*v+s_1^*v+\lambda v}.
\]

As a final step we will again use the fact that when $p_1=p_0=p$ we have $s_{0}^*(\lambda,p,p,q) = (1-q) s^*(\lambda,p)$ and $s_{1}^*(\lambda,p,p,q) = q s^*(\lambda,p)$. Plugging in these identities to the above line gives us the following final formula for $s_{1y}^*(\lambda,p,p,q)$:
\[\boxed{s_{1y}^*(\lambda,p,p,q)=
\frac{\lambda q^2{s^*(\lambda,p)}^2v'(p)v(p)(\varepsilon+s^*(\lambda,p)v(p)) }{[(\varepsilon+s^*(\lambda,p)v(p))^2+\lambda\varepsilon v(p)][\varepsilon+s^*(\lambda,p)v(p)+\lambda v(p)]} - 
\frac{\lambda qs^*(\lambda,p)v'(p)}{\varepsilon+s^*(\lambda,p)v(p)+\lambda v(p)}
.}\]

\hfill

Note that this also gives us the following identity:
\[s_{1y}^*(\lambda,p,p,q)=
\frac{q}{1-q}s_{0y}^*(\lambda,p,p,q) - 
\frac{\lambda qs^*(\lambda,p)v'(p)}{\varepsilon+s^*(\lambda,p)v(p)+\lambda v(p)}.
\]

\subsection{Derivation of \tps{$D^{LR}_{0x}(\lambda,p,p,q)$}, \tps{$D^{LR}_{0y}(\lambda,p,p,q)$}, \tps{$D^{LR}_{1x}(\lambda,p,p,q)$} and \tps{$D^{LR}_{1y}(\lambda,p,p,q)$}, the demand derivatives in LR settings\label{sec: LR_demand_derivs}}

First, note that by using equations (\ref{eq: exp_avail_list}) and (\ref{eq: LR_Demand_Functions}) we can express the listing-side demand functions as linear functions of the steady-state mass of available listings as follows:
\begin{equation}
\begin{aligned}
\label{eq: Demand_lin_mass}
    D^{LR}_0(\lambda,p,p,q) &= \rho - \frac{s_0^*(\lambda,p,p,q)}{1-q}, \\
    D^{LR}_1(\lambda,p,p,q) &= \rho - \frac{s_1^*(\lambda,p,p,q)}{q}.
\end{aligned}
\end{equation}

In Appendix \ref{sec: s0y_s1y_deriv} we found closed form solutions for $s_{0y}^*(\lambda,p,p,q)$ and $s_{1y}^*(\lambda,p,p,q)$. Using these formulas along with equations (\ref{eq: Demand_lin_mass}) we can write $D^{LR}_{0y}(\lambda,p,p,q)$ as follows:
\[\boxed{D^{LR}_{0y}(\lambda,p,p,q) = -\frac{\lambda q{s^*(\lambda,p)}^2v'(p)v(p)(\varepsilon+s^*(\lambda,p)v(p))}{[(\varepsilon+s^*(\lambda,p_0)v(p))^2+\lambda\varepsilon v(p)][\varepsilon+s^*(\lambda,p_0)v(p)+\lambda v(p)]}.}\]

And similarly $D^{LR}_{1y}(\lambda,p,p,q)$ as such:
\[\boxed{
    D^{LR}_{1y}(\lambda,p,p,q) = 
-\frac{\lambda q{s^*(\lambda,p)}^2v'(p)v(p)(\varepsilon+s^*(\lambda,p)v(p)) }{[(\varepsilon+s^*(\lambda,p)v(p))^2+\lambda\varepsilon v(p)][\varepsilon+s^*(\lambda,p)v(p)+\lambda v(p)]} + 
\frac{\lambda s^*(\lambda,p)v'(p)}{\varepsilon+s^*(\lambda,p)v(p)+\lambda v(p)}.
}\]

\quad\\

Note that if we examine our equation for $\widehat{e_p}(p,q)$ (\ref{eq: exp_elast}), we can use our derivations of $D^{LR}_{1y}(\lambda,p,p,q)$ and $D^{LR}_{1y}(\lambda,p,p,q)$ to solve for $\widehat{e_p}^{LR}(p,q)$, giving us that:
\begin{equation}
\label{eq: LR_exp_elast}
    \widehat{e_p}^{LR}(p,q,\lambda)=\frac{p\lambda s^*(\lambda,p)v'(p)}{(\varepsilon+s^*(\lambda,p)v(p)+\lambda v(p))(\rho-s^*(\lambda,p))}.
\end{equation}
\hfill\break

Next we need to solve for $D^{LR}_{0x}(\lambda,p,p,q)$ and $D^{LR}_{1x}(\lambda,p,p,q)$. We will do so again using equations (\ref{eq: Demand_lin_mass}) but first we need to define $s_{0x}^*(\lambda,p,p,q)$ and $s_{1x}^*(\lambda,p,p,q)$. Note that by symmetry in the setup we must have that $s_{0x}^*(\lambda,p,p,q) = s_{1y}^*(\lambda,p,p,(1-q))$ and $s_{1x}^*(\lambda,p,p,q) = s_{0y}^*(\lambda,p,p,(1-q))$. Using these facts allows us to write the following intermediate formula for $s_{0x}^*(\lambda,p,p,q)$:
\[s_{0x}^*(\lambda,p,p,q) = \frac{\lambda (1-q)^2{s^*(\lambda,p)}^2v'(p)v(p)(\varepsilon+s^*(\lambda,p)v(p)) }{[(\varepsilon+s^*(\lambda,p)v(p))^2+\lambda\varepsilon v(p)][\varepsilon+s^*(\lambda,p)v(p)+\lambda v(p)]} - 
\frac{\lambda (1-q)s^*(\lambda,p)v(p)'}{\varepsilon+s^*(\lambda,p)v(p)+\lambda v(p)},\]
and similarly for $s_{1x}^*(\lambda,p,p,q)$:
\[s_{1x}^*(\lambda,p,p,q) =  \frac{\lambda q(1-q){s^*(\lambda,p)}^2v'(p)v(p)(\varepsilon+s^*(\lambda,p)v(p))}{[(\varepsilon+s^*(\lambda,p_0)v(p))^2+\lambda\varepsilon v(p)][\varepsilon+s^*(\lambda,p_0)v(p)+\lambda v(p)]}.\]

Using the above closed form solutions along with equations (\ref{eq: Demand_lin_mass}) we can write $D^{LR}_{0x}(\lambda,p,p,q)$ as follows:
\[\boxed{
    D^{LR}_{0x}(\lambda,p,p,q) = 
-\frac{\lambda (1-q){s^*(\lambda,p)}^2v'(p)v(p)(\varepsilon+s^*(\lambda,p)v(p)) }{[(\varepsilon+s^*(\lambda,p)v(p))^2+\lambda\varepsilon v(p)][\varepsilon+s^*(\lambda,p)v(p)+\lambda v(p)]} + 
\frac{\lambda s^*(\lambda,p)v(p)'}{\varepsilon+s^*(\lambda,p)v(p)+\lambda v(p)}.
}\]

And similarly $D^{LR}_{1x}(\lambda,p,p,q)$ as such:
\[\boxed{D^{LR}_{1x}(\lambda,p,p,q) = -\frac{\lambda (1-q){s^*(\lambda,p)}^2v'(p)v(p)(\varepsilon+s^*(\lambda,p)v(p))}{[(\varepsilon+s^*(\lambda,p_0)v(p))^2+\lambda\varepsilon v(p)][\varepsilon+s^*(\lambda,p_0)v(p)+\lambda v(p)]}.}\]

\subsection{Derivation of \tps{$s^*_x(\lambda,p,p,q)$} and \tps{$s^*_y(\lambda,p,p,q)$} in CR settings \label{sec: CR_s_deriv}}

First we will derive $s^*_x(\lambda,p,p,q)$. We begin by restating equation (\ref{eq: CR_Market_Bal}):
\[\rho-s^*(\lambda,p_0,p_1,q)=\lambda \left[ q \frac{s^*(\lambda,p_0,p_1,q) v(p_1)}{\varepsilon+s^*(\lambda,p_0,p_1,q)v(p_1)} +(1-q) \frac{s^*(\lambda,p_0,p_1,q) v(p_0)}{\varepsilon+s^*(\lambda,p_0,p_1,q)v(p_0)} \right].
\]

 Next, we will differentiate both sides by $p_0$. Note that in writing this section only we will drop all arguments from the function $s^*$ as a shorthand notation. We proceed as follows:
\[\begin{aligned}
    -s^*_x&=\lambda q \frac{(\varepsilon+s^*v(p_1))(s^*_x v(p_1)) - (s^* v(p_1))(s^*_xv(p_1))}{(\varepsilon+s^*v(p_1))^2} \\
&\;\;\;+\lambda(1-q) \frac{(\varepsilon+s^*v(p_0))(s^*_x v(p_0)+s^* v'(p_0)) - (s^* v(p_0))(s^*_x v(p_0)+s^* v'(p_0))}{(\varepsilon+s^*v(p_0))^2} ,\\
-s^*_x&=\varepsilon\lambda q \frac{s^*_x v(p_1) }{(\varepsilon+s^*v(p_1))^2} +\varepsilon\lambda(1-q) \frac{s^*_x v(p_0)+s^* v'(p_0) }{(\varepsilon+s^*v(p_0))^2} .
\end{aligned}
\]

Our next step is to evaluate this quantity at $p_0=p_1=p$. Note that when we do this we will also drop the argument from the value function $v$ because it is implied. When we do this we can simplify as follows:
\[\begin{aligned}
    -s^*_x=&\varepsilon\lambda q \frac{s^*_x v }{(\varepsilon+s^*v)^2} +\varepsilon\lambda(1-q) \frac{s^*_x v+s^* v' }{(\varepsilon+s^*v)^2},\\
    -s^*_x(\varepsilon+s^*v)^2 &= \varepsilon\lambda qs^*_x v + \varepsilon\lambda(1-q)s^*_x v +\varepsilon\lambda(1-q)s^* v'\\
    -s^*_x(\varepsilon+s^*v)^2 &= \varepsilon\lambda s^*_x v +\varepsilon\lambda(1-q)s^* v'\\
    -s^*_x[(\varepsilon+s^*v)^2+ \varepsilon\lambda v] &= \varepsilon\lambda(1-q)s^* v',\\
\end{aligned}\]
which by dividing across simplifies to our final formula for $s^*_x(\lambda,p,p,q)$:
\[\boxed{s^*_x(\lambda,p,p,q) = -\frac{\varepsilon\lambda(1-q)s^*(\lambda,p,p,q) v'(p)}{(\varepsilon+s^*(\lambda,p,p,q)v(p))^2+ \varepsilon\lambda v(p)}.}\]

\hfill \break

Next, we will follow a similar process to derive $s^*_y(\lambda,p,p,q)$. We will again begin at equation (\ref{eq: CR_Market_Bal}), this time differentiating both sides by $p_1$. Note that in writing this section only we will once again drop all arguments from the function $s^*$ as a shorthand notation. We proceed as follows:
\[\begin{aligned}
    -s^*_y&=\lambda q \frac{(\varepsilon+s^*v(p_1))(s^*_y v(p_1)+s^* v'(p_1)) - (s^* v(p_1))(s^*_y v(p_1)+s^* v'(p_1))}{(\varepsilon+s^*v(p_1))^2} \\
&\;\;\;+\lambda(1-q) \frac{(\varepsilon+s^*v(p_0))(s^*_y v(p_0)) - (s^* v(p_0))(s^*_yv(p_0))}{(\varepsilon+s^*v(p_0))^2} ,\\
-s^*_y&=\varepsilon\lambda q \frac{s^*_y v(p_1)+s^* v'(p_1) }{(\varepsilon+s^*v(p_1))^2} +\varepsilon\lambda(1-q) \frac{s^*_y v(p_0) }{(\varepsilon+s^*v(p_0))^2} .
\end{aligned}
\]

Our next step is to evaluate the above quantity at $p_0=p_1=p$. Note that when we do this we will again drop the argument from the value function $v$ because it is now implied. When we do this we can simplify as follows:
\[\begin{aligned}
    -s^*_y&=\varepsilon\lambda q \frac{s^*_y v+s^* v' }{(\varepsilon+s^*v)^2} +\varepsilon\lambda(1-q) \frac{s^*_y v }{(\varepsilon+s^*v)^2},\\
    -s^*_y(\varepsilon+s^*v)^2 &=
    \varepsilon\lambda qs^*_y v +\varepsilon\lambda qs^* v' +
    \varepsilon\lambda(1-q)s^*_y v, \\
    -s^*_y(\varepsilon+s^*v)^2 &=
    \varepsilon\lambda s^*_y v +\varepsilon\lambda qs^* v' , \\
    -s^*_y[(\varepsilon+s^*v)^2+ \varepsilon\lambda v] &= \varepsilon\lambda qs^* v' , \\
\end{aligned}\]
which by dividing across simplifies to our final formula for $s^*_y(\lambda,p,p,q)$:
\[\boxed{s^*_y(\lambda,p,p,q) = -\frac{\varepsilon\lambda q s^*(\lambda,p,p,q) v'(p)}{(\varepsilon+s^*(\lambda,p,p,q)v(p))^2+ \varepsilon\lambda v(p)}.}\]

\subsection{Derivation of \tps{$D^{CR}_{0x}(\lambda,p,p,q)$, $D^{CR}_{0y}(\lambda,p,p,q)$, $D^{CR}_{1x}(\lambda,p,p,q)$ and $D^{CR}_{1y}(\lambda,p,p,q)$}, the demand derivatives in CR settings\label{sec: demand_der_CR}}

These derivations will use both the closed-form formulas for $s^*_x(\lambda,p,p,q)$ and $s^*_y(\lambda,p,p,q)$ that were derived in appendix \ref{sec: CR_s_deriv}, as well as equation (\ref{eq: CR_Demand_Functions}). We begin by restating equations (\ref{eq: CR_Demand_Functions}):
\[
\begin{aligned}
D^{CR}_0(\lambda,p_0,p_1,q)&=\lambda \frac{s^*(\lambda,p_0,p_1,q)v(p_0)}{\varepsilon+s^*(\lambda,p_0,p_1,q)v(p_0)},\\
D^{CR}_1(\lambda,p_0,p_1,q)&=\lambda \frac{s^*(\lambda,p_0,p_1,q)v(p_1)}{\varepsilon+s^*(\lambda,p_0,p_1,q)v(p_1)}.
\end{aligned}
\]

We will first solve for $D^{CR}_{0x}(\lambda,p,p,q)$. Taking the derivative of the first of equations (\ref{eq: CR_Demand_Functions}) with respect to $p_0$ we get the following:

\[\begin{aligned}
    D^{CR}_{0x}(\lambda,p_0,p_1,q)&=\lambda \frac{(\varepsilon+s^*(\lambda,p_0,p_1,q)v(p_0))(s^*_x(\lambda,p_0,p_1,q)v(p_0)+s^*(\lambda,p_0,p_1,q)v'(p_0))}{(\varepsilon+s^*(\lambda,p_0,p_1,q)v(p_0))^2}\\
    &\;\;\;- \lambda \frac{(s^*(\lambda,p_0,p_1,q)v(p_0))(s^*_x(\lambda,p_0,p_1,q)v(p_0)+s^*(\lambda,p_0,p_1,q)v'(p_0))}{(\varepsilon+s^*(\lambda,p_0,p_1,q)v(p_0))^2},\\
    D^{CR}_{0x}(\lambda,p_0,p_1,q)&=\varepsilon\lambda \frac{s^*_x(\lambda,p_0,p_1,q)v(p_0)+s^*(\lambda,p_0,p_1,q)v'(p_0)}{(\varepsilon+s^*(\lambda,p_0,p_1,q)v(p_0))^2}.\\
\end{aligned}\]

Next, we will evaluate at $p_0=p_1=p$:
\[\begin{aligned}
    D^{CR}_{0x}(\lambda,p,p,q)&=\varepsilon\lambda \frac{s^*_x(\lambda,p,p,q)v(p)+s^*(\lambda,p,p,q)v'(p)}{(\varepsilon+s^*(\lambda,p,p,q)v(p))^2},\\
    D^{CR}_{0x}(\lambda,p,p,q)&=\varepsilon\lambda \frac{s^*_x(\lambda,p,p,q)v(p)}{(\varepsilon+s^*(\lambda,p,p,q)v(p))^2}
    +\varepsilon\lambda \frac{s^*(\lambda,p,p,q)v'(p)}{(\varepsilon+s^*(\lambda,p,p,q)v(p))^2}.\\
\end{aligned}\]

Plugging in our formula for $s^*_x(\lambda,p,p,q)$ gives us our final formula for $D^{CR}_{0x}(\lambda,p,p,q)$:
\[\boxed{
    D^{CR}_{0x}(\lambda,p,p,q)=-\frac{\varepsilon^2\lambda^2(1-q)s^*(\lambda,p,p,q) v'(p)v(p)}
    {[\varepsilon+s^*(\lambda,p,p,q)v(p)]^4+ \varepsilon\lambda v(p)[\varepsilon+s^*(\lambda,p_0,p_1,q)v(p_0)]^2}
    +
    \frac{\varepsilon\lambda s^*(\lambda,p,p,q)v'(p)}{(\varepsilon+s^*(\lambda,p,p,q)v(p))^2}.
    }\]

\hfill\break

We will next solve for $D^{CR}_{0y}(\lambda,p,p,q)$. Taking the derivative of the first of equations (\ref{eq: CR_Demand_Functions}) with respect to $p_1$ we get the following:
\[\begin{aligned}
    D^{CR}_{0y}(\lambda,p_0,p_1,q)&=\lambda \frac{(\varepsilon+s^*(\lambda,p_0,p_1,q)v(p_0))(s^*_y(\lambda,p_0,p_1,q)v(p_0))}{(\varepsilon+s^*(\lambda,p_0,p_1,q)v(p_0))^2}\\
    &\;\;\;- \lambda \frac{(s^*(\lambda,p_0,p_1,q)v(p_0))(s^*_y(\lambda,p_0,p_1,q)v(p_0))}{(\varepsilon+s^*(\lambda,p_0,p_1,q)v(p_0))^2}\\
    D^{CR}_{0y}(\lambda,p_0,p_1,q)&=\varepsilon\lambda \frac{s^*_y(\lambda,p_0,p_1,q)v(p_0)}{(\varepsilon+s^*(\lambda,p_0,p_1,q)v(p_0))^2}\\
\end{aligned}\]

All we need to do now is plug in $p_0=p_1=p$, as well as our formula for $s^*_y(\lambda,p,p,q)$. This allows us to write our final formula for $D^{CR}_{0y}(\lambda,p,p,q)$ as follows:
\[\boxed{D^{CR}_{0y}(\lambda,p,p,q)=-\frac{\varepsilon^2\lambda^2 q s^*(\lambda,p,p,q) v'(p)v(p)}{[\varepsilon+s^*(\lambda,p,p,q)v(p)]^4+ \varepsilon\lambda v(p)[\varepsilon+s^*(\lambda,p_0,p_1,q)v(p_0)]^2}.}\]

We now move on to the treatment demand $D^{CR}_1$. We will first solve for $D^{CR}_{1x}(\lambda,p,p,q)$. Taking the derivative of the second of equations (\ref{eq: CR_Demand_Functions}) with respect to $p_0$ we get the following:

\[\begin{aligned}
    D^{CR}_{1x}(\lambda,p_0,p_1,q)&=\lambda \frac{(\varepsilon+s^*(\lambda,p_0,p_1,q)v(p_1))(s^*_x(\lambda,p_0,p_1,q)v(p_1))}{(\varepsilon+s^*(\lambda,p_0,p_1,q)v(p_1))^2}\\
    &\;\;\;- \lambda \frac{(s^*(\lambda,p_0,p_1,q)v(p_1))(s^*_x(\lambda,p_0,p_1,q)v(p_1))}{(\varepsilon+s^*(\lambda,p_0,p_1,q)v(p_1))^2}\\
    D^{CR}_{1x}(\lambda,p_0,p_1,q)&=\varepsilon\lambda \frac{s^*_x(\lambda,p_0,p_1,q)v(p_1)}{(\varepsilon+s^*(\lambda,p_0,p_1,q)v(p_1))^2}\\
\end{aligned}\]

All we need to do now is plug in $p_0=p_1=p$, as well as our formula for $s^*_x(\lambda,p,p,q)$. This allows us to write our final formula for $D^{CR}_{1x}(\lambda,p,p,q)$ as follows:
\[\boxed{
D^{CR}_{1x}(\lambda,p,p,q)=
-\frac{\varepsilon^2\lambda^2(1-q)s^*(\lambda,p,p,q) v'(p)v(p)}{[\varepsilon+s^*(\lambda,p,p,q)v(p)]^4+ \varepsilon\lambda v(p)[\varepsilon+s^*(\lambda,p_0,p_1,q)v(p_0)]^2}
}\]

\hfill\break

Finally, we will solve for $D^{CR}_{1y}(\lambda,p,p,q)$. Taking the derivative of the second of equations (\ref{eq: CR_Demand_Functions}) with respect to $p_1$ we get the following:
\[\begin{aligned}
    D^{CR}_{1y}(\lambda,p_0,p_1,q)&=\lambda \frac{(\varepsilon+s^*(\lambda,p_0,p_1,q)v(p_1))(s^*_y(\lambda,p_0,p_1,q)v(p_1)+s^*(\lambda,p_0,p_1,q)v'(p_1))}{(\varepsilon+s^*(\lambda,p_0,p_1,q)v(p_1))^2}\\
    &\;\;\; - \lambda \frac{(s^*(\lambda,p_0,p_1,q)v(p_1))(s^*_y(\lambda,p_0,p_1,q)v(p_1)+s^*(\lambda,p_0,p_1,q)v'(p_1))}{(\varepsilon+s^*(\lambda,p_0,p_1,q)v(p_1))^2},\\
    D^{CR}_{1y}(\lambda,p_0,p_1,q)&= \varepsilon\lambda \frac{s^*_y(\lambda,p_0,p_1,q)v(p_1)+s^*(\lambda,p_0,p_1,q)v'(p_1)}{(\varepsilon+s^*(\lambda,p_0,p_1,q)v(p_1))^2}\\
\end{aligned}\]

Next, we will evaluate at $p_0=p_1=p$:
\[\begin{aligned}
    D^{CR}_{1y}(\lambda,p,p,q)&=\varepsilon\lambda \frac{s^*_y(\lambda,p,p,q)v(p)+s^*(\lambda,p,p,q)v'(p)}{(\varepsilon+s^*(\lambda,p,p,q)v(p))^2},\\
    D^{CR}_{1y}(\lambda,p,p,q)&=\varepsilon\lambda \frac{s^*_y(\lambda,p,p,q)v(p)}{(\varepsilon+s^*(\lambda,p,p,q)v(p))^2}
    +\varepsilon\lambda \frac{s^*(\lambda,p,p,q)v'(p)}{(\varepsilon+s^*(\lambda,p,p,q)v(p))^2}.\\
\end{aligned}\]

Our final step is to plug in our formula for $s^*_y(\lambda,p,p,q)$ which gives us our final formula for $D^{CR}_{1y}(\lambda,p,p,q)$:
\[\boxed{D^{CR}_{1y}(\lambda,p,p,q) = 
-\frac{\varepsilon^2\lambda^2qs^*(\lambda,p,p,q) v'(p)v(p)}{[\varepsilon+s^*(\lambda,p,p,q)v(p)]^4+ \varepsilon\lambda v(p)[\varepsilon+s^*(\lambda,p_0,p_1,q)v(p_0)]^2}
+ \frac{\varepsilon\lambda s^*(\lambda,p,p,q)v'(p)}{(\varepsilon+s^*(\lambda,p,p,q)v(p))^2}}\]

\quad \\

Note that as was the case for LR, if we examine our equation for $\widehat{e_p}(p,q)$ (\ref{eq: exp_elast}), we can use our derivations of $D^{CR}_{1y}(\lambda,p,p,q)$ and $D^{CR}_{1y}(\lambda,p,p,q)$ to solve for $\widehat{e_p}^{CR}(p,q)$, giving us that:
\begin{equation}
\label{eq: CR_exp_elast}
    \widehat{e_p}^{CR}(p,q,\lambda)=\frac{p\varepsilon\lambda s^*(\lambda,p,p,q)v'(p)}{(\varepsilon+s^*(\lambda,p,p,q)v(p))^2(\rho-s^*(\lambda,p))}.
\end{equation}
\clearpage

\section{Proofs}
\label{app:proofs}

\subsection{Proof \tps{$D(\lambda,p)$} satisfies Assumption \ref{as: demand}}
\label{sec: two_side_demand_assumption_4}

Our first step is to calculate the derivative of $s^*(\lambda,p)$ with respect to $p$. We can do this by differentiating the market balance equation (\ref{eq: market_balance}) as follows:
\begin{equation}
\notag
\begin{aligned}
    \frac{\partial}{\partial p}\left(\rho-s^*(\lambda,p)\right)&=\frac{\partial}{\partial p}\left(\lambda\frac{s^*(\lambda,p)v(p)}{\varepsilon+s^*(\lambda,p)v(p)}\right)
    \\
    -s_p^*(\lambda,p)&=\lambda
    \frac{[\frac{\partial}{\partial p}(s^*(\lambda,p)v(p))](\varepsilon+s^*(\lambda,p)v(p)) - (s^*(\lambda,p)v(p))[\frac{\partial}{\partial p}(s^*(\lambda,p)v(p))]}
    {(\varepsilon+s^*(\lambda,p)v(p))^2}
    \\
    -s_p^*(\lambda,p)&=\lambda
    \frac{\varepsilon(s_p^*(\lambda,p)v(p)+s^*(\lambda,p)v'(p))}
    {(\varepsilon+s^*(\lambda,p)v(p))^2}
    \\
    -s_p^*(\lambda,p)(\varepsilon+s^*(\lambda,p)v(p))^2 & = \varepsilon\lambda s_p^*(\lambda,p)v(p)+\varepsilon\lambda s^*(\lambda,p)v'(p)
    \\
    s_p^*(\lambda,p)  &= -\frac{ \varepsilon\lambda s^*(\lambda,p) v'(p)}
    {(\varepsilon+s^*(\lambda,p)v(p))^2 + \varepsilon\lambda v(p)}.
    \\
\end{aligned}
\end{equation}

From the market balance equation (\ref{eq: market_balance}) we can write $D(\lambda,p)=\rho-s^*(\lambda,p)$. And from there; $\frac{\partial}{\partial p}D(\lambda ,p) =\frac{\partial}{\partial p} (\rho-s^*(\lambda,p))=-s^*_p(\lambda,p)$. If we inspect the quotient in $s_p(\lambda,p)$ we can note that it's numerator is always negative. This is because $\varepsilon,\lambda$, and $s^*(\lambda,p)$ are all constrained to be positive, while $v'(p)<0$ by assumption \ref{as: valuation_function}. The denominator on the other hand is always positive because it is the sum of a square, and the product of $\varepsilon,\lambda$ and $v(p)$, all of which are positive. Since the whole quotient is negated, we must have that $s_p^*(\lambda,p)> 0$ and thus, that $\frac{\partial}{\partial p}D(\lambda ,p)< 0$. We can therefore conclude that the market demand function $D(\lambda,p)$ in the mean-field logit choice model satisfies Assumption \ref{as: demand}.

\quad\\
\subsection{Proof that \tps{$D_0^{LR}(\lambda,p_0,p_1,q)$} and \tps{$D_1^{LR}(\lambda,p_0,p_1,q)$} satisfy Assumption \ref{as: control_treat_demand}}
\label{sec: LR_assmption_6}
To prove this result we will first show that $s^*_0(\lambda,p_0,p_1,q)$ is non-increasing in $p_1$, and that $s^*_1(\lambda,p_0,p_1,q)$ is non-increasing in $p_0$.\\

First, assume for the sake of contradiction that $s^*_0(\lambda,p_0,p_1,q)$ is increasing in $p_1$. Recall the first LR balance equation (\ref{eq: exp_avail_list}):
\[(1-q)\rho-s_0^*(\lambda,p_0,p_1,q)=\lambda \frac{s_0^*(\lambda,p_0,p_1,q)v(p_0)}{\varepsilon+s_0^*(\lambda,p_0,p_1,q)v(p_0)+s_1^*(\lambda,p_0,p_1,q)v(p_1)}.\]

The LHS of this equation is decreasing in $p_1$ by initial assumption, which implies that the RHS must be as well. Looking at the RHS, we see that the numerator is increasing in $p_1$ which means that for the full quotient to be decreasing in $p_1$, the denominator must be increasing in $p_1$ strictly faster than the numerator is. This tells us that $s_1^*(\lambda,p_0,p_1,q)v(p_1)$ must be strictly increasing in $p_1$, and since $v(p_1)$ is strictly decreasing in $p_1$ by Assumption $\ref{as: valuation_function}$, we must have that $s_1^*(\lambda,p_0,p_1,q)$ is increasing in $p_1$.\\

We have now shown that $s^*_0(\lambda,p_0,p_1,q)$ increasing in $p_1$ implies $s_1^*(\lambda,p_0,p_1,q)$ is increasing in $p_1$ too. To see why this is a contradiction, consider the sum of the two LR balance equations (\ref{eq: exp_avail_list}):
\[\rho-s_0^*(\lambda,p_0,p_1,q)-s_1^*(\lambda,p_0,p_1,q)=
\lambda \frac{s_0^*(\lambda,p_0,p_1,q)v(p_0)+
s_1^*(\lambda,p_0,p_1,q)v(p_1)}{\varepsilon+s_0^*(\lambda,p_0,p_1,q)v(p_0)+s_1^*(\lambda,p_0,p_1,q)v(p_1)}.\]

Since both $s^*_0(\lambda,p_0,p_1,q)$ and $s^*_1(\lambda,p_0,p_1,q)$ are increasing in $p_1$, the LHS is decreasing in $p_1$ so the RHS must be as well. This implies that $s_0^*(\lambda,p_0,p_1,q)v(p_0)+
s_1^*(\lambda,p_0,p_1,q)v(p_1)$ must be decreasing in $p_1$. 

Now, consider the second of the LR balance equations (\ref{eq: exp_avail_list}):
\[q\rho-s_1^*(\lambda,p_0,p_1,q)=
\lambda \frac{
s_1^*(\lambda,p_0,p_1,q)v(p_1)}{\varepsilon+s_0^*(\lambda,p_0,p_1,q)v(p_0)+s_1^*(\lambda,p_0,p_1,q)v(p_1)}.\]

Examining the RHS, we know that the numerator $s_1^*(\lambda,p_0,p_1,q)v(p_1)$ must be increasing in $p_1$, and the denominator $s_0^*(\lambda,p_0,p_1,q)v(p_0)+
s_1^*(\lambda,p_0,p_1,q)v(p_1)$ must be decreasing in $p_1$. This implies that the RHS must be increasing in $p_1$. The LHS however is decreasing in $p_1$, hence we arrive at a contradiction; implying that our initial assumption must have been false. We have thus proven that $s^*_0(\lambda,p_0,p_1,q)$ is non-increasing in $p_1$. It is not hard to see that by symmetry, we can extend this argument to state that $s^*_1(\lambda,p_0,p_1,q)$ is non-increasing in $p_0$.\\

Our final step is to note that by using the LR balance equations (\ref{eq: exp_avail_list}), we can rewrite the listing side control and treatment demand definitions (\ref{eq: LR_Demand_Functions}) as follows:
\[
\begin{aligned}
    D_0^{LR}(\lambda,p_0,p_1,q) &= \rho - \frac{s^*_0(\lambda,p_0,p_1,q)}{1-q},\text{ and}\\
D_1^{LR}(\lambda,p_0,p_1,q) &= \rho - \frac{s^*_1(\lambda,p_0,p_1,q)}{q}
.\\
\end{aligned}
\]

We then simply use our first result---that $s^*_0(\lambda,p_0,p_1,q)$ is non-increasing in $p_1$ and $s^*_1(\lambda,p_0,p_1,q)$ is non-increasing in $p_0$---to state that $D_0^{LR}(\lambda,p_0,p_1,q)$ is non-decreasing in $p_1$, and that $D_1^{LR}(\lambda,p_0,p_1,q)$ is non-decreasing in $p_0$, concluding our proof that $D_0^{LR}(\lambda,p_0,p_1,q)$ and $D_1^{LR}(\lambda,p_0,p_1,q)$ satisfy Assumption \ref{as: control_treat_demand}.
\newline

\subsection{Proof that \tps{$D_0^{CR}(\lambda,p_0,p_1,q)$} and \tps{$D_1^{CR}(\lambda,p_0,p_1,q)$} satisfy Assumption \ref{as: control_treat_demand}}
\label{sec: CR_assmption_6}

Our first step in this proof is going to be to demonstrate that $s^*(\lambda,p_0,p_1,q)$ is non-decreasing in $p_1$. 
Consider the following proof by contradiction: assume that $s^*(\lambda,p_0,p_1,q)$ is strictly decreasing in $p_1$. Recall that the CR market balance equation (\ref{eq: CR_Market_Bal}) is written as follows:
$$\rho-s^*(\lambda,p_0,p_1,q)=\lambda \left[ q \frac{s^*(\lambda,p_0,p_1,q) v(p_1)}{\varepsilon+s^*(\lambda,p_0,p_1,q)v(p_1)} +(1-q) \frac{s^*(\lambda,p_0,p_1,q) v(p_0)}{\varepsilon+s^*(\lambda,p_0,p_1,q)v(p_0)} \right].
    $$

Because we have assumed that $s^*(\lambda,p_0,p_1,q)$ is strictly decreasing in $p_1$ the LHS is strictly increasing in $p_1$, implying that the RHS must be as well. Since all terms in the RHS are positive and additive, for the RHS to be increasing in $p_1$ it must be that at least one of it's quotients is increasing in $p_1$. The first quotient ($\frac{s^*(\lambda,p_0,p_1,q) v(p_1)}{\varepsilon+s^*(\lambda,p_0,p_1,q)v(p_1)}$), is decreasing in $p_1$ because both $s^*(\lambda,p_0,p_1,q)$ and $v(p_1)$ are decreasing in $p_1$ by assumption, and $\frac{a-c}{b-c}<\frac{a}{b}\;\forall\;a,b,c>0$ whenever $a<b$. Similarly, if we consider the second quotient ($\frac{s^*(\lambda,p_0,p_1,q) v(p_0)}{\varepsilon+s^*(\lambda,p_0,p_1,q)v(p_0)}$) we know that $s^*(\lambda,p_0,p_1,q)$ is decreasing in $p_1$ by assumption and $v(p_0)$ is constant so we can again use that fact that $\frac{a-c}{b-c}<\frac{a}{b}\;\forall\;a,b,c>0$ whenever $a<b$ to say that the second quotient is decreasing in $p_1$. This of course implies that the RHS is decreasing in $p_1$, which contradicts our initial assumption, proving that $s^*(\lambda,p_0,p_1,q)$ is non-decreasing in $p_1$. 

Using symmetry in the market balance equation we can expand this argument to state that  $s^*(\lambda,p_0,p_1,q)$ is also non-decreasing in $p_0$. \\

Now, recall customer side control and treatment demand definitions (\ref{eq: CR_Demand_Functions}):

$$
\begin{aligned}
D^{CR}_0(\lambda,p_0,p_1,q)&=\lambda \frac{s^*(\lambda,p_0,p_1,q)v(p_0)}{\varepsilon+s^*(\lambda,p_0,p_1,q)v(p_0)}, \text{ and}\\
D^{CR}_1(\lambda,p_0,p_1,q)&=\lambda \frac{s^*(\lambda,p_0,p_1,q)v(p_1)}{\varepsilon+s^*(\lambda,p_0,p_1,q)v(p_1)}.
\end{aligned}$$

Since $s^*(\lambda,p_0,p_1,q)$ is non-decreasing in both $p_0$ and $p_1$, we can use the fact that $\frac{a}{b}<\frac{a+c}{b+c}\;\forall\;a,b,c>0$ whenever $a<b$ to state that $D^{CR}_0(\lambda,p_0,p_1,q)$ is non-decreasing in $p_1$ and $D^{CR}_1(\lambda,p_0,p_1,q)$ is non-decreasing in $p_0$, completing our proof that $D^{CR}_0(\lambda,p_0,p_1,q)$ and $D^{CR}_1(\lambda,p_0,p_1,q)$ satisfy Assumption \ref{as: control_treat_demand}.
\newline

\subsection{Proof that \tps{$\widehat{e_p}(p,q)$} is monotonically decreasing in \tps{$p$} for $LR$}
\label{sec: hat_ep_LR}
We will start with equation (\ref{eq: exp_elast}) which gave us the following definition of $\widehat{e_p}(p,q)$:
\[\widehat{e_p}(p,q)=\frac{p( D_{1y}(p,p,q) -  D_{0y}(p,p,q) )}{D(p)}.\]

If we now use the equations for $D_{1y}(p,p,q) $ and $  D_{0y}(p,p,q)$ that we derive in appendix \ref{sec: LR_demand_derivs} we see that we rewrite this as follows:
\[
 \widehat{e_p}(p,q) = \frac{p s^*(\lambda, p) v'(p)}{ (\epsilon + s^*(\lambda, p) v(p) + \lambda v(p))( \rho - s^*(\lambda, p))}. 
\]

Note that this can be rewritten as:
\begin{equation}
\label{eq:ehat_LR_arranged}
 -\widehat{e_p}(p,q) = \frac{-pv'(p)}{v(p)}\times s^*(\lambda, p)\times\frac{1 }{ (\epsilon/v(p) + s^*(\lambda, p) + \lambda)( \rho - s^*(\lambda, p))}.
\end{equation}

Looking at each term in equation (\ref{eq:ehat_LR_arranged}) we can see that the first two terms are monotonically increasing in $p$, $-pv'(p)/v(p)$ by assumption \ref{as: valuation_function}, and $s^*(\lambda, p)$ by proof \ref{sec: two_side_demand_assumption_4}. We can now focus our attention to denominator of the third term:
\[(\epsilon/v(p) + s^*(\lambda, p) + \lambda)( \rho - s^*(\lambda, p)).\]

If we use equation (\ref{eq: market_balance}) we can rewrite this as follows:
\[\left( \frac{\epsilon}{v(p)} + s(p)+ \lambda \right) \left( \frac{ \lambda s(p) v(p)} { \epsilon + v(p) s(p)} \right),\]
\begin{equation}
    \label{eq: intermediate_LR_ehat_den}= \frac{ \lambda s(p) \epsilon + \lambda v(p) s(p)^2 + \lambda^2 s(p) v(p)}{\epsilon + v(p) s(p)}.
\end{equation}

Note now that by rearranging equation (\ref{eq: market_balance}) we get the following identity:
\[  s(p)^2 v(p) + \lambda s(p) v(p) + \epsilon s(p) = \rho (\epsilon + v(p) s(p)). \]

If we substitute this value into equation (\ref{eq: intermediate_LR_ehat_den}) we get that (\ref{eq: intermediate_LR_ehat_den}) simplifies to $\lambda\rho$. This means our third term in equation (\ref{eq:ehat_LR_arranged}) is a constant, and since the other two terms were monotonically increasing, it must be that the product as a whole is. Since $-\widehat{e_p}(p,q)$ is monotonically increasing, $\widehat{e_p}(p,q)$ must be monotonically decreasing, concluding our proof. \qedsymbol
\newline

\subsection{Proof that \tps{$\widehat{e_p}(p,q)$} is monotonically decreasing in \tps{$p$} for $CR$}
in CR settings:\label{sec: hat_ep_CR}
We will start with equation (\ref{eq: exp_elast}) which gave us the following definition of $\widehat{e_p}(p,q)$:
\[\widehat{e_p}(p,q)=\frac{p( D_{1y}(p,p,q) -  D_{0y}(p,p,q) )}{D(p)}.\]

If we now use the equations for $D_{1y}(p,p,q) $ and $  D_{0y}(p,p,q)$ that we derive in appendix \ref{sec: demand_der_CR} we see that we rewrite this as follows:
\[\widehat{e_p}(p,q) = \frac{p \lambda \epsilon s^*(\lambda, p) v'(p)}{ (\epsilon + s^*(\lambda, p) v(p))^2 ( \rho - s^*(\lambda, p))}.\]

If we use equation (\ref{eq: market_balance}) we can rewrite the above equation as the follows:
\begin{equation}
\label{eq:ehat_CR_arranged}
 -\widehat{e_p}(p,q) = \frac{-p v'(p)}{v(p)}\times\frac{\epsilon}{ \epsilon + s^*(\lambda, p) v(p)}
\end{equation}

We already know the first term, $-pv'(p)/v(p)$, is monotonically increasing by assumption \ref{as: valuation_function}. This leaves us with the second term to deal with. Observe that by using equation (\ref{eq: market_balance}) again we can rewrite this second term as:
\[ \frac{\epsilon }{ \epsilon + s^*(\lambda, p) v(p)} = \frac{\epsilon +s^*(\lambda, p) v(p)}{ \epsilon + s^*(\lambda, p) v(p)} - \frac{s^*(\lambda, p) v(p)}{ \epsilon + s^*(\lambda, p) v(p)} = 1 - \frac{\rho - s^*(\lambda, p)}{\lambda}. \]
The far right side equation is monotonically increasing in $p$, since $s^*(p)$ is monotonically increasing in $p$ by proof \ref{sec: two_side_demand_assumption_4}. Since both terms in \ref{eq:ehat_CR_arranged} are monotonically increasing, we have proven $-\widehat{e_p}(p,q)$ is as well, and therefore that $\widehat{e_p}(p,q)$ is monotonically decreasing in $p$, concluding our proof. \qedsymbol
\newline

\subsection{Proof that \tps{$e_p(p)$} is monotonically decreasing in \tps{$p$}}
\label{sec: ep_mono_proof}

Recall that the price elasticity of demand is defined as $e(p) = pD'(p)/D(p)$. Recall that by using equation (\ref{eq: market_balance}) we can state that:
\[ D(\lambda, p) = \rho - s^*(\lambda, p) = \lambda \frac{s^*(\lambda,p)v(p)}{\varepsilon+s^*(\lambda,p)v(p)}. \]

This allows us to rewrite the price elasticity of demand as follows:
\[e(p) = \frac{-p s^*_p(\lambda,p)}{\rho - s^*(\lambda, p)} = \frac{-p s^*_p(\lambda,p)(\varepsilon+s^*(\lambda,p)v(p))}{\lambda s^*(\lambda,p)v(p)}\]

Now, using the formula for $s^*_p(\lambda,p)$ that we find in proof \ref{sec: two_side_demand_assumption_4}, we can substitute to get the following:
\[e(p) = \frac{p(\varepsilon+s^*(\lambda,p)v(p))\varepsilon\lambda s^*(\lambda,p) v'(p)}{\lambda s^*(\lambda,p)v(p)[(\varepsilon+s^*(\lambda,p)v(p))^2 + \varepsilon\lambda v(p)]},\]
which simplifies to:
\[e(p) = \frac{p(\varepsilon+s^*(\lambda,p)v(p))\varepsilon v'(p)}{v(p)[(\varepsilon+s^*(\lambda,p)v(p))^2 + \varepsilon\lambda v(p)]},\]
\[-e(p) = \frac{-pv'(p)}{v(p)}\times \frac{(\varepsilon+s^*(\lambda,p)v(p))\varepsilon }{(\varepsilon+s^*(\lambda,p)v(p))^2 + \varepsilon\lambda v(p)},\]
\begin{equation}
\label{eq: ep_arranged}
-e(p) = \frac{-pv'(p)}{v(p)}\times \frac{\varepsilon }{\varepsilon+s^*(\lambda,p)v(p) + \varepsilon(\rho-s^*(\lambda,p))/s^*(\lambda,p)},
\end{equation}

Where the denominator in the last line was simplified using equation (\ref{eq: market_balance}). Inspecting equation (\ref{eq: ep_arranged}) we see that the first term in monotonically increasing by assumption \ref{as: valuation_function}, this leaves just the denominator of the second term to worry about:
\begin{equation}
\label{eq: ep_right_deno}
    \varepsilon+s^*(\lambda,p)v(p) + \varepsilon\frac{(\rho-s^*(\lambda,p))}{s^*(\lambda,p)}.
\end{equation}

Considering first $\varepsilon+s^*(\lambda,p)v(p)$, we can observe that in appendix \ref{sec: hat_ep_CR} we proved that
\[\frac{\varepsilon}{\varepsilon+s^*(\lambda,p)v(p)}\]
was monotonically increasing which in turn proves that $\varepsilon+s^*(\lambda,p)v(p)$ is monotonically decreasing. Now looking at the second term in equation (\ref{eq: ep_right_deno}), $\varepsilon(\rho-s^*(\lambda,p))/s^*(\lambda,p)$, we see that the numerator is monotonically decreasing in $p$, and the denominator is monotonically increasing in $p$ (both by the proof in appendix \ref{sec: two_side_demand_assumption_4}), and so can conclude that the whole term is monotonically decreasing in $p$. 

This means that the denominator of the second term in equation (\ref{eq: ep_arranged}) is monotonically decreasing in $p$, proving that the second term as a whole is monotonically increasing. Since both terms are monotonically increasing we have proven that $-e(p)$ is too, proving that $e(p)$ is monotonically decreasing in $p$. \qedsymbol
\newline

\subsection{Proof of Theorems \ref{thm: LR_interval} and \ref{thm: CR_interval}}\label{sec: interval_proof}

In this section we show that regardless of experimental setup, and for any market balance parameter $\lambda$, there is a non-empty, closed and convex set $P_{\lambda}\subseteq [c,\infty)$ such that the sign of the canonical estimator for $GTE_\pi(\lambda,p)$ at price $p$ will be wrong if and only if $p\in P_{\lambda}$.\\

First, recall the two conditions for the sign of the canonical estimator to be wrong: condition (a) states that we must have $GTE_\pi(\lambda, p)\geq 0 $, while condition (b) states that we must have $Bias_\pi(p,q)\geq GTE_\pi(\lambda,p)$. In the canonical case $\widehat{GTE}_\pi(p,q) = GTE_\pi(\lambda,p) - Bias_\pi(p,q)$, so when these two conditions are met, the sign of $\widehat{GTE}_\pi(p,q)$ is non-positive while the sign of $GTE_\pi(\lambda,p)$ is non-negative, meaning the estimated sign is wrong.\\

The outline of our proof will go as follows. Take $P^{(a)}_{\lambda}$ and $P^{(b)}_{\lambda}$ to be the set of prices $p\in [c,\infty)$ for which conditions (a) and (b) respectively are met, and let $P_{\lambda}$ be their intersection.
\begin{enumerate}
    \item We first show that there exits $\overline p \in P^{(a)}_{\lambda}$ such that $p\in P^{(a)}_{\lambda}$ if and only if $p\leq\overline p$.
    \item We then show that there exits $\underline p\in P^{(b)}_{\lambda}$ such that $p\in P^{(b)}_{\lambda}$ if and only if $p\geq\underline p$.
    \item Finally, we will show that $\underline p\leq \overline p$.
\end{enumerate}

Steps 1 and 2 prove that $P_{\lambda}$ is convex and closed, but not that it is non-empty. Step 3 shows that there exists some $p'$ such that $\underline p\leq p'\leq \overline p$, implying $p'\in P^{(a)}_{\lambda}$ and $p'\in P^{(b)}_{\lambda}$, which of course means $p'\in P_{\lambda}$, proving $P_{\lambda}$ is non-empty and concluding our proof. We will now prove each of these steps in detail.\\

We will start first with step 1. In section \ref{sec: bias_analysis} we demonstrate that condition (a) is satisfied at price $p'$ if and only if:
\begin{align}
\frac{p'-c}{p'} \leq -\frac{1}{e_p(p')} \tag{a}
\end{align}

Recall also our formula for $e_p(p)$: 
\[e_p(p) = \frac{-p s^*_p(\lambda,p)}{\rho - s^*(\lambda, p)}.\]

Note that by appendix \ref{sec: two_side_demand_assumption_4} we have that over $[c,\infty)$, $s^*(\lambda, p)$ is continuous and $s^*(\lambda, p)\in (0,\rho)$. This means that $e_p(p)$ is continuous and as shown in Appendix \ref{sec: ep_mono_proof}, is monotonically decreasing. This means the function
\[f_a(p) = \frac{p'-c}{p'}+\frac{1}{e_p(p')}\]
is continuous and strictly increasing over $[c,\infty)$. Note that $f_a(c)<0$ and yet $\lim_{p\to\infty}f_a(p) = 1$ so there exists some $p'\in[c,\infty)$ such that $f_a(p')>0$. Using the intermediate value theorem and the definition of strict monotonicity, we can then say that there exists a unique $\overline p $ such that $f_a(\overline p)=0$. Note that condition (a) is satisfied at a price $p$ if and only if $f_a(p)\leq 0$. This means that (a) is satisfied at $\overline p$, meaning $\overline p \in P^{(a)}_{\lambda}$. It also means that condition (a) is satisfied for all $p<\overline p$. We can therefore write $P^{(a)}_{\lambda}$ as $P^{(b)}_{\lambda} = [c,\overline p]$, proving step 1.\\

We now move on to step 2. In section \ref{sec: bias_analysis} we demonstrate that condition (b) is satisfied at price $p'$ if and only if:
\begin{align}
-\frac{1}{\widehat{e_p}(p')}\leq\frac{p'-c}{p'}.\tag{b}
\end{align}

Recall our formulas for $\widehat{e_p}(p)$ in both the LR and CR settings (equations \ref{eq: LR_exp_elast} and \ref{eq: CR_exp_elast} respectively): 
\[\widehat{e_p}^{LR}(p,q,\lambda) = \frac{p s^*(\lambda, p) v'(p)}{ (\epsilon + s^*(\lambda, p) v(p) + \lambda v(p))( \rho - s^*(\lambda, p))}.\]
\[\widehat{e_p}^{CR}(p,q,\lambda) = \frac{p \lambda \epsilon s^*(\lambda, p) v'(p)}{ (\epsilon + s^*(\lambda, p) v(p))^2 ( \rho - s^*(\lambda, p))},\]

By assumption \ref{as: valuation_function} and appendix  \ref{sec: two_side_demand_assumption_4}, we know that $v(p)$, $v'(p)$, and $s^*(\lambda,p)$ are all continuous and that over $[c,\infty)$, $s^*(\lambda,p)\in(0,\rho)$. This means that over $[c,\infty)$, $\widehat{e_p}(p,q)$ is continuous for LR and CR. In appendix sections \ref{sec: hat_ep_LR} and \ref{sec: hat_ep_CR} we prove that $\widehat{e_p}(p)$ is monotonically decreasing in both CR and LR settings--it is also negative. This means the function
\[f_b(p) = \frac{p'-c}{p'}+\frac{1}{\widehat{e_p}(p')}\]
is continuous and strictly increasing over $[c,\infty)$. Using the intermediate value theorem and the definition of strict monotonicity, we can then say that there exists a unique $\underline p $ such that $f_b(\underline p)=0$.  This means that (b) is satisfied at $\underline p$, meaning $\underline p \in P^{(b)}_{\lambda}$. It also means that condition (b) is satisfied for all $p>\underline p$ by monotonicity. We can therefore write $P^{(b)}_{\lambda}$ as $P^{(b)}_{\lambda} = [\underline p, \infty)$, proving step 2 for LR and CR. \\

Finally, we address step 3. We aim to show that $\underline p \leq \overline p$. We show this via proof by contradiction. Imagine $\underline p > \overline p$. This would imply that there exists a $p'\in (\overline p, \underline p)$, where at $p'$ neither condition holds. This means we must have $GTE_\pi(\lambda,p')< 0$ and $Bias_\pi(p',q)< GTE_\pi(\lambda, p')$. This though implies that $Bias_\pi(p',q)<0$, which is a contradiction because under Assumptions \ref{as: diff_D} and \ref{as: control_treat_demand} Bias is non-negative.

 We can therefore conclude that $\underline p \leq \overline p$, which (as argued previously) establishes that $P^{(a)}_{\lambda}\cap P^{(b)}_{\lambda}\neq \emptyset$. This concludes our proof that $P_{\lambda}$ is non-empty, convex, and closed in both the LR and CR experiments.  \qedsymbol

\end{APPENDICES}
\end{document}